\pdfoutput=1

\documentclass[twocolumn]{aastex631}

\usepackage{graphicx}	
\usepackage{amsmath}	
\usepackage{soul}
\usepackage[normalem]{ulem}
\usepackage{float}
\usepackage{multirow}
\usepackage[dvipsnames]{colortbl}
\usepackage[caption=false]{subfig}
\usepackage{wrapfig}
\usepackage{longtable}
\usepackage{hyperref}
\usepackage[normalem]{ulem}
\usepackage{comment}

\graphicspath{{./}{Figures/}}

\newcommand{\mandalamassrange}{$10^{7.53}\leq M_\ast/M_\odot\leq 10^{9.06}$}
\newcommand{\innergrad}{$0 \leq R/R_e \leq 1$}
\newcommand{\outergrad}{$0.75 \leq R/R_e \leq 1.5$}
\newcommand{\medgrad}{$\text{med}\left(\{\nabla \mathcal{G}\}_i\right)$}
\newcommand{\notpnotfivetnotpone}{$\sim0.05\text{–}0.1 \text{ dex/}R_e$}
\newcommand{\notpfivetoone}{$0.5\text{–}1 \text{ dex/}R_e$}
\newcommand{\lesssimnotpone}{$\lesssim0.1 \text{ dex/}R_e$}

\begin{document}

\title{The MaNGA Dwarf Galaxy Sample (MaNDala): stellar profiles and gradients characterization}

\author[0000-0001-9553-8230]{M. Cano-Díaz}
\affiliation{Cátedras CONAHCYT. Universidad Nacional Autónoma de México, Instituto de Astronomía, AP 70-264, 04510, Ciudad de México, México}

\author[0000-0002-0170-5358]{A. Rodríguez-Puebla}
\affiliation{Universidad Nacional Autónoma de México, Instituto de Astronomía, AP 70-264, 04510, Ciudad de México, México}

\author[0000-0002-4216-7138]{A.C. Robleto-Orús}
\affiliation{Centro de Investigación de Astrofísica y Ciencias Espaciales (CIACE), Universidad Nacional Aut\'onoma de Nicaragua, Managua (UNAN-Managua), C.P. 663, Managua, Nicaragua}

\author[0000-0002-3461-2342]{V. \'Avila-Reese}
\affiliation{Universidad Nacional Autónoma de México, Instituto de Astronomía, AP 70-264, 04510, Ciudad de México, México}

\author[0000-0001-9601-7779]{H.M. Hern\'andez-Toledo}
\affiliation{Universidad Nacional Autónoma de México, Instituto de Astronomía, AP 70-264, 04510, Ciudad de México, México}

\author[0000-0001-8694-1204]{J. A. V\'azquez-Mata}
\affiliation{Universidad Nacional Autónoma de México, Instituto de Astronomía, AP 70-264, 04510, Ciudad de México, México}

\author[0000-0002-0523-5509]{O. Valenzuela}
\affiliation{Universidad Nacional Autónoma de México, Instituto de Astronomía, AP 70-264, 04510, Ciudad de México, México}

\author[0000-0002-3724-1583]{J.J. Gonz\'alez}
\affiliation{Universidad Nacional Autónoma de México, Instituto de Astronomía, AP 70-264, 04510, Ciudad de México, México}

\author[0000-0002-9790-6313]{H.J. Ibarra-Medel}
\affiliation{Universidad Nacional Autónoma de México, Instituto de Astronomía, AP 70-264, 04510, Ciudad de México, México}

\author{J.C. Clemente-González}
\affiliation{Universidad Nacional Autónoma de México, Instituto de Astronomía, AP 70-264, 04510, Ciudad de México, México}

\author[0009-0007-5026-7861]{L. Pavón-Alvarez}
\affiliation{Facultad de Ciencias, Universidad Nacional Aut\'onoma de M\'exico, Circuito Exterior S/N, 04510, CDMX, M\'exico}



\begin{abstract}

We derived robust radial profiles and inner/outer gradients of various stellar population (SP) properties for a sample of 124 bright dwarf galaxies, \mandalamassrange , from the MaNDala sample, using integral field spectroscopy observations. Given the complex structure of dwarf galaxies, we explored four different methods to derive SP radial profiles: two based on concentric elliptical rings, and other two exploiting the spatially resolved nature of the data. For each method, we applied four approaches to calculate the following inner (\innergrad) and outer (\outergrad) gradients: luminosity- and mass-weighted age and stellar metallicity, dust attenuation, $D_{n4000}$ index, stellar mass and star formation rate (SFR) surface densities, and specific SFR. While the PSF has a minor impact on the SP gradients, the choice of methodology for characterizing radial profiles significantly affects them. At fixed property, differences in inner gradients from concentric rings methods are typically \notpnotfivetnotpone, while outer gradients can reach \notpfivetoone, relative to the median of all gradients of that property, \medgrad. Spatially resolved methods yield smaller differences, \lesssimnotpone. For some SP gradients, such as $\nabla_\text{SFR}$, the dispersion among the methods is comparable to \medgrad{}. While it is not possible to select a single preferred method for determining SP gradients, we suggest to use \medgrad{} for each SP property. The resulting median age and metallicity suggest that, overall, bright dwarfs experienced moderate inside-out formation, accompanied by significant early SF from low-metallicity gas with outward radial migration of old SPs. The derived SP gradients provide strong constraints on feedback mechanisms in dwarf galaxies.

\end{abstract}

\keywords{Dwarf galaxies(416) --- Galaxy properties(615) --- Galaxy spectroscopy(2171) }


\section{Introduction} \label{sec:intro}

 Dwarf galaxies (DGs) are challenging objects in extragalactic astrophysics. On one side, their low surface brightness nature make them difficult objects to be observed, particularly beyond our Local Group. On the other, due to their low surface densities and weak gravitational potentials, they are expected to be the most sensitive galaxies to UV background, feedback and environmental effects, making it difficult to understand their true and complex structure \citep[see e.g.][]{Mateo+1998,Weinberg+2015,Colin+2015,Bullock+2017}.  Moreover, understanding the radial distribution of their physical properties, from observational studies, is not trivial. However pursuing these type of studies is important in order to deduce their formation and evolutionary paths \citep[e.g.][]{Koleva+2011,Taibi2022}. 

Radial variations in stellar population, SP, properties, such as age, metallicity, and star formation (SF), provide valuable insights into their formation and evolutionary processes. These variations reveal the processes that have shaped the distribution of stars and provide evidence for both internal and external influences on galaxy evolution. For example, gradients in age and metallicity offer clues about the the spatial formation history of galaxies (e.g., inside-out or outside-in growth modes) and on the {\it in-situ} and {\it ex-situ} assembly of the galaxy SPs \citep[e.g.,][]{Tortora+2010,Oyarzun+2019,Avila-Reese+2023,Cannarozzo+2023}. Additionally, feedback process, such as supernova explosions and stellar winds, in addition of bursty SF histories, produce fluctuations in the inner gravitational potential, a redistribution of the gas, stars, and dark matter. These effects are particularly significant in galaxies with shallow potentials wells, such as in dwarf galaxies \citep[see e.g.,][]{Governato+2010,diCintio+2014,Somerville_Dave2015,Hopkins+2023} and dependent on the nature of dark matter \citep[][]{Herpich+2014,Colin+2015,Governato+2015,Gonzalez-Samaniego+2016}. Environmental effects can also influence SP properties gradients via tidal interactions and ram pressure stripping. 

There is general consensus that the gradients of low-mass galaxies are more diverse than those of massive ones and tend to be flatter, or even positive in the case of the age \citep[][see for more references \citealp{Riggs+2024}]{Tortora+2010}. Moreover, while massive galaxies generally follow the classical inside-out formation scenario, low-mass and dwarf galaxies exhibit observational evidence suggesting an apparent outside-in formation scenario \citep[see for a discussion,][]{Riggs+2024}. 

However, despite observational efforts, characterizing and quantifying SP property gradients in dwarf galaxies remains challenging due to their irregular, clumpy morphologies and bursty SF histories. Typically when using observations of massive galaxies, gradients can be measured by tracing the galactocentric distances of inner regions of the galaxies. If the property can be measured from well-resolved individual regions, such as HII regions, or via the Integral Field Spectroscopy (IFS) technique, a radial profile is constructed. The individual points are then either fitted linearly or binned radially before fitting \citep[e.g., ][among many others and references therein]{Kewley2010, Magrini2016, Belfiore2017,BarreraBallesteros2023}. Another alternative is to perform the difference between two radial points in order to obtain a measurement of the gradients \citep{GonzalezDelgado2015,Avila-Reese+2023}. For large galaxy samples, a single gradient that characterize a whole population of galaxies can be obtained by stacking the radial profiles \citep[see e.g., ][among others]{Wang2019,Parikh2021}.

All the mentioned methodologies can be applied to dwarf galaxies, provided with spatially resolved observations. For example, \citet{Grossi2020} used IFS observations of two Virgo dwarf galaxies to derive metallicity maps, which were then radially binned, to obtain radial profiles. These profiles were eventually linearly fitted to determine their gradients. Similarly, \citet{Cai2021} used a sample of 60 low-mass AGN-hosting galaxies with IFS observations from the Mapping Nearby Galaxies at APO \citep[MaNGA][]{Bundy2015} project, with stellar masses $\le 5 \times 10^9 M_{\odot}$, for which they obtained radially binned profiles of age and metallicity, performing linear fits in three radial regions, inner, intermediate and outer region for each of them. Also, they explored a stacking approach to compute a single gradient for their sample in three different radial regions. Using a slightly different approach, \citet{Taibi2018} and \citet{HermosaMunoz2020} accounted for the non-linear nature of the metallicity radial profiles for a sample of Local Group dwarf galaxies. Instead of direct linear fits, they applied a non-linear Gaussian process regression fit \citep{Pedregosa2011} to smooth the profiles before fitting them linearly to derive final gradients.

Since there is not a unique method to derive galaxy gradients, it is important to acknowledge the possibility that the very methodology to derive the gradients may be biased. Moreover, knowing that the dwarf galaxies are bursty and clumpy in nature, these biases should be analyzed in more detail. In this work, the first of a series on dwarf-galaxies gradients, we explore different methodologies to derive radial profiles and gradients, for a set of local dwarf galaxies, the \textbf{MaN}GA \textbf{D}warf G\textbf{ala}xy Sample \citep[MaNDala, ][]{CanoDiaz2022}. The sample of 136 dwarf galaxies, observed as part of the final data release \citep[DR17,][]{Abdurro'uf2022} of MaNGA, covers morphologies from irregulars to early-types, stellar masses in the $10^8-10^9M_{\odot}$ range, with a median stellar mass of $M_\ast\sim10^{8.97}M_{\odot}$,
redshift of $z\sim 0.019$, in a diversity of environments. MaNGA was part of the Sloan Digital Sky Survey IV \citep[SDSS IV, ][]{Blanton2017}, which observed over 10,000 nearby galaxies using the IFS technique in the optical wavelength range. This sample is, to our knowledge, the largest public sample of dwarf galaxies observed with the IFS technique. The MaNGA collaboration as well as their independent research groups have made available a series of dataproducts from which a first analysis is performed in order to retrieve physical information, in particular coming from the SPs of the galaxies and/or from their ionized gas component. In this work, we use the fossil record SP data products from a Value Added Catalogue (VAC) of the SDSS collaboration \citep{sanchez18a,Sanchez2022} to estimate the projected radial profiles of ten stellar properties, and derive inner and outer gradients for the MaNDala sample.

Section \ref{Sec:SampleData} of this work describes the MaNDala sample and its data. Sections \ref{Sec:RadialProfilesMethods} and \ref{Sec:grad_methods} present, respectively, the different methods to derive the radial profiles and gradients. Section \ref{Sec:MethodSelection} discusses the resulting profiles and gradients for the MaNDAla sample. Finally, sections \ref{Sec:InVstGradiens}, \ref{Sec:discussion} and \ref{Sec:conclusions} present our main results, discussion and conclusions respectively. This work adopts a cosmology of $H_0 = 70$ km/s/Mpc, $\Omega_{M}$ = 0.3, and $\Omega_{\Lambda}$ = 0.7.

\section{The Dwarf Galaxy Sample: MaNDala}\label{Sec:SampleData}

\subsection{Sample}\label{Sec:sample}
MaNDala consists of 136 local bright dwarf galaxies ($0.0002 < z < 0.033$ and $10^{7.53} < M_{*} < 10^{9.06} M_{\odot}$) with median values of  $z=0.019$ and $M_\ast = 9.33 \times 10^{8} M_{\odot}$, respectively. Even though most of the galaxies in the sample lay in a small $M_{*}$ range, there is not a priori bias for this, instead this occurs due to the dwarf galaxies available in the final MaNGA sample. The galaxies are located in environments that ranges from isolated to denser environments and display a wide range of morphological types, but when separated in two main groups the majority of them are classified as late types ($\sim$81\%), while the rest ($\sim$19\%) are early types \citep[see Section 5.1 in][we notice, however, that a new and more detailed morphological classification is currently in the making and will be reported elsewhere]{CanoDiaz2022}. Also this sample is dominated by star-forming galaxies ($\sim$92\%), while $\sim$4\% are classified as passive and the remaining $\sim$4\% as transitioning between these two stages \citep[see details about this classification in Section 5.4 in][]{CanoDiaz2022}. 

The MaNDala galaxies were selected as targets by the MaNGA SDSS IV project, either from its parent sample or as ancillary targets. The spatially resolved spectroscopic data and data products are therefore available for all the objects in the sample. The MaNGA data covers up to 1.5 or 2.5 effective radii ($R_{e}$, according to the NSA catalogue \citealp{Blanton2011}), for most of the objects, using a set of different sizes of integral field units made from fiber bundles, which range from 19 to 127 fibers, where each of these fibers has a size of 2$\arcsec$ \citep{Drory15}. In particular to characterize the sample through the MaNGA data, we made use of the dataproducts provided by the Pipe3D VAC \citep{sanchez18a}. Our MaNDala sample also comprises archival photometric data in the $g$, $r$ and $z$ bands provided by the Dark Energy Spectroscopic Instrument (DESI) Legacy Imaging Surveys \citep{Dey2019}, through their ninth data release (DR9). The detailed description of the sample selection, data and basic characterization of MaNDala is given in \cite{CanoDiaz2022}. Finally, access to the results of the sample characterization using both, MaNGA and DESI data sets, is available in the form of a SDSS VAC\footnote{\url{https://data.sdss.org/sas/dr17/manga/mandala/}}. Access to the documentation of the VAC is also available through the SDSS webpage and through our own website\footnote{\url{https://mandalasample.wordpress.com/}}.

This work focuses on the study of inner and outer gradients of several stellar and emission properties, hereafter refer to as just SP,  from the MaNDala sample. As the inner region of the galaxies, we define the 0 to 1 $R_{e}$ interval in the  $r$ photometric band \citep{CanoDiaz2022}. However, since we use our own estimations of the effective radii for MaNDala, some of these values may differ from those in the NSA Catalogue, and hence, for some galaxies the spatial coverage lays bellow $R_{e}$. We cut the MaNDala sample to maintain only the galaxies that are covered up to at least 0.95 $R_{e}$, using our own estimation of the $R_{e}$ values. This leaves us with a final sample of 124 galaxies, which represents $\sim91\%$ of the sample. In this work we do not explore the effects of morphology or environment on the SP gradients, since we intend to pursue this in detail in a following work.

\subsection{Spectroscopic and Photometric Data}\label{Sec:data}

For this work we use the dataproducts provided by the \textsc{Pipe3D} VAC, as described in \citet{CanoDiaz2022}. These dataproducts are publicly available and were derived using the \textsc{pyPipe3D} spectral fitting code \citep{Lacerda2022}, which is an update to the \textsc{Pipe3D} code \citep{Sanchez16a, Sanchez16b}. To derive the stellar properties of the galaxies, \textsc{pyPipe3D} performs a spectral fit to find the best SP Synthesis model for each analyzed galaxy, for which it adopts a Salpeter Initial Mass Function \citep[IMF;][]{Salpeter1955}, a Cardelli attenuation law \citep{Cardelli1989}, and makes use of a new stellar library called \textsc{MaStar\textunderscore sLOG}, which is based on the MaNGA Stellar library \citep[MaStar;][]{Yan2019}. This new library has a sampling of 273 Single Stellar Populations (SSPs), conformed by 39 ages and 7 metallicities \citep[see][for the detailed description]{Sanchez2022}. Once the best SP Synthesis models have been found, \textsc{pyPipe3D} subtracts them from the original spectra of the working galaxy in order to isolate the ionized gas spectra, which is then fitted using single Gaussian components for the emission lines \citep{Sanchez2022}. 

An important step in the data analysis performed by \textsc{pyPipe3D} is the spatial binning. In order to perform the stellar continuum fit in the best way possible a high signal-to-noise ratio (S/N) is needed. Since the MaNGA data have a spatially resolved nature, it is not possible to have a constant S/N over the entire coverage of the Integral Field Units. In order to palliate this, \textsc{pyPipe3D} performs a spatial binning, which is the so called continuum segmentation \citep[for a complete description of this procedure see Section 3.3 of ][]{Sanchez16b}, adding as many adjacent spaxels as needed (if their intensities differ by less than a given percentage), starting by finding the pixel with the higher intensity, and calculating how many of the adjacent pixels are needed, to obtain a final tessella with S/N$\geq 50$ \citep{Sanchez16b, Sanchez2022}; those pixels already showing this value of S/N are considered as single-pixel tesellas (the output S/N in each tesella is measured using the co-added spectra in the wavelength range of 4500 - 5500 Angstroms). This binning procedure is performed before the SSPs fitting. Once the best model is derived, a ``dezonification'' process \citep[first introduced by][]{CidFernandes2013} is applied to recover the original spatial configuration of the data, which is then used to derive the gaseous component of the spectra \citep{Sanchez16b,Sanchez2022}. This process is achieved using the stored information of the position of the pixels that were used to construct the tessellas. For a full description of the implementation of this procedure in the \textsc{Pipe3D} code refer to Section 3.4.4 in \citet{Sanchez16b}. As a result, the stellar properties derived by \textsc{pyPipe3D} display the effects of the binning, with visible tessellations presenting identical values. In contrast, the maps of the gaseous component do not show signs of the binning, as each pixel retains its unique own value.


 In this work, we employ resolved maps of various stellar properties, to derive the radial profiles. A list of these properties and their nomenclature is provided in Table \ref{Table:SP_definitions}. For all the properties, the maps are already available, except for those of $\Sigma_{SFR}$ and sSFR, which we derive based on the best fitting SSPs determination from \textsc{pyPipe3D} and by the $H\alpha$ luminosity converted into SFRs according to the \citet{Kennicutt98} conversion factor. To achieve the above, we retrieve the following maps: the luminosity fraction contribution for several age-metallicity SSPs (specifically for the 7 available metallicities and for ages up to 33 Myr), as well as the maps of $H\alpha$ flux and EW (EW$_{H\alpha}$), see Appendix \ref{Append:SFRs} for details.

\begin{table}
\small
\centering
    \begin{tabular}{ l c c}
    \hline
    \hline
Stellar Property & Nomenclature & Units \\
\hline
Luminosity weighted age & Age$_{LW}$ & $yr$ \\
Mass weighted age & Age$_{MW}$ & $yr$ \\
\shortstack{Luminosity weighted \\ metallicity} & $Z_{LW}$ & \shortstack{normalized\\to $Z_{\odot}$} \\ 
\shortstack{Mass weighted \\ metallicity} & $Z_{MW}$ & \shortstack{normalized\\to $Z_{\odot}$}\\ 
Dust attenuation & $A_V$ & mag \\ 
$D_{n4000}$ index& $D_{n4000}$ & $\AA$ \\ 
Mass-to-light ratio & $M/L$ & $M_{\odot}/L_{\odot}$ \\ 
Stellar surface density & $\Sigma_{*}$ & $M_{\odot}$ $kpc^{-2}$\\ 
\shortstack{SSP based SFR\\surface density} & $\Sigma_{SFR_{SSP}}$ & $M_{\odot}$ $yr^{-1}$ $kpc^{-2}$\\ 
SSP based sSFR & sSFR$_{SSP}$ & $yr^{-1}$ \\ 
\shortstack{$H_{\alpha}$ based SFR\\surface density } & $\Sigma_{SFR_{H\alpha}}$ & $M_{\odot}$  $yr^{-1}$ $kpc^{-2}$ \\ 
$H_{\alpha}$ based sSFR & sSFR$_{H\alpha}$ & $yr^{-1}$ \\ 
\hline  
\hline  
    \end{tabular}
\caption{Summary of the galaxy properties used in this study, for which gradients are obtained.}\label{Table:SP_definitions}
\end{table}

In the present study, we take advantage of our previous photometric analysis for all MaNDala galaxies, as presented in \citet{CanoDiaz2022}, to derive radial profiles for the galaxy properties described previously. This photometric analysis used the $grz$-band optical imaging from the DESI \citep{Dey2019} where projected
azimuthally-averaged surface brightness (SB) profiles were extracted using the IRAF\footnote{IRAF (Image Reduction and Analysis Facility) is distributed by the National Optical Astronomy Observatory, which is operated by the Association of
Universities for Research in Astronomy, Inc., under cooperative agreement
with the National Science Foundation.} image analysis tools. The $r$-band images were chosen as fiducial for the extraction of the isophotal profiles and then applied to the other $gz$-band images to obtain uniform multi-band photometry,  including geometric (ellipticity and position angle) profiles and color gradients. From the DESI imaging we probed $r$ -band SBs down to (on average) 27.02 mag arcsec$^{2}$. Due to the greater depth of DESI compared to the NSA photometric analysis, based on SDSS, some differences between our half-light radius and that reported in the NSA catalog are expected.

In total, in this paper we study 10 galaxy SP properties, with $\Sigma_{SFR}$ and sSFR profiles estimated in two different ways. Next, we describe the details on the characterization of the radial profiles in the MaNDALA sample.

\section{Radial profiles derivation methods}\label{Sec:RadialProfilesMethods}

An important fraction of dwarf galaxies exhibit irregular and clumpy morphologies, in contrast to normal and massive galaxies which exhibit more regular or weakly fluctuating shapes \citep[see e.g.][]{Meyer2014,2017Ann}. Therefore, when analyzing the spatially resolved properties of their adjacent bins, determining the best way to characterize the radial distributions is far from being trivial. As we will discuss below, two main approaches can be adopted to derive radial distributions from the projected maps of the different SP properties. 

The first method is to perform elliptical radial  bins and characterize each SP property, within each bin. Such characterization can be done either by using a statistical representative value such as the \emph{mean} or the \emph{median} value of all the pixels within the radial bin. Alternatively, another option is by integrating all the values within each elliptical radial bin, Section \ref{Sec:CollapsedProfiles} below. Both approaches intend to collapse all the spatially resolved information of the profiles in a single number per radial bin, and for this reason we will refer to this type of profiles as collapsed radial profiles. The second method is to \emph{fully} take advantage of the resolved nature of the data and simply derive resolved radial profiles exploiting the information of each individual region of the data. 

Both methods have advantages and disadvantages and can actually lead to different results. For example, collapsed radial profiles help at smoothing the possibly noisy nature of the data, leading to much easier to fit radial profiles. However proceeding this way leads to some information loss. On the other hand using fully spatially resolved radial profiles allows the utilization of all the data points. Nevertheless, if the profiles are inherently noisy, outliers may bias the information, making it more challenging to fit the profiles accurately, and thus their gradients.

We now outline the methodologies employed to explore the two approaches to derive the radial distributions and their gradients for all the galaxy properties under study. The spatially resolved nature of the MaNDala sample allows to estimate gradients that cover the inner part of the galaxies, $0<R/R_e<1$, as well as their outer parts, $0.75<R/R_e<1.5$. The reason to select this last coverage is because the MaNGA data covers either up to 1.5 or up to 2.5 $R_{e}$ for different galaxies, so a convenient solution is to select a radii range that covers only up to the smallest radial coverage provided by our data set.\footnote{Notice that for an exponential disk, $R=1.5\;R_e$ implies that we are integrating $\sim 72\%$ of the total light, and even more for stellar mass. As we will see later, MaNDala galaxies have an average mass-to-light ratio of $M/L\sim2.8$.} The lower limit of the range was chosen to be similar in length to that chosen for the inner gradients, but with as little overlap as possible.

In our final sample, there are 15 galaxies that are spatially resolved below $R<1.45\;R_{e}$ (recall that we are using the half-light radius based on our DESI photometry analysis, see also Section \ref{Sec:sample}). Two of them are spatially resolved up to $R_{e}$, so they are discarded for the external analysis only. For the remaining 13 galaxies, their external radii are within the range between $1.02\lesssim{}R/R_e\lesssim{}1.43$. We consider that even if their external gradients do not meet our definitions, including them will not significantly alter our conclusions, as they represent only 10.5\% of the entire sample.

Regarding the effect of the PSF, we found that accounting it in our analysis does not affect the estimation of either the radial profiles or the gradients, except for one of the methods that will be introduced in Section \ref{Sec:grad_methods}. However, we notice that it does not affect the conclusions obtained in this paper, once we have set the values of our fiducial gradient (see details in Section \ref{Sec:MethodsComparisson}). While in the remaining sections of this paper we do not discuss the effect of the PSF, Section \ref{Sec:PSF_impact} discusses in detail its impact in our derived quantities. In general, we found that the effect of the PSF is only minor for all the quantities derived from this paper.

\subsection{Collapsed radial profiles}\label{Sec:CollapsedProfiles}

 To obtain collapsed radial profiles for each SP property we define a set of concentric, projected elliptical rings for each galaxy in our sample. The computation of the elliptical rings takes into account the photometric isophotal and geometrical information of the MaNDala sample, which is available with the photometric isophotal analysis we performed in \citet{CanoDiaz2022}, based on DESI archival data. Specifically, each ring  has been derived by means of the geometrical information provided by the isophotal analysis: ellipticity ($\epsilon$) and Position Angle \citep[P. A.; for these and other details regarding the isophotal analysis, such as the center determination of our dwarf galaxies refer to Section 4.1 in][]{CanoDiaz2022}. However since DESI and MaNGA have different spatial resolutions (the former is smaller than the latter) we have collapsed this information into the median values of the inclination angle ($i$) and P.A. for as many DESI isophotes that can be fitted into a elliptical ring that has a width of approximately one radius of the MaNGA average PSF ($R_{PSF}$ $\approx$ 1.25$\arcsec$). This means that the projected elliptical rings used in this analysis follow the internal geometry of the galaxies, and have a width of  $\sim R_{PSF}$. 

\begin{figure*}
\centering
    \subfloat{%
	   \includegraphics[width=0.21\textwidth]{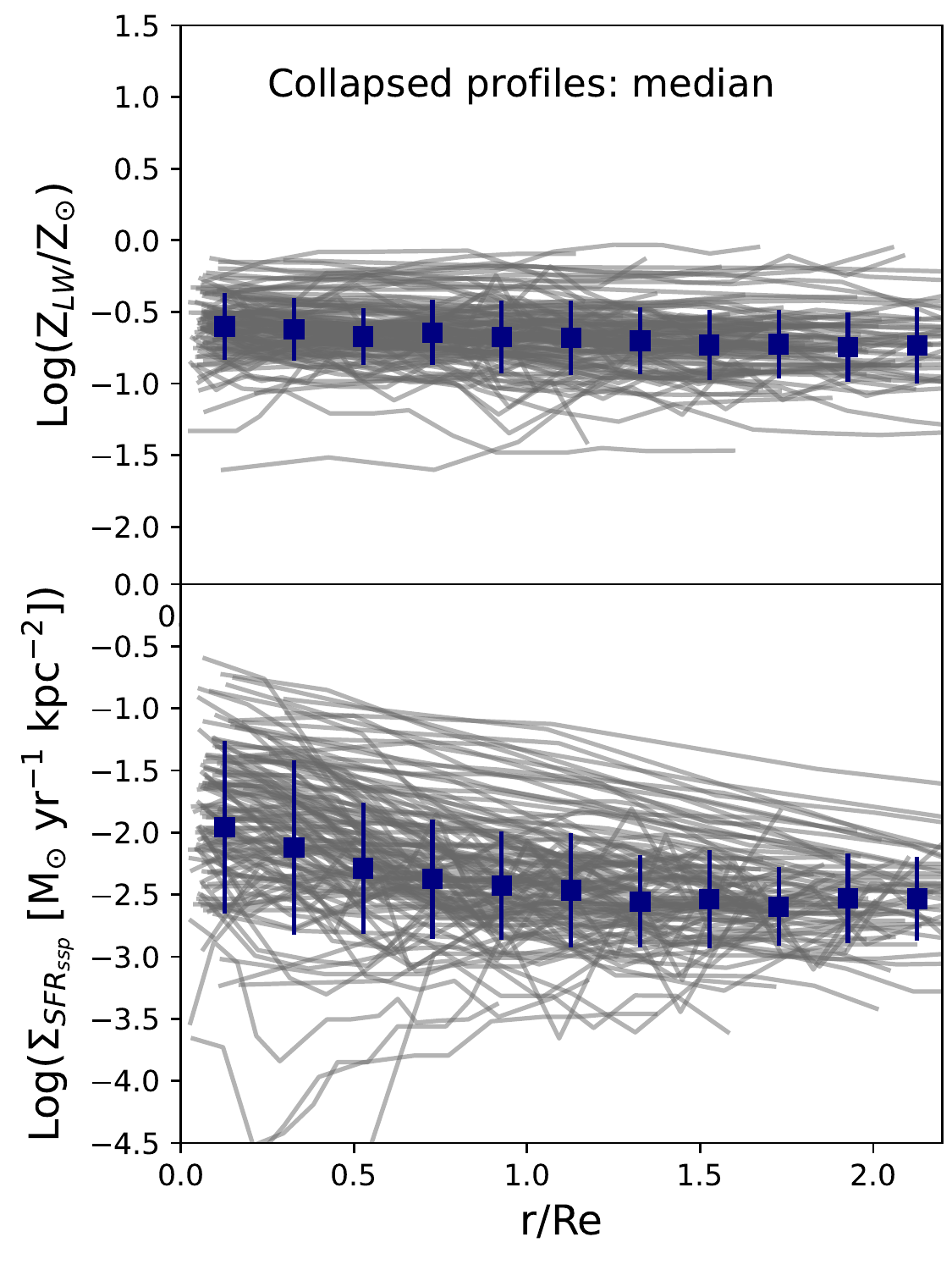}
    }\qquad
    \subfloat{%
	   \includegraphics[width=0.21\textwidth]{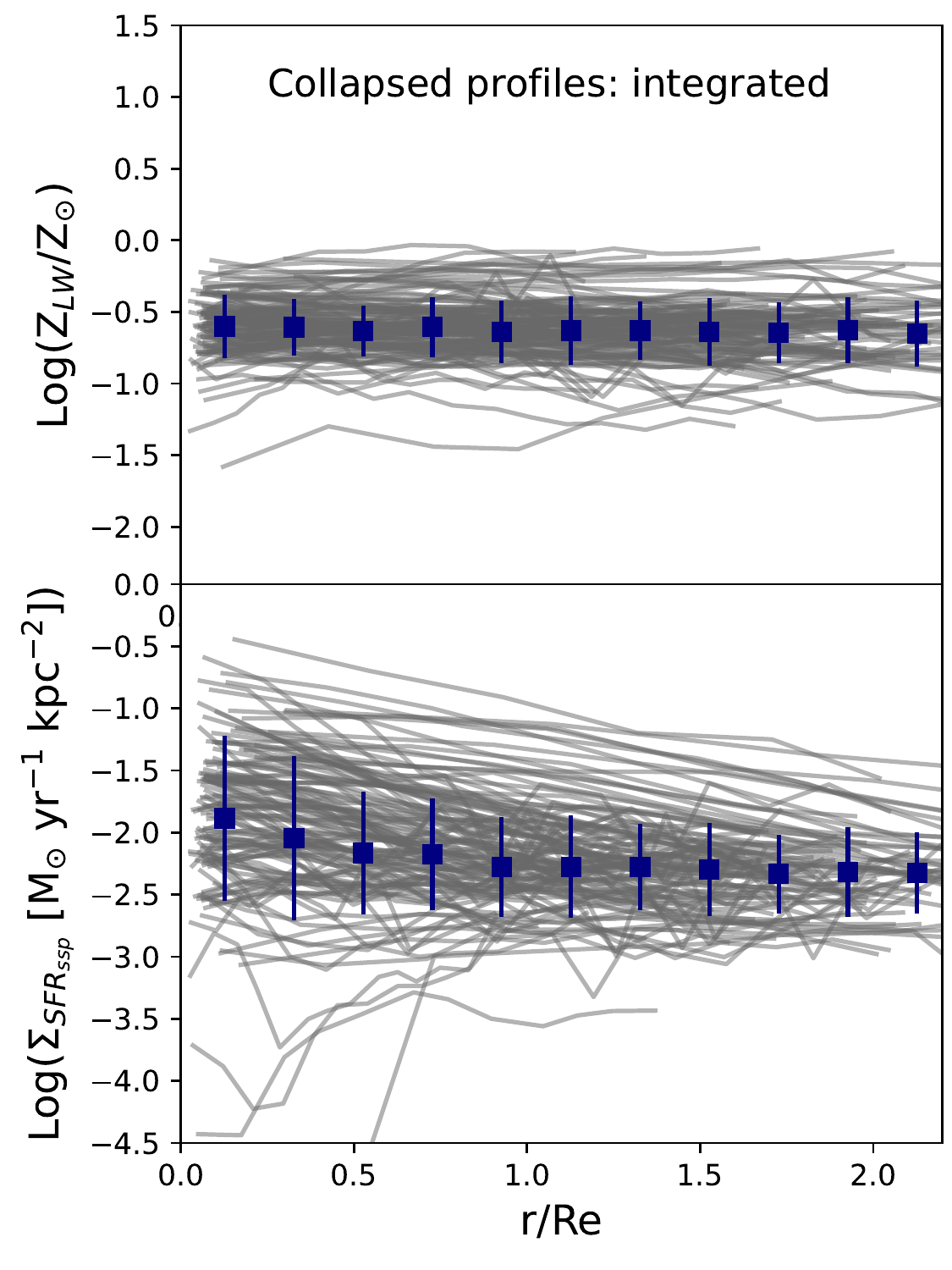}
     }\qquad
    \subfloat{%
	   \includegraphics[width=0.21\textwidth]{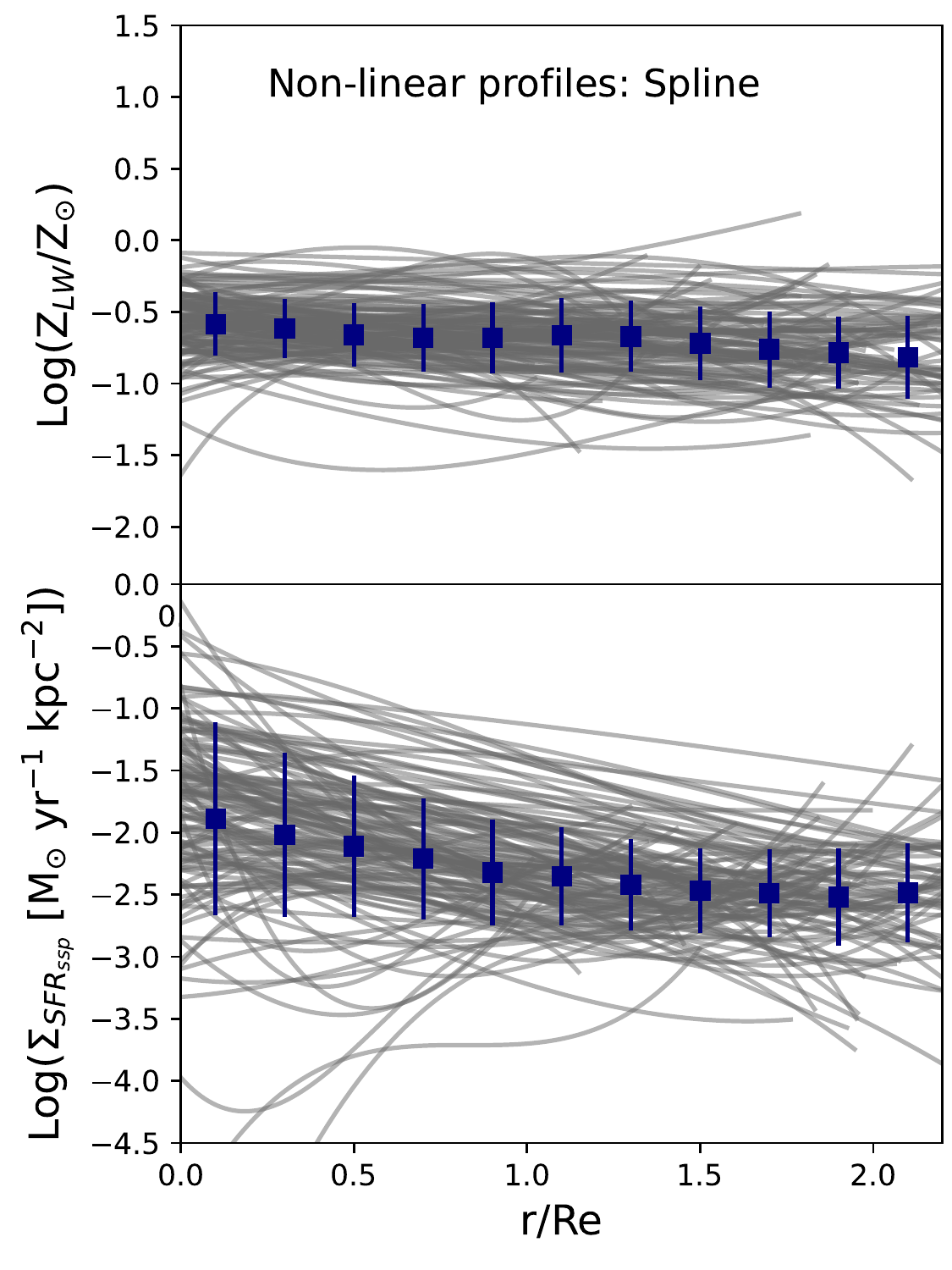}
    }\qquad
    \subfloat{%
	   \includegraphics[width=0.21\textwidth]{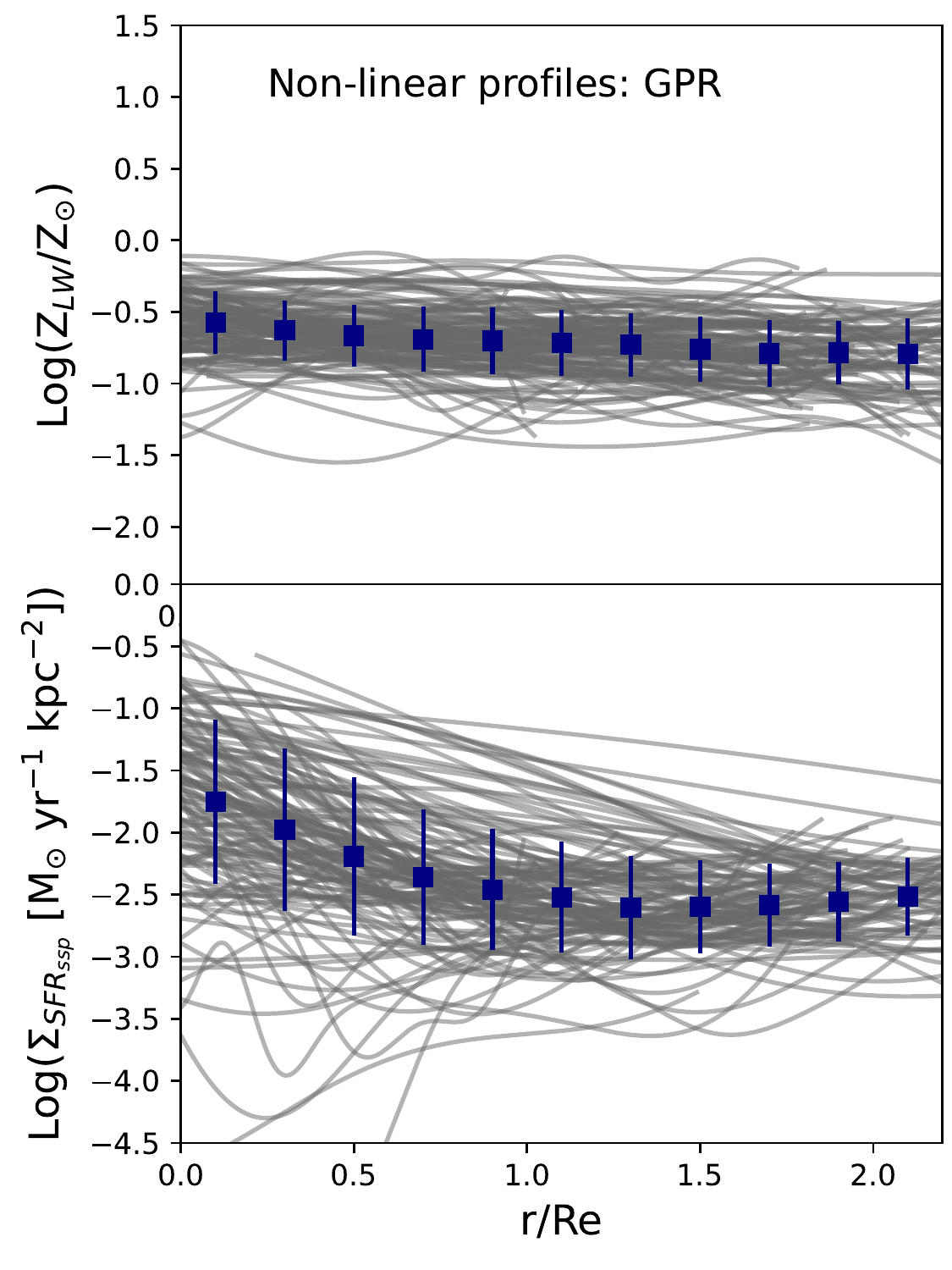}
     }\qquad
\caption{From left to right: the profiles of two properties, as examples, for the entire MaNDala working sample derived with the collapsed median, collapsed integrated, non-linear Spline and non-linear GPR methods, shown in gray lines. On top, with blue squares, the median profiles for each method are shown, made from bins of equal size of 0.2 $R_{e}$, where the error bars represent $1\sigma$ of the distribution within each bin.}
\label{Fig:example_all_profiles} 
\end{figure*}

\subsubsection{Statistical characterization of elliptical rings} \label{Sec:ProfilesMedians}

For each galaxy in our final sample, we calculate the median of the points within each predefined ring for the SP property we are studying. We associate the standard deviation within each ring as the error bar. Using the median values and the middle points of the elliptical rings as a function of $R_{e}$, we derive the radial bins and obtain a median profile for all the SP properties of the MaNDala galaxies. We also experiment deriving equivalent profiles but using the mean of each ring instead of the median, for which we obtained very similar results. However, we decided to keep only the median profiles since the median is less sensitive to outlying values. In the leftmost panel of Figure \ref{Fig:example_all_profiles}, we show all the profiles for the $Z_{LW}$ and $SFR_{ssp}$ properties, derived with this method in the grey solid lines. For the full set of profiles see Figure \ref{Fig:all_profiles_medianprof} in Appendix \ref{Append:AllProfiles}.

\subsubsection{Integration of elliptical rings}\label{Sec:ProfilesIntegrated}

In this section we describe the integration of all the values of the studied properties within each defined ring. This is done as follows:

 In the case of regular intensive properties (ages, metallicities, $A_{V}$, $D_{n4000}$ and  $M/L$), each ring is integrated and weighted by the fraction of mass contained in each pixel, $f_{*}$, in a given bin in the following way:
\begin{equation}
\label{eq:WeightIntRings}
 X = \sum_{0}^{n} x_{i} \cdot f_{*,i} = \frac{\sum_{0}^{n} x_{i}\Sigma_{*_{i}}}{\sum_{0}^{n} \Sigma_{*_{i}}},
\end{equation}
where $x$ represents any of the intensive properties, $i$ denotes each pixel within the given ring, $n$ is the total number of pixels in the ring. Notice that we opted to weight by $\Sigma_{*}$ so the effect of young populations are minimized and do not bias the integrated value within each ring. In the case of the extensive variables ($\Sigma_{*}$ and $\Sigma_{SFR}$) only a simple integration of the values of all the pixels within a ring is performed. Finally, we define $\text{sSFRs} = \Sigma_{SFR}/\Sigma_{*}$. For all cases we associate as error bar for each integration the value of the standard deviation within each ring, we also use the middle points of the elliptical rings in terms of the $R_{e}$ to complete the radial bins definition in this case. In the second panel from left to right of Figure \ref{Fig:example_all_profiles} we show all the profiles derived with this method for the $Z_{LW}$ and $SFR_{ssp}$ properties. For the full set of profiles see Figure \ref{Fig:all_profiles_integratedprof} in Appendix \ref{Append:AllProfiles}.

\subsection{Non-linear radial profiles}

\subsubsection{Cubic splines}\label{Sec:ProfilesSplines}

We use a cubic spline data interpolator as functional forms to describe the non-linear nature of each of the spatially resolved profiles of the properties of interest. In order to address the segmentation issue, we implement the following strategy. First, we assign weights to the data points in the radial distributions:

\begin{equation}
\label{eq:WeightSplines}
 W = \frac{1}{N_{\rm pix}},
\end{equation}
where $N_{\rm pix}$ is the total number of pixels 
within each segment, i.e. the weights $W$ are the inverse of the fraction of pixels contained in the segment to which each pixel belongs to (i.e., not including background nor masked pixels). The weights are then divided by their sum for all segments, so that the sum of these now normalized weights is equal to one. The more pixels are contained in the segment, the smaller will be the weight assigned to the data points within it.

We apply a 2D-Gaussian kernel density estimation to the radial distribution of each property, including the previously calculated weights, to obtain a probability density distribution for each spatially resolved profile. We use the \textsc{stats.gaussian\_kde} implementation from the \textsc{scipy} libraries \citep{scipy}. In this way, regions of higher density correspond with the position of points with larger weights (i.e. belonging to smaller segments). 

We then fit the cubic splines to the probability density distribution obtained  using the \textsc{UnivariateSpline} package from \textsc{Scipy}, resulting in a smoothed non-linear profile (leftmost bottom panel of Figure \ref{Fig:example_profiles}).

After performing a visual inspection of the spline interpolation results, we discarded some galaxies from the analysis because the spline interpolation did not reasonably follow the radial distribution of the data. Furthermore, the interpolation failed for some properties of a given galaxy but not for others, meaning that the final number of galaxies analyzed by this method is not the same for all properties, as it is the case with the collapsed profiles (Sec. \ref{Sec:CollapsedProfiles}). For properties where there is good data sampling (e.g. ages and metallicities) the loss of galaxies that cannot be interpolated with the splines ranges from 2 to 8 galaxies (1.6 - 6.4\% out the total sample). Meanwhile for the properties in which the data sampling is poorer ($\Sigma_{SFR}$ and sSFR traced by $H\alpha$) the loss increases up to 11 galaxies (8.9\% of the total sample). For the uncertainty treatment, we do not directly use the uncertainties of each data point. This information is implicit in the way the weights $W$ are defined, since larger weights result from smaller segments, and therefore from pixels associated with spaxels with higher S/N. 

In the third panel from left to right of Figure \ref{Fig:example_profiles} we show all the profiles for the $Z_{LW}$ and $SFR_{ssp}$ properties, derived with this method. For the full set of profiles see Figure \ref{Fig:all_profiles_splineprof} in Appendix \ref{Append:AllProfiles}

\subsubsection{Gaussian Process Regression fit}\label{Sec:ProfilesGPR}

For this approach, we employ the full spatially resolved potential of the data, for which only the values of all the pixels for each studied property and their projected galactocentric distance are needed. To obtain the fits for the gradients, we use a method that addresses the non-linear nature of the profiles and helps mitigating the spatial segmentation effects introduced by \textsc{pyPipe3D} (see Section \ref{Sec:data}). The effect of this segmentation is visible in the radial distributions causing many pixels at different radii within any map get assigned the same value. This effect was found in the SP properties profiles, as explained in Section \ref{Sec:data}, but not in the ionized gas ones. For regions far from the centre (which usually contain the pixels with the lowest S/N ratios) the segments are large enough to introduce a bias that affects the fitting procedures and, in consequence, the gradients. Therefore, it is important to account for the size of the segments. In the left side panels of Figure \ref{Fig:example_profiles}, the black circles represent the data from individual pixels, which in this case are for the $Z_{LW}$ and the $\Sigma_{*}$ properties. The effect of the segmentation in the radial distributions of both properties is seen as conglomerates of the same values but at different radii. This is the effect of the aforementioned spatial segmentation, rather than a natural behaviour of the spatial distributions. 

Following the procedure introduced by \citet{HermosaMunoz2020} and \citet{Taibi2022} to derive gradients in Local Group dwarf galaxies, we perform a Gaussian Process Regression (GPR) fit, a Bayesian approach that models the resolved radial profiles and provides a posterior probability distribution of each profile. As explained in \citet{Taibi2022}, the advantage of this method is that it does not require any prior assumption about the shape of the profiles to derive accurate solutions. For this work we also adopt the usage of the \textsc{Python} based package \textsc{GaussianProcessRegressor} provided by the \textsc{scikit-learn} library \citep{Pedregosa2011}. 

The GPR fit does not allow weighting the data, meaning that it does not correct for the spatial binning effect visible in the profiles. However, it requires selecting a kernel or a combination of kernels as input. Following \citet{HermosaMunoz2020} and \citet{Taibi2022}, we use a white noise and a radial basis function (Gaussian-like) kernel. Finally, we have discarded points from the spatially resolved profiles that have uncertainties $>70\%$ or equal to 0.

We implemented a three-step GPR iteration strategy to refine the priors in the fitting routine (initially a guess for the white noise level, the length scale of the radial basis function, and their variation limits). This strategy was designed to work automatically:
\begin{itemize}
    \item In the first iteration, we use a set of constant, empirically selected priors for all the galaxies, for the ten radial properties we are analyzing,
    
    \item The \textsc{GaussianProcessRegressor} routine provides a $R^{2}$ score as the goodness-of-fit, called the coefficient of determination,\footnote{See the definition of the coefficient of determination in the routine documentation: \url{https://scikit-learn.org/stable/modules/generated/sklearn.gaussian_process.GaussianProcessRegressor}} with 1 being the ideal value. If the $R^{2}$ score value is $<0.75$ for any profile, the process enters into a second iteration using the posteriors from the first iteration as the updated priors,

    \item If the posteriors from the second iteration hit either the upper and lower variation limits, the process then enters a third and final iteration expanding these limits by two orders of magnitude.
\end{itemize}

As for uncertainties, we implemented a Bootstrap strategy in our automated routine. After determining the final iteration for each profile, we repeat the final fit for a set of 90 randomly selected samples, each with a size equal to the original data. We also tested a larger bootstrap using 150 samples but the size of the uncertainties remain the same as the first one, when performing the rounding of the values. We estimate the uncertainty as the standard deviation of the distances between the original fit and the bootstrapped fits at each radius. In the rightmost panel of Figure \ref{Fig:example_all_profiles} we show all the profiles derived with this method for the $Z_{LW}$ and $SFR_{ssp}$ properties. For the full set of profiles see Figure \ref{Fig:all_profiles_gprprof} in Appendix \ref{Append:AllProfiles}.

\begin{figure*}
  \centering
    \includegraphics[width=1\textwidth]{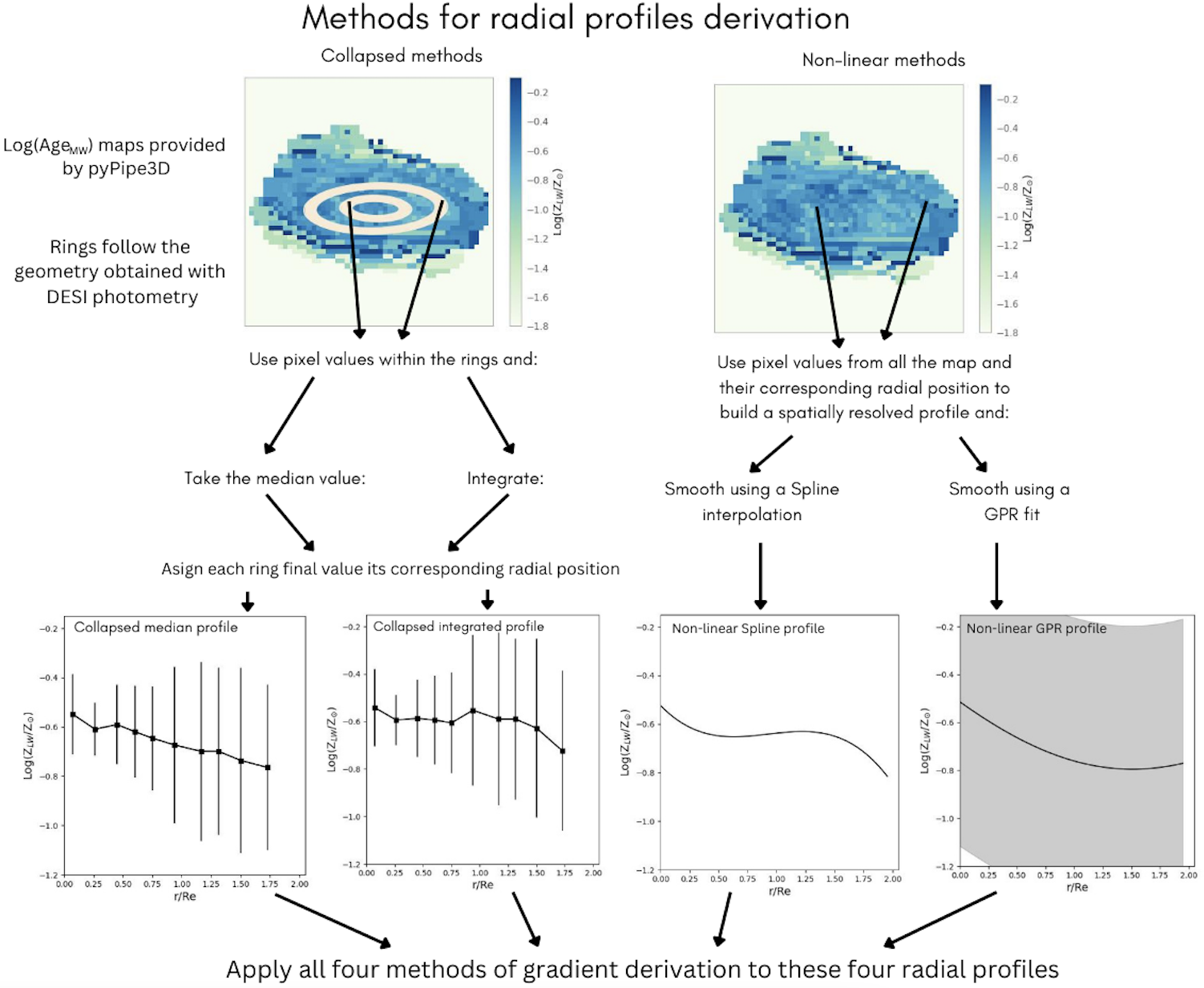}
  \caption{\label{Fig:RadProfile_Methods} An illustration that shows the main points behind the four different methodologies to derive the radial profiles presented in Section \ref{Sec:RadialProfilesMethods}. We use as example the galaxy manga-7815-6101, for which we present its $Z_{LW}$ map. In the left side of the figure the strategies to implement the collapsed methods are shown, while in the right side the same is visible but for the non-linear methods.}
\end{figure*}

As a summary, Figure \ref{Fig:RadProfile_Methods} illustrates the four methods employed to derive the radial profiles. Finally, at this point is important to mention that even though the usage of IFS dataproducts in which a spatial binning was performed, such as those provided by \textsc{Pipe3D}, an unnatural external flattening of the radial profiles and in consequence to the gradients is introduced. However this effect is expected to be small in properties inferred by archaeological methods such as the Age \citep{Ibarra-Medel+2019}. Due to this we assume that the effect of the spatial binning is not dominant when measuring the gradients, even in the external regions.

\section{Gradients derivation methods}\label{Sec:grad_methods}

Once the radial profiles are derived using a given method, the SP properties gradients discussed in Section \ref{Sec:data} can be estimated following various strategies. In other words, for each of the four radial profile characterization methods described in the preceding section, we apply different approaches to compute SP gradients. In this section, we explore three methods. The units of the gradients are given in $dex$ $R_e^{-1}$ except for $A_V$ and $D_{n4000}$ which are given in $mag$ $R_e^{-1}$ and $\AA$ $R_e^{-1}$ respectively.

\subsection{Slope of linear fits} \label{Sec:linear_fits}

A linear fit can be performed to each radial profile, whether collapsed or spatially resolved to estimate the gradients for each property, which is approximated by the slope of the fit. That is, we assume the following functional form for the segment of the profile that will be fitted:
\begin{equation}
    \mathcal{G}(x) \equiv \log G(x) = G_0 + \nabla_{i} x,
    \label{eq:linear_fit_grads}
\end{equation}
where $G(x)$ stands for the galaxy property,\footnote{In the case of $A_V$ and $D_{n4000}$ we assume that $A_V = A_{V,0}+ \nabla_{i} x$ and $D_{n4000} = D_{n4000,0}+ \nabla_{i} x$.} $\nabla_{i}$ is the slope or the gradient, $i$ refers to either `in' or `out' and $x$ is the radial bin normalized by $R_e$: $x = R/R_e$. This method has been widely used in previous works in the literature \citep[e.g. ][]{Belfiore2017,SanchezMenguiano2018,Ferreras2019,Parikh2021,BarreraBallesteros2023}. 

In our case, the linear fits are performed using least-squares optimization routines based on the Levenberg-Marquardt algorithm \citep{Press+1992}. In this paper we compute linear fits for inner and outer gradients as: 
\begin{itemize}
    \item Inner gradients, $\nabla_{in}$: between 0 and $R_e$,
    \item Outer gradients, $\nabla_{out}$: between $0.75R_e$ and $1.5R_e$.
\end{itemize}
These linear fits consider the errors derived from the SP properties profiles, see Section \ref{Sec:RadialProfilesMethods}.

In the case of the Splines method, the linear model produces the slope and intercept of the fitted models and the covariance matrix for these parameters. We obtain the errors for the fitted parameters (i.e, gradients and intercepts) by taking the square root of the corresponding elements of the diagonal of the covariance matrix. 

It is important to remark that in some cases, the linear fits cannot be obtained over the collapsed profiles, due to the small number of bins and pixels available in some galaxies. The number of lost galaxies in these analyses vary from property to property. To perform the linear fits, we use galaxies resolved enough by first removing rings with less than 2 pixels and fitting only those properties with at least 3 data points. For example, in the case of the ages and metallicities, the loss is of 27 galaxies (21.8\% of the total sample) when fitting the inner region, but this loss grows up to 71 galaxies (57.3\%) for the outer region. However for the properties in which less points are available ($\Sigma_{SFR}$ and sSFR traced by $H\alpha$), the loss of galaxies in the inner region results in 32 galaxies (25.8\%) and in 74 galaxies (59.7\%) for the outer region. In the case of the linear fits performed over the non-linear Spline profiles, besides the galaxies discarded due to an unsuccessful cubic spline interpolation (see details in Section \ref{Sec:ProfilesSplines}), it was necessary to discard a few more galaxies for the outer gradient calculation. The reason for this is that it is not possible to perform the linear fit in this radial region, mainly because the sampling of points is not good enough to obtain a reliable interpolation in these regions. In this case, the galaxy loss for the outer gradients is of 2 galaxies for the properties with a good sampling, except for the case of $\Sigma_{SFR}$ and sSFR based on the SSPs, in which the loss is of 3 galaxies (1.6-2.4\% respectively). However, the amount of losses goes up to 7 galaxies (5.6\%) for the worst sampled properties ($\Sigma_{SFR}$ and sSFR derived from $H\alpha$). When adding to these percentages, galaxies that were not able to be interpolated by the Spline method the total amount of loss grows up to 8.8\% for the bulk of properties, and to 14.5\% for the $\Sigma_{SFR}$ and sSFR traced by $H\alpha$. Finally, in the case of the linear fits performed over the GPR smoothed probability distributions, all of them provide a final estimation of the inner gradients for all of our working sample and for all the galaxies considered for the external analysis, meaning that for this method there is no galaxy loss.

A disadvantage of this method is that it assumes a prior parametric functionality in the radial distribution of galaxy properties, which can result in either spurious trends or artificial increasing of the scatter around the distribution of radial gradients. This is particularly evident when the slopes have strong curvatures around the region of interest. The solid light blue lines on the two leftmost panels of Figure \ref{Fig:example_profiles} show examples of linear fits performed over the two aforementioned inner and outer regions, for the four types of radial profiles.

\subsection{Generalized fits} \label{Sec:general_fits}

The surface brightness profiles of galaxies are often well described by a \citet{Sersic_1963} function. That is the case for galaxies in the MaNDala dwarf galaxy sample, exhibiting a diversity of Sersic indices ranging from $n \sim1$ to $\sim5$, as shown in \citet{CanoDiaz2022}. Assuming that the mass-to-light ratio is constant, it is then expected that the mass density profile is also described by a Sersic profile with similar Serisc indices. Additionally, \citet{Cano-Diaz+2019} showed that the spatially resolved star-forming main sequence of late-type galaxies follows $\Sigma_\text{SFR} \propto \Sigma_\ast^{0.94}$, consistent with a Sersic profile. Moreover, \citet{Rodriguez-Puebla+2017} showed that the SF history of small/dwarf galaxies based on the semi-empirical modelling of the galaxy-halo connection is approximately constant. That will imply that the radial profile of stellar ages should display a small curvature consistent with a sub-exponential behaviour. Thus, a way to generalize the gradient profiles described by Eq. (\ref{eq:linear_fit_grads}) is to assume a more flexible function that allows for either sub- or supra-exponential behavior. The generalized function form is given by
\begin{equation}
    \mathcal{G}(x) = G_0 + \frac{G_1}{\ln 10} x^n.
    \label{eq:general_fit_grads}
\end{equation}
As is evident by comparing Eq. (\ref{eq:general_fit_grads}) with Eq. (\ref{eq:linear_fit_grads}) both are equal when $n=1$ (exponential profile) and thus $\nabla = G_1 / \ln 10$. The above function has a slope that depends on the radius given by:
\begin{equation}
    \frac{d \mathcal{G}(x)}{dx} = \frac{nG_1}{\ln 10}x^{n-1} = \frac{n}{x}\left(\mathcal{G}(x) - G_0\right),
    \label{eq:grad_general_fit}
\end{equation}
which can be used as a proxy for the local gradient, of the radial property $\mathcal{G}(x)$.

Similarly to the preceding section, we used a least-square optimization routine based on the Levenberg-Marquardt algorithm. In the case of these more generalized fits, we use the full radial profiles by requiring that at least 4 radial bins per profile are available. To determine the best-fit parameters, we run the Levenberg-Marquardt algorithm for $N=100$ iterations, each time updating the best-fit model based on the previous one. To ensure convergence, particularly for $n$, we impose the following condition: if the error in any parameter exceeds the best-fit parameter value itself or if $\chi>40$ then we set $n=1$. Otherwise, $n$ treated as a free parameter, fitted alongside the other parameters. Finally, notice that the generalized fits have the advantage of fitting the entire radial profile, unlike the linear fits, which were defined over different radial ranges.

Although Eq. (\ref{eq:general_fit_grads}) is more flexible and allows a more accurate and richer description of the data, it has a similar drawback to the linear fits: it assumes a prior function for the radial profiles.

 The two leftmost panels of each row in Figure \ref{Fig:example_profiles}, with solid red lines, show examples of these generalized fits. Errors associated with the gradients derived through Eq. \ref{eq:grad_general_fit} are not provided in this case, but we anticipate that they will depend on the error of the Sersic index, $n$, and the zero point, $G_0$.

\subsection{Difference between two radial points}\label{Sec:diff_2points}

An alternative way to obtain the gradients from the radial profiles, either in the collapsed ones or in the smoothed spatially resolved ones, is by performing the difference between two radial points in the observed profiles. This procedure has also been explored in the literature before \citep[e.g.][]{GonzalezDelgado2015,Avila-Reese+2023} with the advantage of not making any assumptions on the radial profile of the galaxies.

In this work we explore this methodology for the inner regions (between the first bin/point of the collapsed/non-linear profiles and $R_{e}$), and outer regions ($0.75<R/R_e<1.5$). In general, the way to compute these gradients is the following:

\begin{equation}
\label{eq:DiffGradients}
 \nabla_{G} = \frac{\mathcal{G}(x_{out})-\mathcal{G}(x_{in})}{x_{out}-x_{in}},
\end{equation}
where, as above, $\mathcal{G}(x)$ is any radial profile of the SP properties used in this work, while $x_{in}$ correspond to the innermost radial region for the gradients, and $x_{out}$ to the outermost radial region. Since the collapsed profiles and smoothed spatially resolved profiles are not continuous, we use the $x_{in}$ and $x_{out}$ values that are closest to the adopted inner and outer radial ranges.

For the inner region it is important to consider the possible effect of the PSF. A more detailed analysis on this is presented in Section \ref{Sec:PSF_impact}. Meanwhile, we derive these gradients regardless if the first radial bin/point is inside the radius of the PSF (notice that for the collapsed methodology explained in Section \ref{Sec:CollapsedProfiles} it is possible that the first radial bin remains $\sim 0.1$ underneath the $R_{PSF}$, while for the spatially resolved profiles, there are pixels that lay below the $R_{PSF}$).

In this approach, we do not assign uncertainty values to derive the gradients because, depending on the type of radial profile, the corresponding error bars are different. For the collapsed profiles, the error bars are the standard deviation of the elliptical rings; for the non-linear spline profiles, there are no error bars for individual points; and for the non-linear GPR profiles, the error bars are derived from a bootstrap method. These facts do not permit a uniform estimate of the error bars. 

In the case of the generalized fits (Section  \ref{Sec:general_fits} ) we also compute a gradient based on two points given by

\begin{equation}
    \nabla_{G,\text{fit}} = \frac{G_1}{\ln 10}\frac{x_{out}^n-x_{in}^n}{x_{out}-x_{in}} =  \left(\frac{1-s^n}{1-s}\right) \frac{d \mathcal{G}(x)}{n \; dx}\bigg|_{x=x_{out}},
\end{equation}
where $s = x_{in}/x_{out}$. The above equation shows that the gradient between two points is related to the local slope of the outer radius of the galaxy profile: $\nabla_{G,\text{fit}} \propto d\mathcal{G}(x_{out})/dx$. In the particular case of $s=0$ and $n=1$ then $\nabla_{G,\text{fit}} = d\mathcal{G}(x_{out})/dx$ that will be also related to the gradient defined by Eq. \ref{eq:linear_fit_grads}; $\nabla = \nabla_{G,\text{fit}}$.

\begin{table*}
\centering
    \begin{tabular}{ l c c c c}
\multicolumn{5}{c}{Gradients Derivation Methods} \\
    \hline
    \hline
   & \multicolumn{2}{c}{\shortstack{Collapsed\\profiles}} & \multicolumn{2}{c}{\shortstack{Spatially Resolved \\profiles}} \\
 & {\shortstack{Median\\Sec.\ref{Sec:ProfilesMedians}}}  & {\shortstack{Integrated\\Sec.\ref{Sec:ProfilesIntegrated}}}  & {\shortstack{Spline\\Sec.\ref{Sec:ProfilesSplines}}}  & {\shortstack{GPR\\Sec. \ref{Sec:ProfilesGPR}}}\\
 \hline
\rowcolor{lightgray}
\multicolumn{1}{c|}{Radial bins} & size $\sim$1 $r_{PSF}$ & \multicolumn{1}{c|}{size $\sim$1 $r_{PSF}$} & No & No\\
\multicolumn{1}{c|}{Weights} & No & \multicolumn{1}{c|}{{\shortstack{Eq. \ref{eq:WeightIntRings} for\\ intensive \\ properties}}}& \shortstack{Eq. \ref{eq:WeightSplines} and \\ uncertainty=0\\or $>70\%$  are out.} & \shortstack{uncertainty=0\\or $>70\%$ are out}\\
\rowcolor{lightgray}
\multicolumn{1}{c|}{Kernel} & No & \multicolumn{1}{c|}{No} & Gaussian & \shortstack{White Noise\\+ Gaussian} \\
\multicolumn{1}{c|}{Errors} & \shortstack{Std. Dev.\\in each bin} & \multicolumn{1}{c|}{\shortstack{Std. Dev.\\in each bin}} & Weights from S/N & Bootstrap \\
\rowcolor{lightgray}
\multicolumn{1}{c|}{{\shortstack{Linear Fit\\  Gradient}}} & \shortstack{L.M.\textsuperscript{\textdagger} with\\errors} & \multicolumn{1}{c|}{\shortstack{L.M. with\\errors}} & \shortstack{L.M. with\\errors} & \shortstack{L.M. with\\errors} \\
\multicolumn{1}{c|}{{\shortstack{2 Points Difference\\ Gradient}}} & {\shortstack{Using innermost \\ and closest bin to \\ 1, 0.75 and 1.5 $R_{e}$ }} & \multicolumn{1}{c|}{{\shortstack{Using innermost \\ and closest bin to \\ 1, 0.75 and 1.5 $R_{e}$ }}} & {\shortstack{Using innermost \\ and closest point to \\ 1, 0.75 and 1.5 $R_{e}$ }} & {\shortstack{Using innermost \\ and closest point to \\ 1, 0.75 and 1.5 $R_{e}$ }} \\
\hline  
\hline  
    \end{tabular}
\caption{Summary of the main characteristics of the four tested methodologies to derive the gradients for the MaNDala sample, explained in Section \ref{Sec:RadialProfilesMethods}. The errors referred to in the second to last row are those described in the previous one. \textsuperscript{\textdagger} L.M. = Levenberg-Marquardt.
}
\label{Table:Methods}
\end{table*}

\subsection{Summary of radial profiles and gradients derivation methods}

As explained before, the final estimation of the gradients for the various properties of the galaxies, depends on two factors: 
\begin{itemize}
    \item The methods used to derive the corresponding radial profiles and,

    \item The methods used to calculate the gradients from these radial profiles.
\end{itemize}

\begin{figure*}
\centering
    \subfloat{%
	   \includegraphics[width=0.21\textwidth]{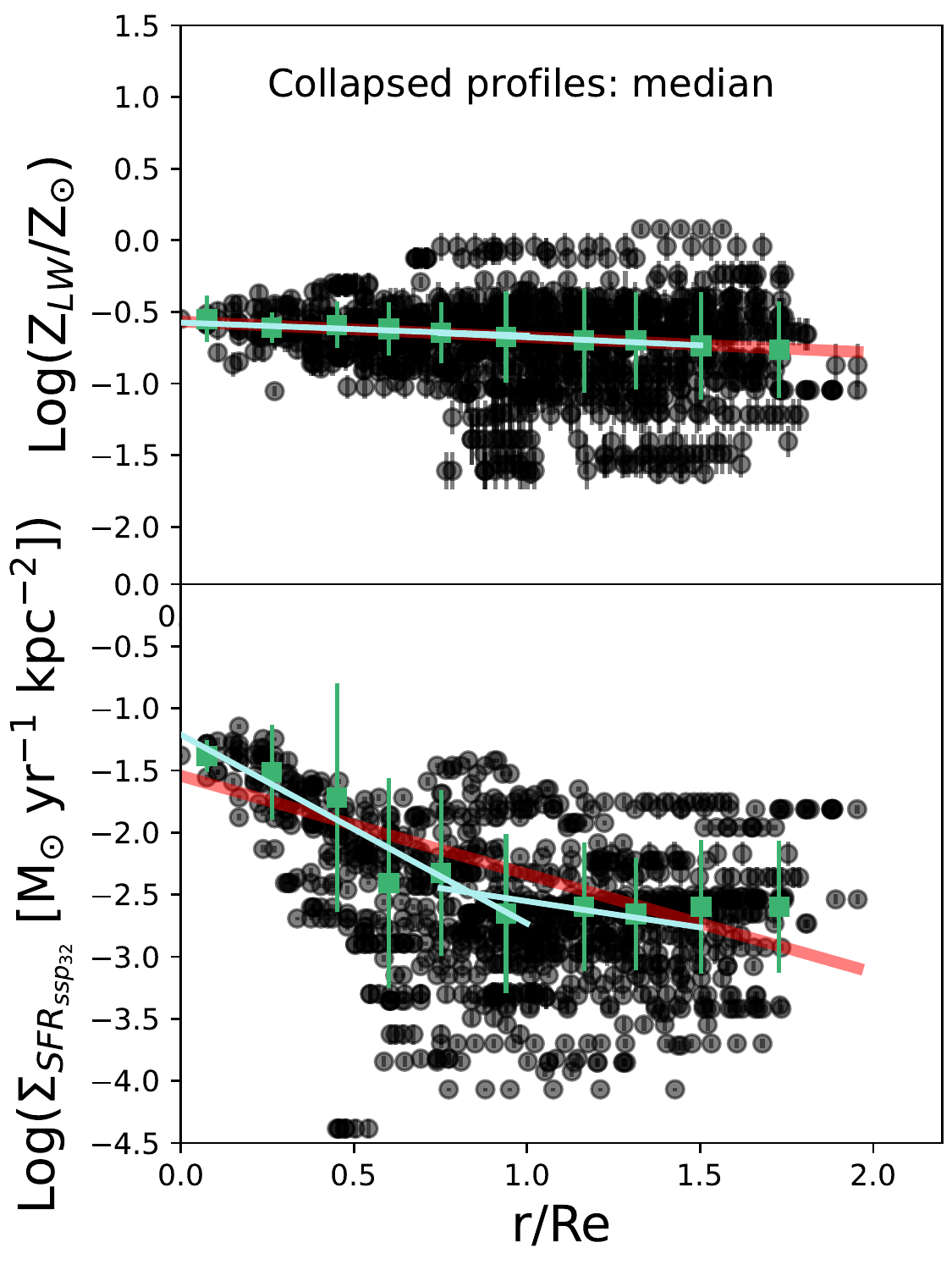}
    }\qquad
    \subfloat{%
	   \includegraphics[width=0.21\textwidth]{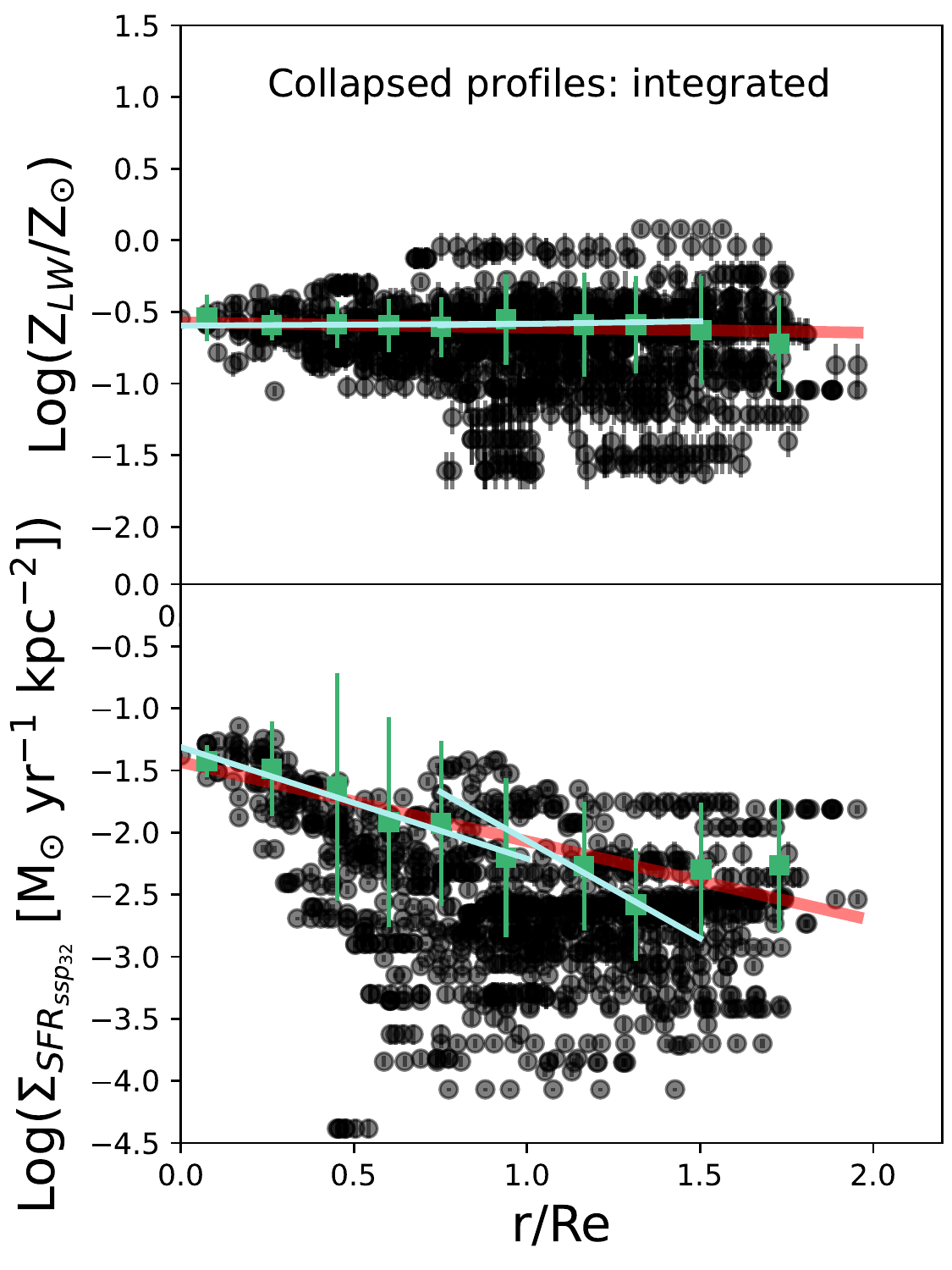}
     }\qquad
    \subfloat{%
	   \includegraphics[width=0.21\textwidth]{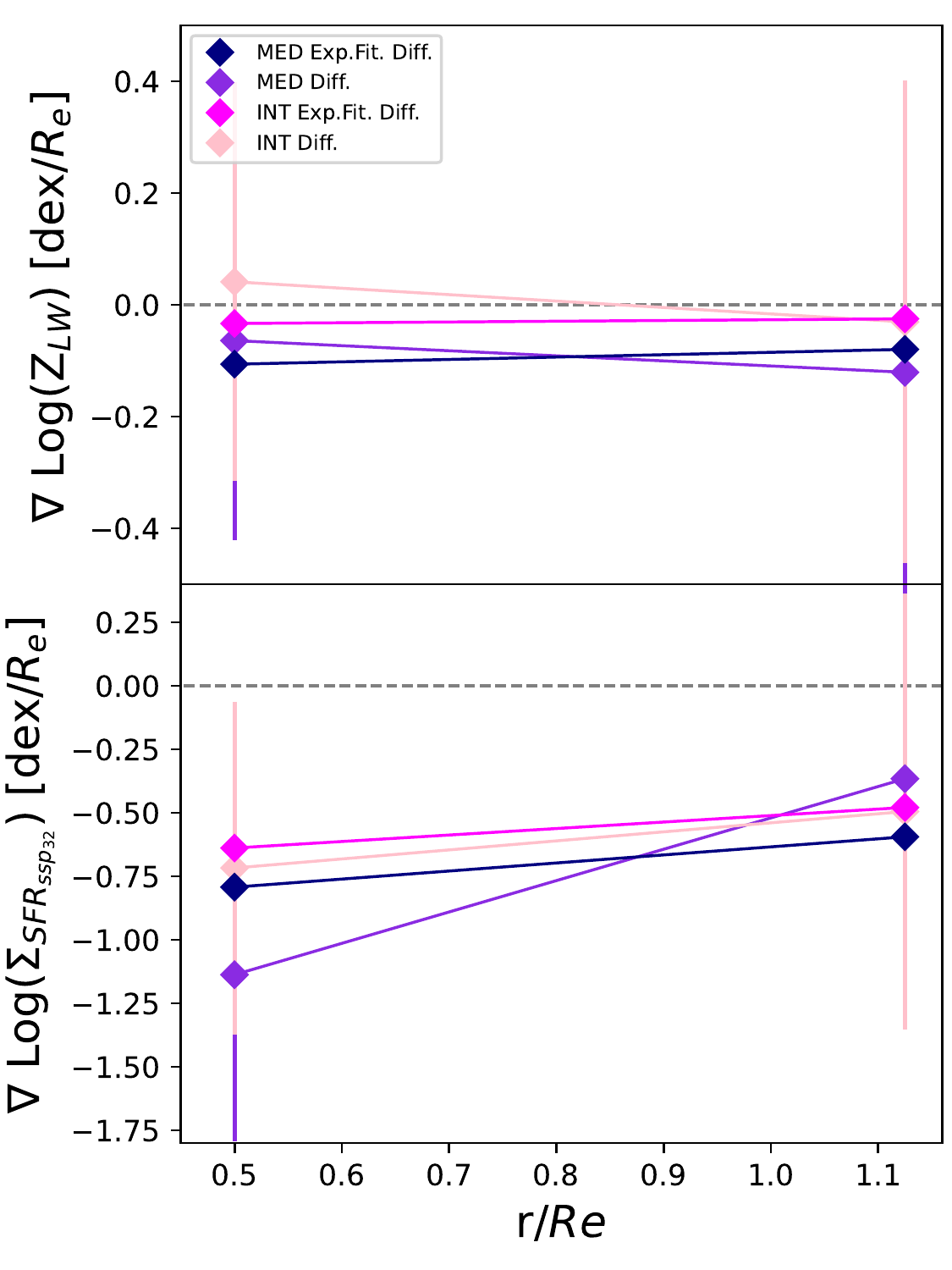}
    }\qquad
    \subfloat{%
	   \includegraphics[width=0.21\textwidth]{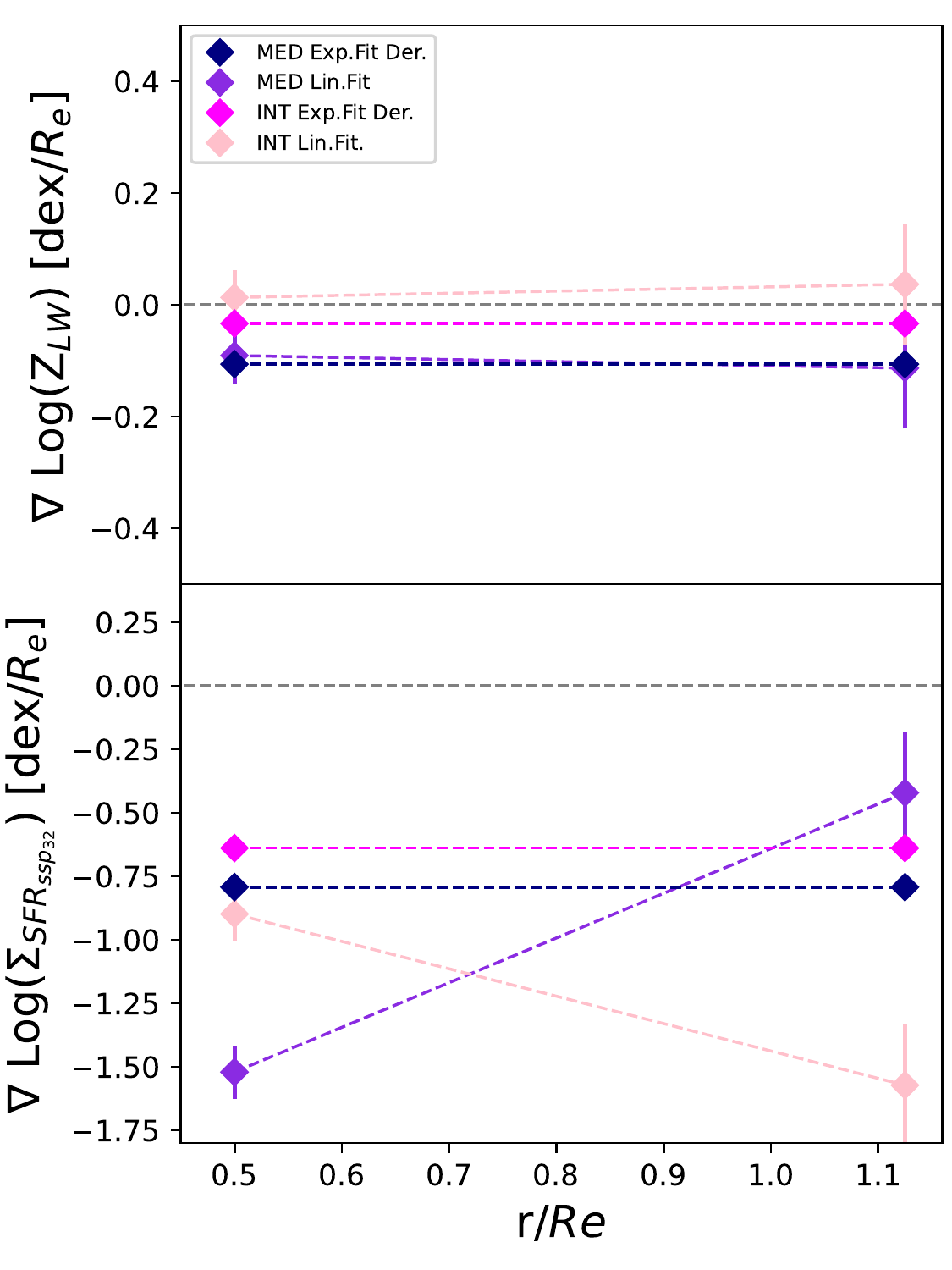}
     }\qquad
    \subfloat{%
	   \includegraphics[width=0.21\textwidth]{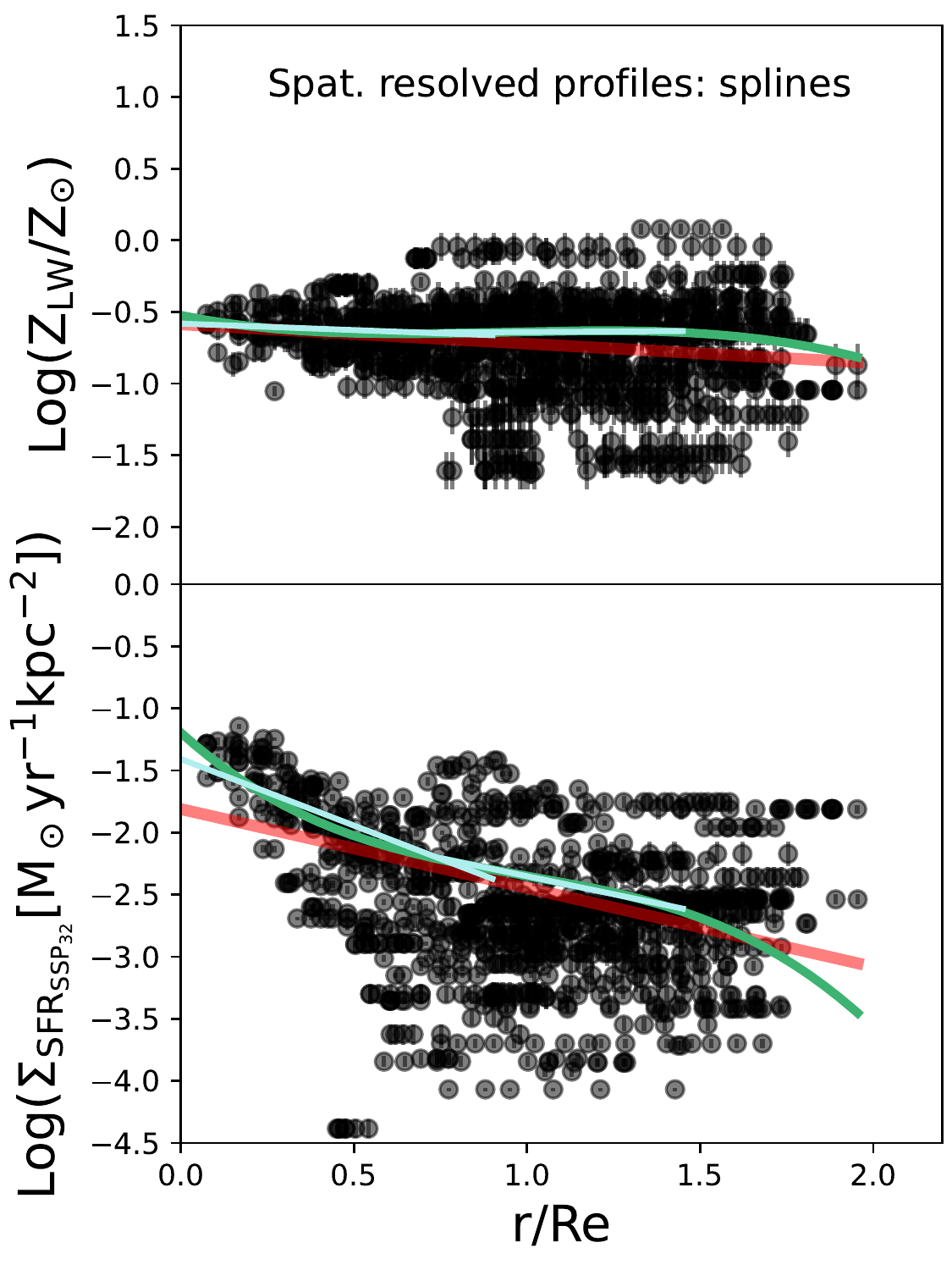}
    }\qquad
    \subfloat{%
	   \includegraphics[width=0.21\textwidth]{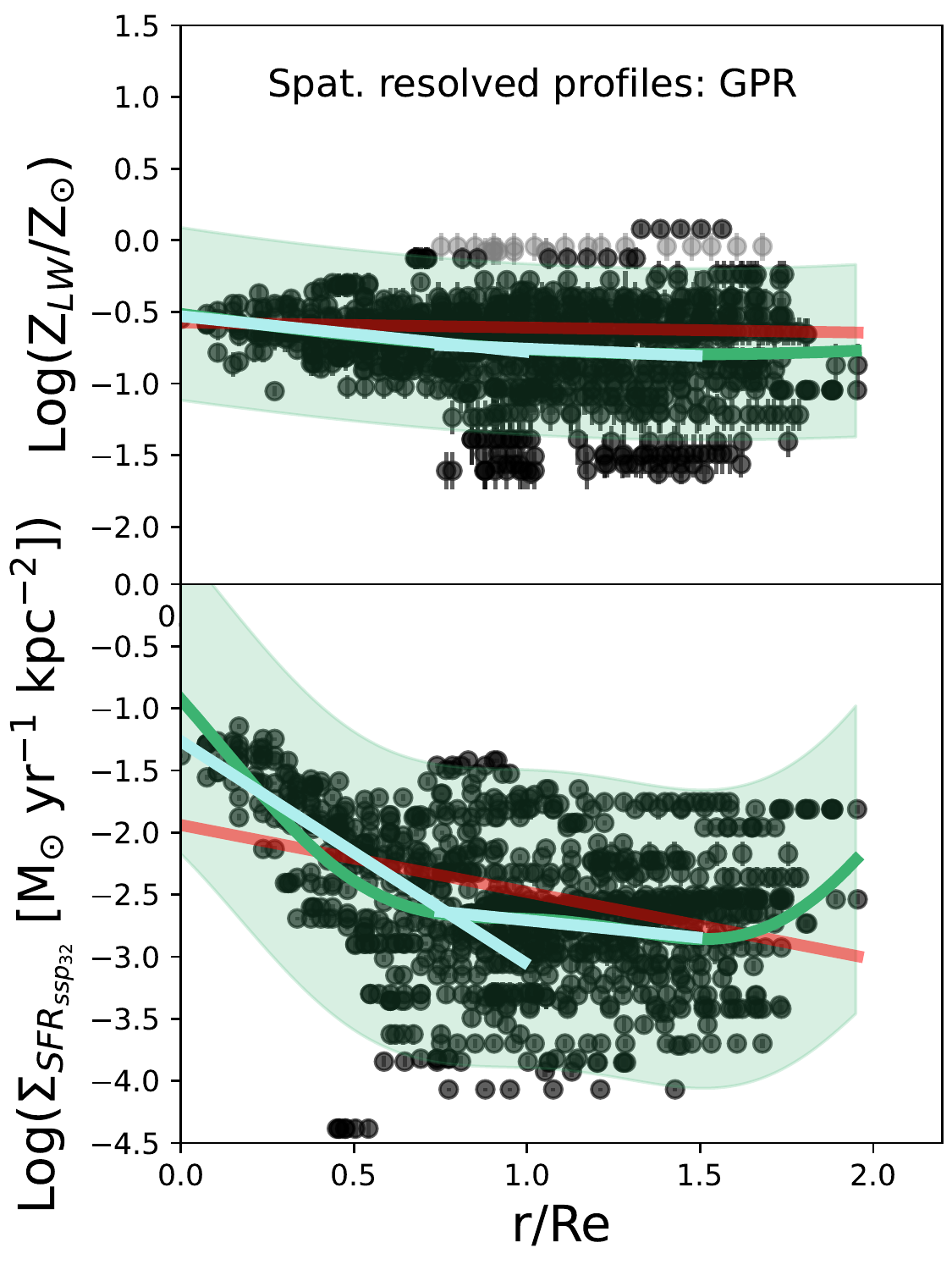}
    }\qquad
    \subfloat{%
	   \includegraphics[width=0.21\textwidth]{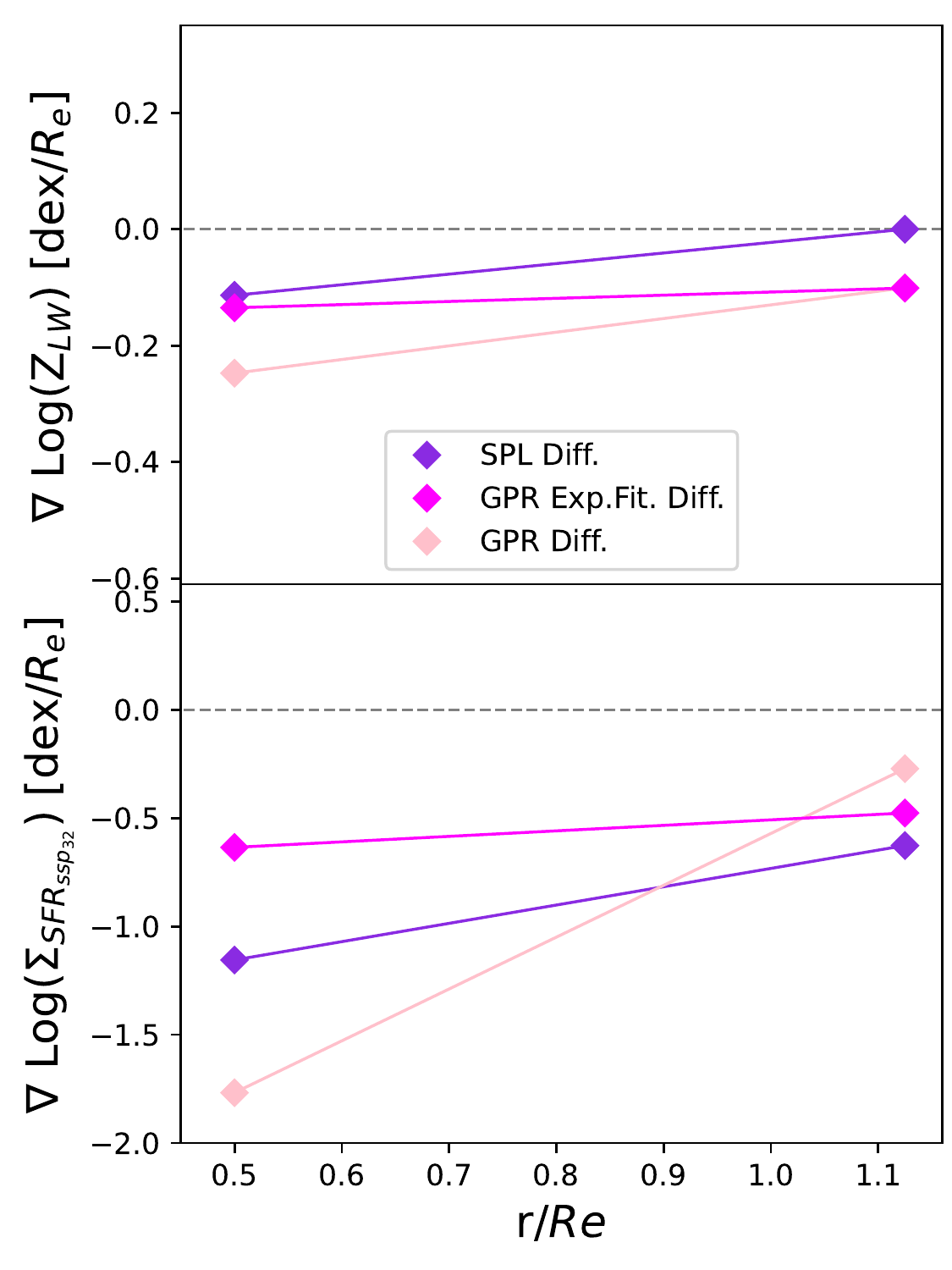}
    }\qquad
    \subfloat{%
	   \includegraphics[width=0.21\textwidth]{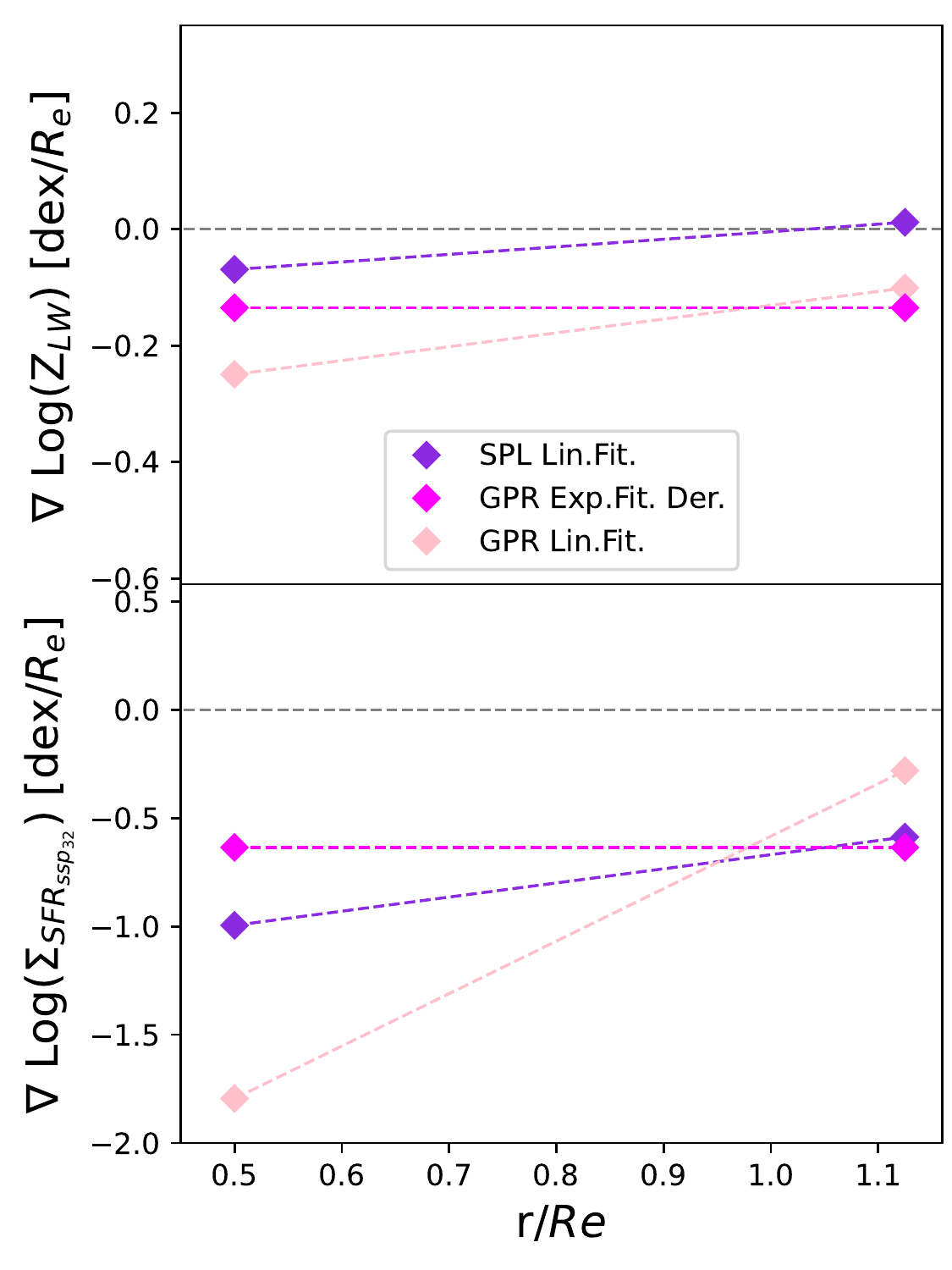}
    }\qquad     
\caption{Two profiles for the galaxy manga-7815-6101 are shown to compare the two collapsed methods for the profiles, as well as their gradients derivation. The black circles in all panels represent the individual data from each pixel in their respective maps. The left hand upper panel shows the case of the median collapsed profile, in both of its plots, the green squares represent the collapsed profile, with its associated error bars which correspond to 2$\sigma$, while the light blue lines show the linear fits performed to these collapsed profile in the two chosen radial ranges (0-1 $R_{e}$ and 0.75-1.5 $R_{e}$). In the next panel we show the same as in the last one but for the collapsed integrated method. In the first left hand bottom panel, we show the result of the non-linear spline interpolation in the green solid line, while the resulting linear fits are shown in the light blue lines. Similarly, in the next panel we show in green line, the posterior probability distribution of the non-linear GPR fit, the green shadow corresponds to its $95\%$ confidence interval, while the light blue lines also show the linear fits in the inner and outer regions. The red solid lines in all these panels represent the generalized fit performed over the previously introduced collapsed and non-linear profiles. In the right hand panels (both up and bottom) we show comparisons of the inner and outer gradients for this galaxy (located at the medium point of their measured radial range), color and shape coded by the method used to obtain them: using solid lines we show the results for $i)$ the differences between two points gradients, which were performed using the exponential fit over the collapsed or non-linear profiles (dark blue and magenta), $ii$ the differences between two points gradients obtained directly over the collapsed and non-linear profiles (purple and light pink). Similarly using dashed lines, we show the results for $iii$ the gradients obtained performing the derivative of the exponential fits performed over the collapsed and non-linear profiles (dark blue and magenta) and $iv)$ those obtained performing linear fits over the collapsed and non-linear profiles (purple and light pink).}
\label{Fig:example_profiles} 
\end{figure*}

The main goal of this paper is to understand how significantly these methods impact the final values of the gradients. To determine the best approach, particularly for dwarf galaxies showing a noisy nature in their radial distributions, such as the dwarf galaxies, we compare all methods to assess any systematic trends between them. This exercise becomes relevant because even \emph{subtle differences} in gradients can lead to different interpretations of the data and, ultimately, the underlying physics driving them.

Figure \ref{Fig:example_profiles} presents a summary of the points discussed above, using as example the galaxy manga-7815-6101, and two of its SP radial distributions: $Z_{LW}$ (intensive property) and $\Sigma_{SFR_{ssp}}$ (extensive property). Here we display the results for different fitting methods to obtain the gradients over the collapsed and non-linear profiles (Appendix \ref{Append:ExampleProfiles} shows the complete set of profiles for the same galaxy). The dots in black in the first two columns of this figure represent the pixels of the spatially resolved radial distribution of the aforementioned SP properties; the error bar associated with each dot is taken from the \textsc{pyPipe3D} dataproducts. The results for the collapsed profiles are shown in the leftmost upper panels (see Sec. \ref{Sec:ProfilesMedians} and \ref{Sec:ProfilesIntegrated}) of Figure \ref{Fig:example_profiles} in green squares, while their error bar represents 2$\sigma$. The result of using a linear fit between 0 and 1 $R_{e}$ and 0.75 and 1.5 $R_{e}$ to these median profiles is shown in light blue solid lines, while the results of the generalized exponential fit is shown in red lines. Similarly in the lower panels, the green solid lines show the non-linear profiles resulting from the Spline interpolation and the posterior probability distribution of the GPR fit, while the green shading shows their 95\% confidence interval. Similarly, the light blue lines show the linear fits performed to these distributions within the mentioned radial ranges, and the red lines show the generalized fit. In the next two panels of both rows, we present the results for the inner and outer gradients using all the methods over the collapsed integrated and non-linear GPR profiles. The solid lines represent the gradients obtained using the difference between two points for the generalized exponential fits (in dark blue and magenta according to the radial profile used) and directly over each of the profiles (in purple and light pink). In dashed lines we show the gradients obtained with the derivative of the generalized exponential fit over the different profiles (in dark blue and magenta) and those obtained with two linear fits over the same profiles (in purple and light pink).

A thorough discussion of this and Figure \ref{Fig:example_all_profiles} will be presented in the next Section, however, a quick inspection shows how the galaxy gradients depend on the profile characterization method employed.

\section{Profiles and gradients from the different methods}
\label{Sec:MethodSelection}

As noted in the Introduction, quantifying SP properties gradients and radial profiles in dwarf galaxies is extremely challenging due to their irregular, clumpy, and bursty nature. Therefore, in the present study, our primary goal is to understand which methods for deriving radial profiles and gradients are most robust in characterizing this unique type of galaxy.

In this section, we start looking for differences among various methods for deriving radial profiles and quantifying their impact on a specific method of obtaining gradients. We then generalize our findings to all the methods explored here.

\subsection{Comparison between different methods for characterizing radial profiles}

To compare the previously explained methods of deriving the radial profiles and the final gradients, we refer back to Figure \ref{Fig:example_profiles} which shows the example of two radial profiles: $Z_{LW}$ and $\Sigma_{SFR_{ssp}}$ for galaxy 7815-6101 (see Figures \ref{Fig:apena_7815-6101_MedProfiles}-\ref{Fig:apena_7815-6101_GPRProfiles} for the full set of profiles of the same galaxy). Overall both methods (collapsing the radial distributions and using non-linear fits agree on the general shape of the profiles. However, differences arise in more complex distribution as the $\Sigma_{SFR_{ssp}}$ profile compared to less scattered ones like the $Z_{LW}$ profile.

Starting with the collapsed profiles (the two leftmost upper side panels of Figure \ref{Fig:example_profiles}), we notice that the median and integrated profiles (green squares) behave similarly in the central parts of the galaxy. However, in both profiles, it is clear that towards the external regions the median profile tends to drop bellow the values reached by the integrated profile. This is a natural behavior when using the median statistical estimator which is less prone to being biased towards larger values in the distribution. Instead, the integrated profiles tend to weight less the individual points in the spatially resolved radial distributions that have lower values in the external parts of the galaxies, which as we have stated before are the ones expected to have the worst S/N. In the non-linear profiles (green solid lines on the two leftmost bottom panels of  Figure \ref{Fig:example_profiles}), the profile $Z_{LW}$, which is less scattered, shows that both techniques yield very similar results with the linear fits (light blue lines) being nearly identical to those from the integrated profile, just slightly less steep. In the case of the more complex profile ($\Sigma_{SFR_{ssp}}$), the differences are larger in the outer regions. The GPR method results in a much flatter linear fit in the outer range compared to the Cubic Spline fit, which is slightly decreasing. The Cubic Spline method, appears to provide an intermediate solution between the two collapsed methods. 

Complementing the profiles plots in Figure \ref{Fig:example_profiles}, the full set of profiles for all the sample is shown in Figure \ref{Fig:example_all_profiles}, and in Appendix \ref{Append:AllProfiles}. Overall, the methods to derive radial profiles are generally consistent. However, the non-linear Spline method introduces some divergence at the outermost parts of the galaxies (either towards negative or positive values), resulting in a significant scatter across the profiles for each property. In later sections, we will show that the divergence has minimal impact in the final derived gradients, since both, inner and outer gradients are measured before the divergence of the profiles occurs. Finally, in these plots we observe that the non-linear GPR method tends to produce profiles with lower values for a given galactocentric distance, similar to the collapsed-median method, while the non-linear spline and the collapsed integrated tend to yield higher values.

As shown in Figure \ref{Fig:example_profiles}, see also figures in Appendix \ref{Append:AllProfiles}, the various methods for deriving radial profiles can lead, as discussed below, to varying results which may influence the computation of SP properties gradients.

\begin{figure*}
    \includegraphics[width=\textwidth]{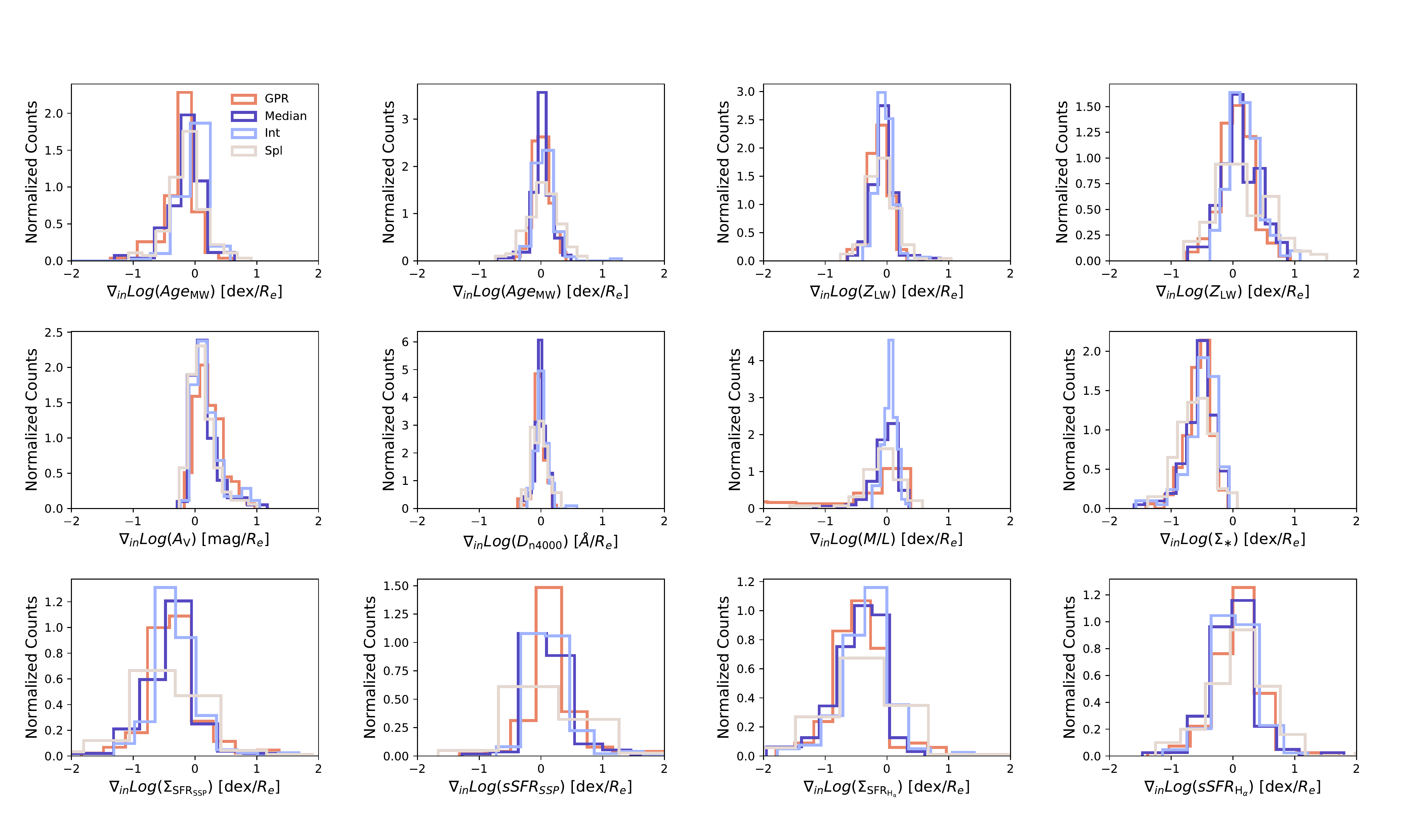}
  \caption{Distribution if the inner gradients derived with the method of the difference between two radial points (see Section \ref{Sec:diff_2points}), for the case of the generalized fits. Color code is used to differentiate the different methods used to derive the radial profiles, as explained in Section \ref{Sec:RadialProfilesMethods}.}
  \label{Fig:GradComparisson_dist_diff_in}
\end{figure*}

\begin{figure*}
    \includegraphics[width=\textwidth]{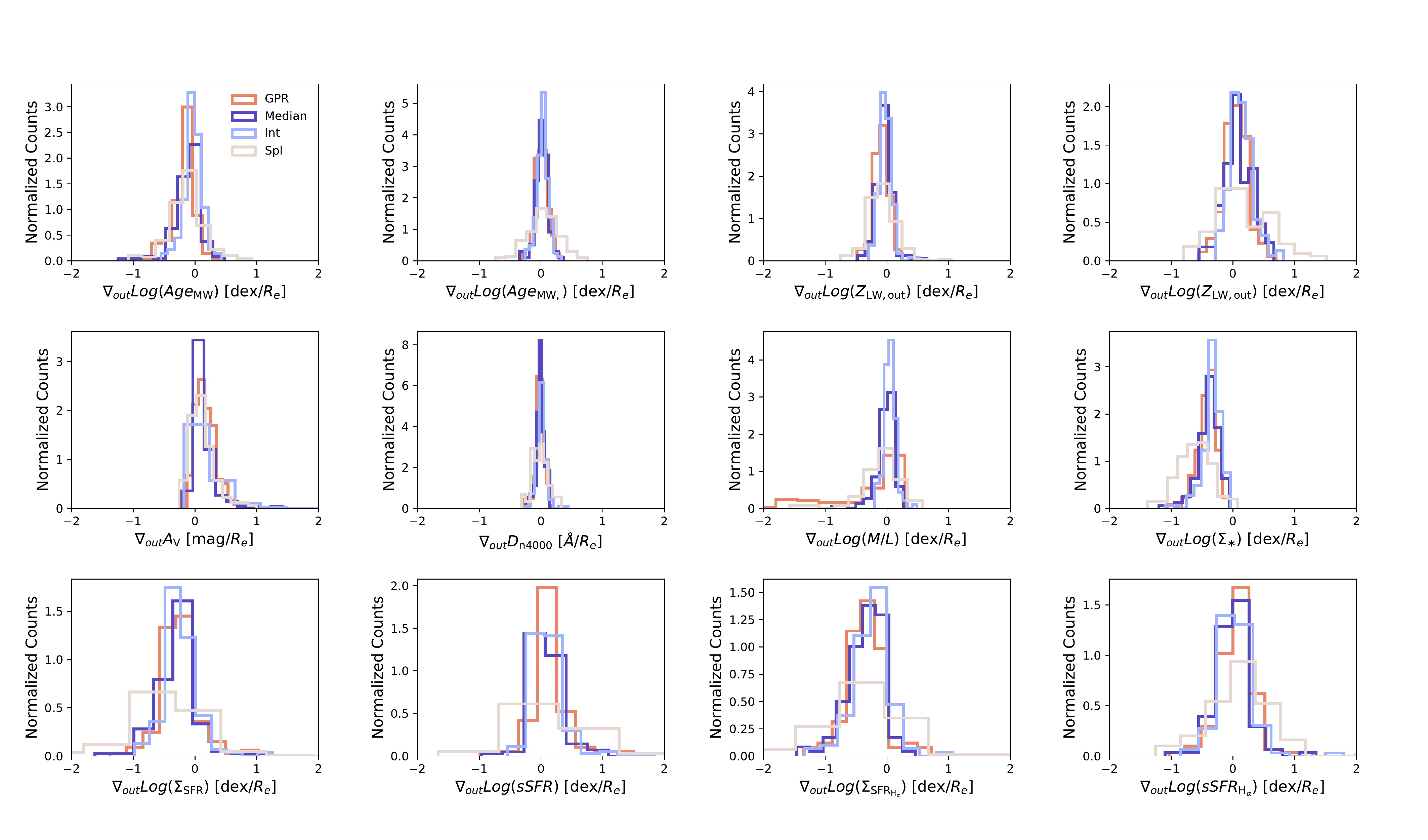}
  \caption{Same as Figure \ref{Fig:GradComparisson_dist_diff_in} but for the outer gradients.}
  \label{Fig:GradComparisson_dist_diff_out}
\end{figure*}

\begin{figure*}
    \includegraphics[width=\textwidth]{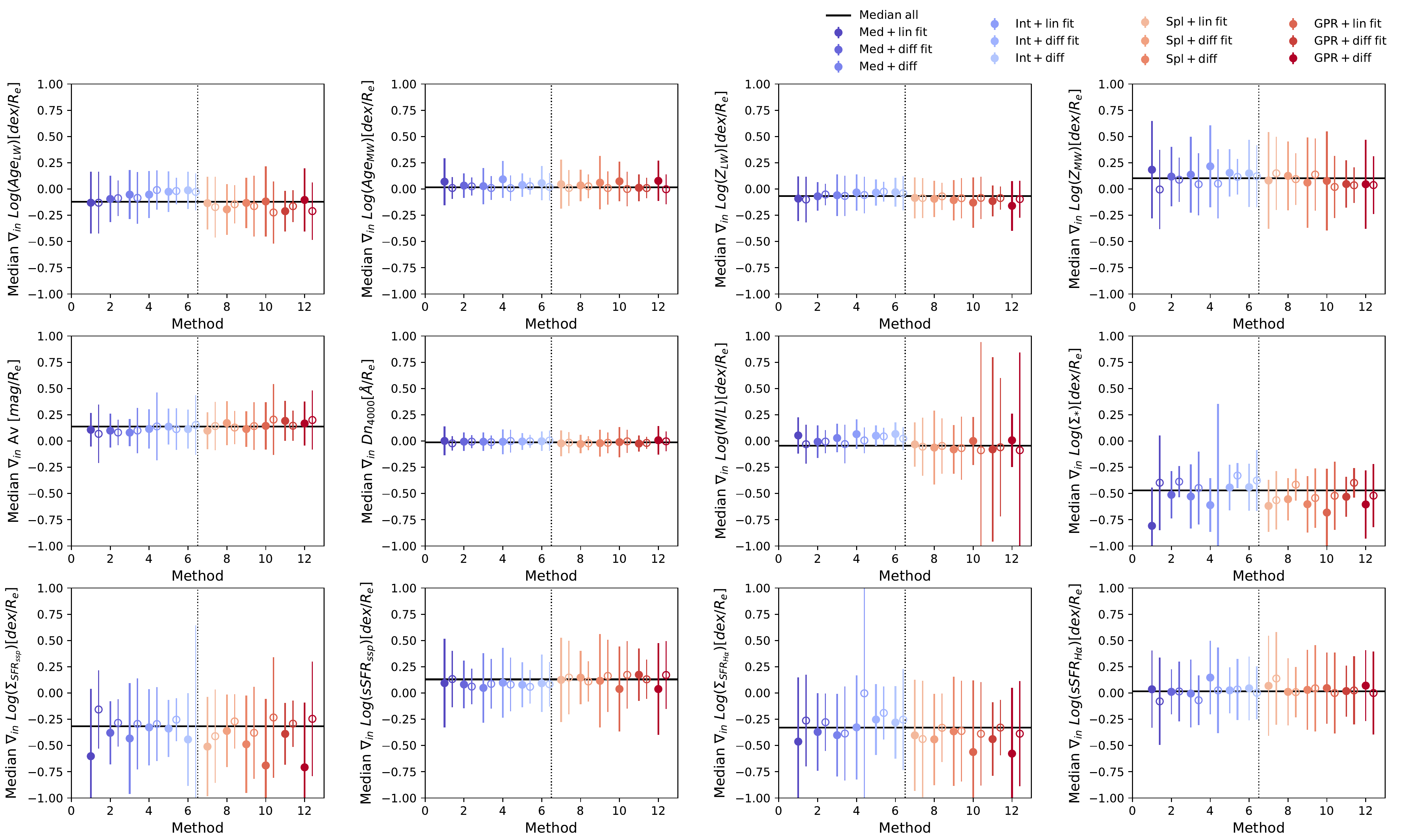}
  \caption{Median values of the full set of gradients for each stellar property. The color code is used to indicate the different methods used to derive the radial profiles and the gradient, as introduced in Section \ref{Sec:RadialProfilesMethods} and \ref{Sec:grad_methods}. (Med=collapsed median profiles, Int=collapsed integrated profiles, Spl=non-linear Spline profiles, GPR=non-linear GPR profiles. lin fit=linear fit gradients, diff fit=difference between two points from the generalized fit gradients, diff=difference between two points from the radial profiles)The filled and empty symbols show the inner and outer gradients respectively, and the error bar correspond to $1\sigma$ of the distribution of each set of gradients. The vertical solid line indicates the limit between the collapsed methods to derive the radial profiles and the non-linear ones.}
  \label{Fig:GradComparisson_grads_vs_median}
\end{figure*}

\subsection{Comparison between different methods for obtaining the gradient}\label{Sec:MethodsComparisson}

We begin our discussion by demonstrating how different methods for deriving radial profiles and gradient definitions affect the results on $\Sigma_{SFR_{ssp}}$ and $Z_{LW}$ from galaxy 7815-6101.

The two rightmost panels of Figure \ref{Fig:example_profiles}, show the final internal and external gradients of galaxy 7815-6101, with different methods color coded. All methods yield a similar result for the simpler profile ($Z_{LW}$), except the collapsed integrated profile, which gives positive gradients, rather than negative. On average, differences are less than $\sim0.1$ dex $R_e^{-1}$ among the methods, but the non-linear spline profiles gives higher outer gradient values. The non-linear GPR profiles yield the lowest values whereas the collapsed median profile gives consistent inner and outer gradients. 

For the complex profile ($\Sigma_{SFR_{ssp}}$), the two-point radial difference and the collapsed median profile result in larger differences between $\nabla_{in}$ and $\nabla_{out}$ similar to the non-linear GPR profile. The example of galaxy 7815-6101 shows the complexity and the large dispersion of quantities derived from dwarf galaxies, which in the case of $\Sigma_{SFR_{ssp}}$ can be as large as $\sim1$ dex $R_e^{-1}$. While this galaxy may not represent the entire sample, we will next explore statistical trends for the whole distribution of galaxies.

Figures \ref{Fig:GradComparisson_dist_diff_in} and \ref{Fig:GradComparisson_dist_diff_out} present the distribution of inner and outer gradients for the two-point difference approach (Section \ref{Sec:diff_2points}). The four profile methods are in reasonable agreement. However, specific quantities, such as the inner gradients of $Age_{LW}$, $Z_{MW}$, $M/L$ and $\Sigma_{SFR}$ and ${sSFR}$, both for SSP and H$_\alpha$, exhibit larger scatter and some disagreement between methods. For the outer gradients, the situation is quite similar. Similar Figures were obtained for linear and generalized fits, Section \ref{Sec:RadialProfilesMethods}, but not shown for the sake of space, they lead to similar conclusions. Finally, detailed analysis of gradient distributions is beyond the scope of this paper, but it will be further explored in Cano-Díaz et al. (in prep.) in terms of galaxy formation and evolution, see also Section \ref{Sec:discussion_grads_vs_mass}.

What is the ``true" gradient in a galaxy? Given all the methods employed in this paper, it is not possible to provide a definitive answer to this question or to determine which method provides the most robust measurement of a gradient. For this reason, we created a metric to compare all the methods objectively. For each galaxy, we compute the median value from all the gradients available for that galaxy:
\begin{equation}
    \text{med}\left(\{\nabla \mathcal{G}\}_i\right),
\end{equation}
where $\mathcal{G}$ represents the galaxy property and the subscript $i$ represents the set of all the gradients available for that galaxy, based on the methods explained in Section \ref{Sec:grad_methods}.

Figure \ref{Fig:GradComparisson_grads_vs_median} shows the median inner and outer gradients obtained for each methodology (filled and empty colored circles, respectively) to visualize the overall differences. The black line represents the overall median of all inner gradients between the methods. For clarity, we omit the results of the derivative method from the general fit gradients, as they closely resemble those obtained from the two-point difference method. These results are also excluded from Figures \ref{Fig:median_offset_psf}-\ref{Fig:median_offset_methods}, described below.

First, we note that the dispersion around the median can be as large as the gradients themselves for most of the properties, see also Table \ref{Table:MedianGrads_Stats}. Most gradients are flat, so even small fluctuations can appear significant. More important, for properties gradients like $Age_{LW}$, $Z_{LW}$, $Z_{MW}$, $A_V$ and  $M/L$ non-linear methods give gradients with opposite sign compared to the collapsed methods, which could drastically change the physical interpretations. Notice that error bars are not related errors in the median estimate but to $68\%$ of the distribution. As for the outer gradients, we obtained similar conclusions. regardless the method, $Age_{MW}$, $Z_{LW}$, and $D_{n4000}$ tend to have flat gradients throughout the galaxy. This is consistent with these SP properties having less complex profiles (see figures in Appendix \ref{Append:AllProfiles}). Also, the non-linear approaches (Spline interpolation and GPR), systematically result in lower gradients, while the collapsed ones result on the highest gradients (except for $sSFR$ and $A_V$). The largest gradient variations are found for the surface density properties ($\Sigma_{*}$ and $\Sigma_{SFR}$), which display more scatter and complex radial profiles (see again Figures in Appendix \ref{Append:AllProfiles}). We conclude that some of the methodologies presented here may introduce changes in the final interpretations of the results However, this is to be expected for dwarf galaxies, whose properties are very irregularly distributed in space. In fact, different methods capture different aspects of this irregular spatial distribution. Given this irregularity, it is difficult to define a single quantity (the gradient) to characterize the spatial distribution. The median gradient of each property calculated here is a compromise in this situation, and its dispersion provides a kind of natural quantification of the intrinsic irregularity mentioned above.

Finally, it is important to note that linear fits cannot always be estimated on all galaxies. As mentioned in Section \ref{Sec:grad_methods}, we lack sufficient data to perform the linear fits using the collapsed methods, leading to loose $\sim22-26\%$ of galaxies in the inner region, and $\sim57-60\%$ for the outer region. For the non-linear methods, this situation improves. The loss of outer gradients using the Spline interpolation method is $\sim4-14.5\%$, while the Spline interpolation method fails in rare cases for the inner gradients. Instead, when using the GPR method, no galaxies are lost, making it a robust method for extracting the inner and outer gradients for the entire sample.

In conclusion, this section demonstrates that certain SP properties are more sensitive to the methodology used for characterizing radial profiles, as shown in Figure \ref{Fig:example_profiles}. Gradients are influenced not only by the profile characterization method, see Figures \ref{Fig:GradComparisson_dist_diff_in} and \ref{Fig:GradComparisson_dist_diff_out}, but also by the method used for estimating the gradients themselves, see Figure \ref{Fig:GradComparisson_grads_vs_median}. Table \ref{Table:MedianGrads_Stats} summarizes how different methods contribute to the gradients showing that the dispersion among methods is comparable to the median value of all methods. Therefore, to provide a robust quantification of a gradient, we opt to use the median value and consider the dispersion between methods as the error bar. In any case, we provide a table with all the gradients derived with all the methods as online material for interested readers. In Appendix \ref{Append:OnlineTable} a description of the table is provided.

\section{The profiles and gradients of dwarf galaxies}\label{Sec:InVstGradiens}

In the preceding Sections we demonstrated that radial profiles and their gradients vary depending on the method used. To minimize this variance, we defined the median gradient across all methods as a robust measure, with the variance between methods providing the corresponding uncertainty due to the natural irregular spatial distribution of properties in dwarf galaxies, see Subsection \ref{Sec:MethodsComparisson}. We now use these gradients to discuss the implications for dwarf galaxies.

Precisely because of the spatial irregularity of dwarf galaxies, a single gradient to characterize the radial profiles is not sufficient. Thus, we measure at least two gradients, one internal and one external (see subsection \ref{Sec:linear_fits}). Our internal gradient, defined at $0\leq R\leq 1 R_e$, is the most widely used in the literature to quantify galaxy gradients, so the external one can be seen as a complement. Using both gradients for several galaxy properties in the $\nabla_{out}$ Vs. $\nabla_{in}$ diagram, we can explore the radial assembly and evolution processes of 124 MaNDala galaxies, currently the largest sample of dwarf galaxies with both IFS and photometric information. Figure \ref{Fig:QuadrantsInVst} helps interpret this diagram, by illustrating different cases. Each quadrant shows different profile shapes within $0\leq R\leq1.5 R_e$ and $0.75\leq R\leq 1.5 R_e$ for the inner and outer gradients. 

Points in the first or third quadrants indicate profiles that consistently increase or decrease. Points falling in the second quadrant indicate profiles decreasing in the inner region and at the same time increasing in the outer region, whereas points falling in the fourth quadrant reflect a profile increasing in the inner region and decreasing in the outer region. Particular cases occur when: a point lies along the one-to-one relation, meaning that gradients do not change between inner and outer regions, and  points near the zero values indicate flat gradients, i.e. the studied galaxy property remains nearly the same over the entire region.

Table \ref{Table:MedianGrads_Stats} presents key statistics of inner and outer gradients, based on the median for all methods. The first column indicates the galaxy property. The next columns show the mean, median, and standard deviation of the distribution of the corresponding inner and outer gradients. The last two columns use the mean and median to determine positions within the quadrants, see Figure \ref{Fig:QuadrantsInVst}. Except for $M/L$, the mean and median are stable between properties. Interestingly, dwarf galaxies tend to avoid quadrant II, favoring quadrant III (decreasing profiles), followed by quadrant I (increasing profiles). That is, decreasing overall profiles are preferred, followed by increasing overall profiles. A notable result is that if a dwarf galaxy has a negative inner profile, its outer profile is also negative.

\begin{figure*}
\centering
    \subfloat{%
	   \includegraphics[width=0.45\textwidth]{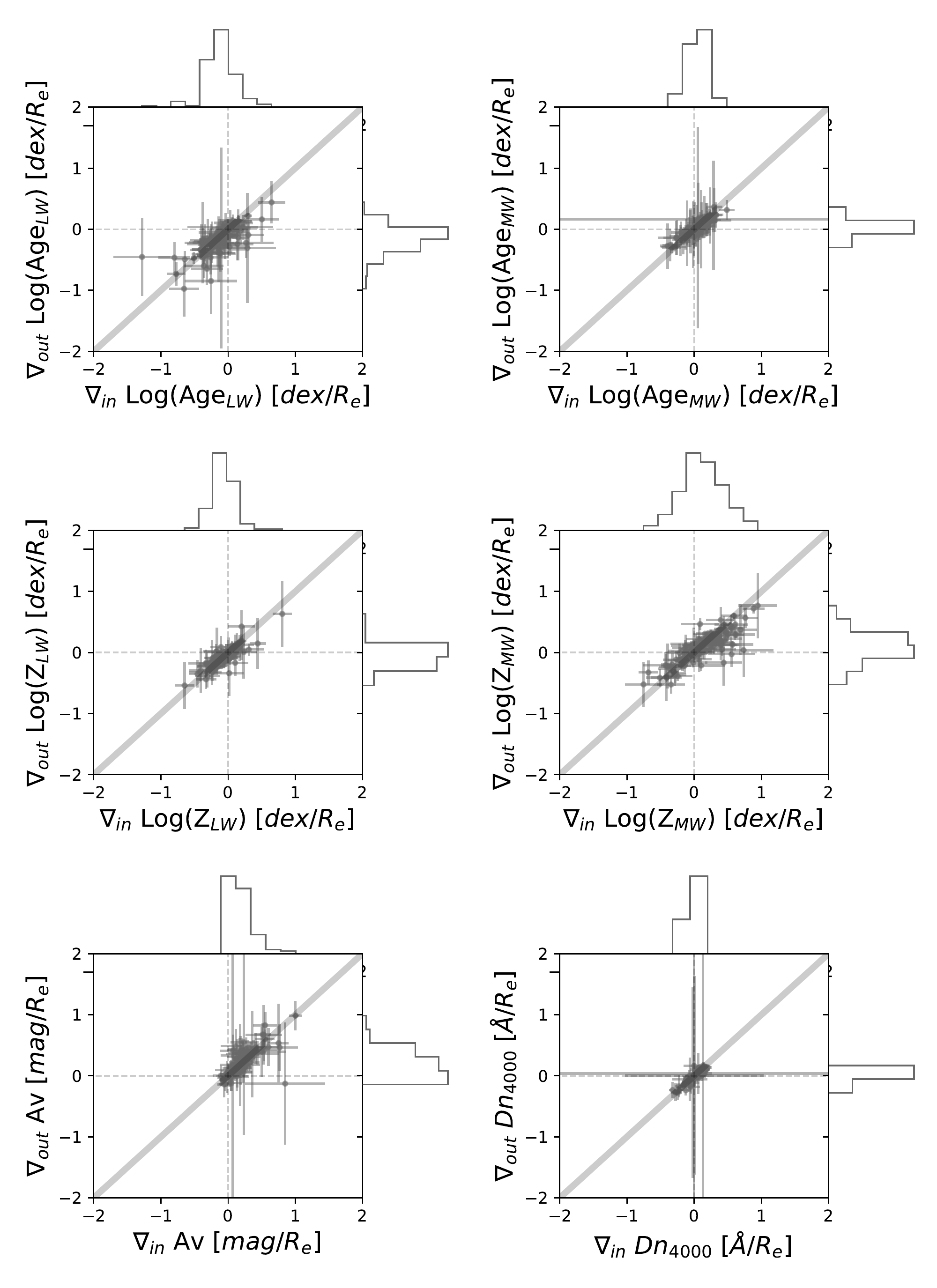}
    }\qquad
    \subfloat{%
	   \includegraphics[width=0.45\textwidth]{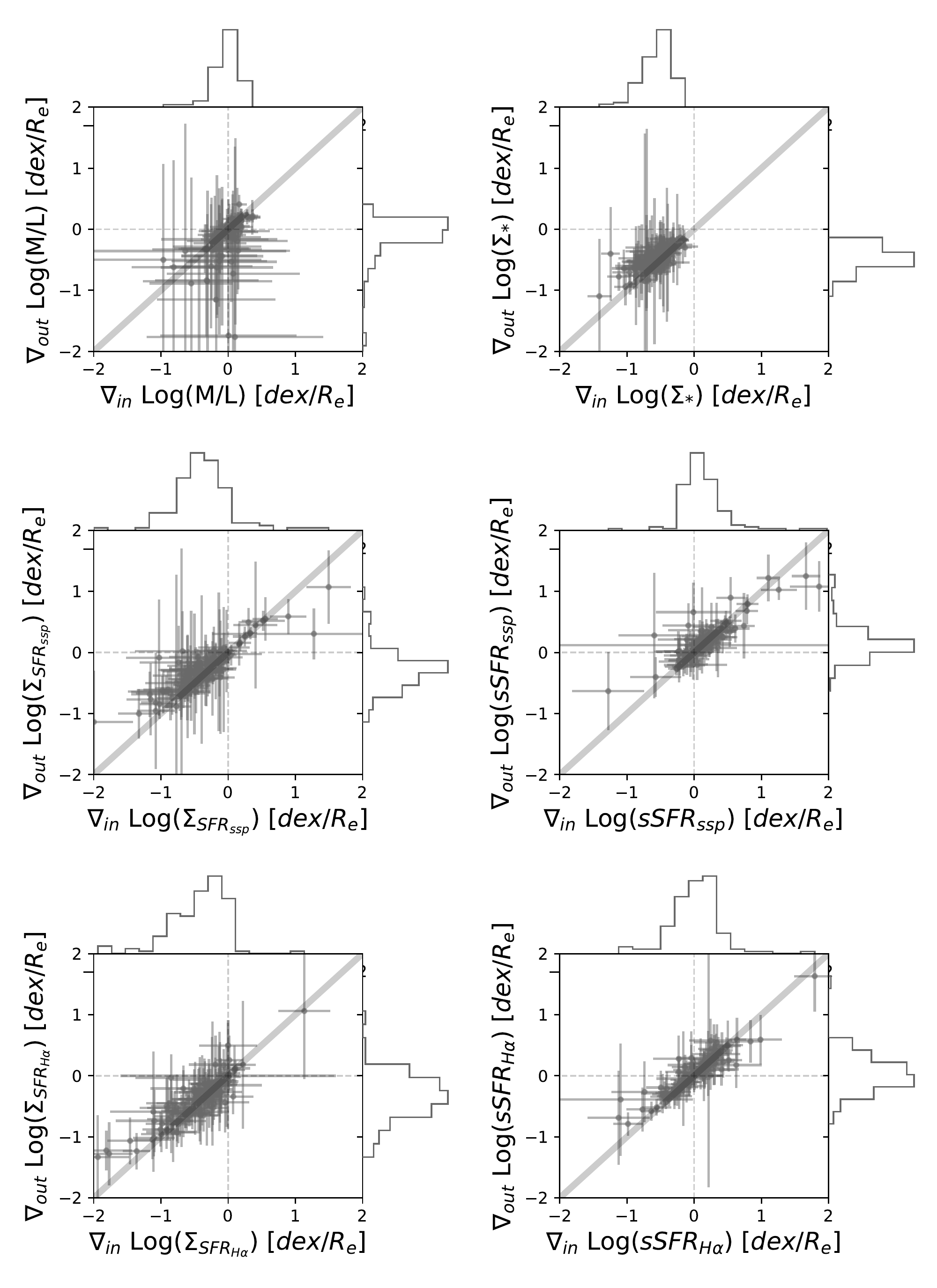}
    }\qquad
\caption{Comparisons between the inner (0-1$R_{e}$) and outer (0.75-1.5$R_{e}$) median gradients, where the error bars represent the $1\sigma$ dispersion among all the methods. The solid line represents the one-to-one relation. On the top and right side of each plot a histogram of the distribution of the gradients in the diagram is provided.} 
\label{Fig:CompareInVsOutGrad}
\end{figure*}

Figure \ref{Fig:CompareInVsOutGrad} shows the $\nabla_{out}$ Vs. $\nabla_{in}$ for each of the ten properties studied throughout the entire dwarf galaxy sample. To interpret these results, we refer to the scheme in Figure \ref{Fig:QuadrantsInVst}. The solid line represents the one-to-one relation, and histograms of the distribution of the data are shown outside the plotting region. Figures \ref{Fig:all_profiles_medianprof}--\ref{Fig:all_profiles_gprprof} in Appendix \ref{Append:AllProfiles} show the radial profiles for all MaNDala galaxies using different methods: collapsed median, collapsed integrated, non-linear spline, and non-linear GPR methods (see Sect. \ref{Sec:RadialProfilesMethods}). The blue squares with error bars show the corresponding medians and 1$\sigma$ distribution in the radial bins. Despite the different methods, the overall trends are similar. Therefore, for the analysis below, the reader may focus solely on Figure \ref{Fig:all_profiles_medianprof}. Looking at these radial profiles and their characterization with the inner and outer gradients shown in Figure \ref{Fig:CompareInVsOutGrad}, complementing them with the statistics of these gradients reported in Table \ref{Table:MedianGrads_Stats}, we can highlight the following results for each of the properties studied here (for a discussion on the implications of these results, see Section \ref{Sec:implications}):

\begin{figure}
  \centering
    \includegraphics[width=0.9\columnwidth]{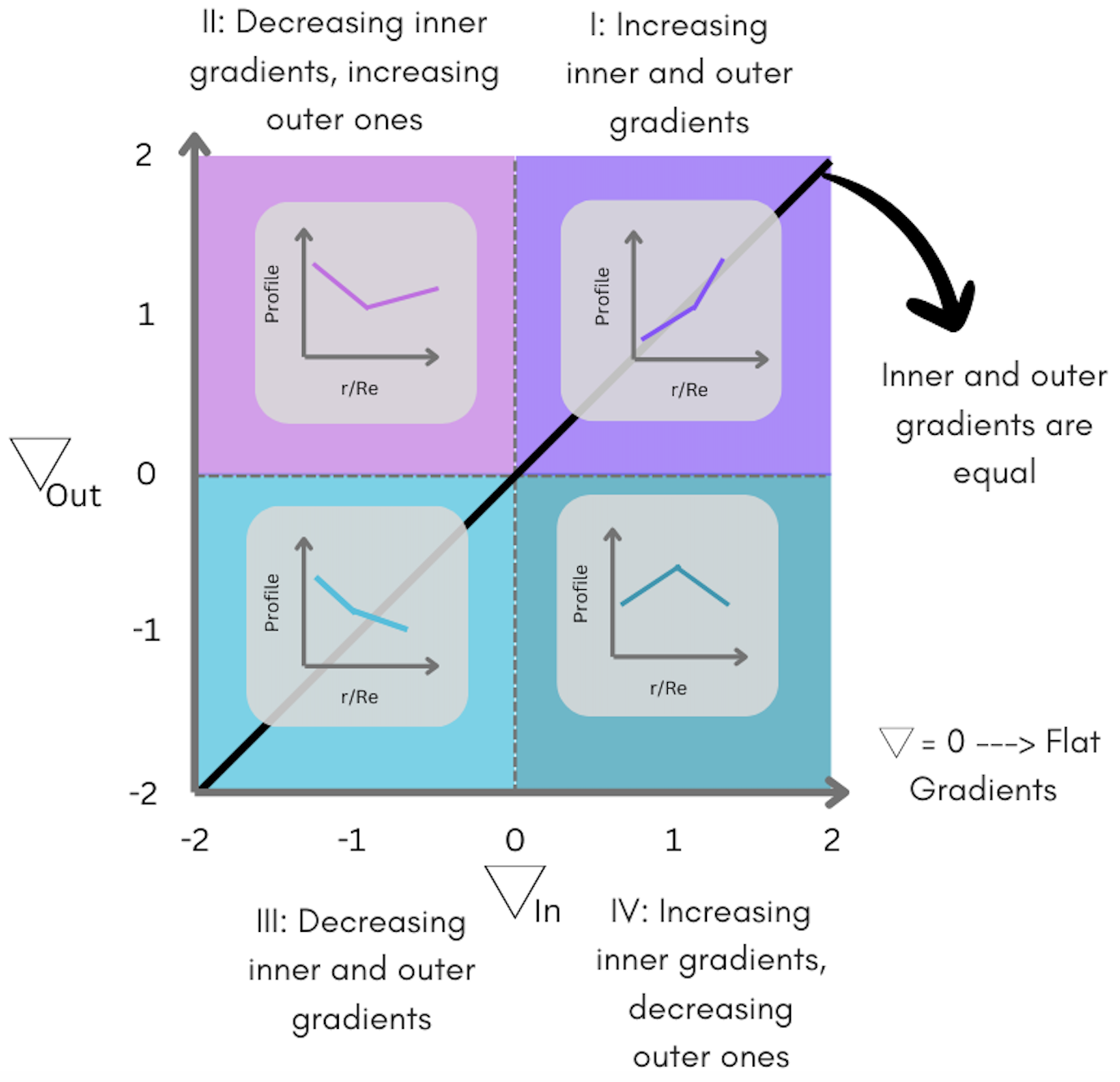}
  \caption{\label{Fig:QuadrantsInVst} An illustration that shows the interpretation of the gradients falling in each of the four quadrants of the $\nabla_{out}$ Vs. $\nabla_{in}$ diagram, along with the cases of gradients falling in the one-to-one relation and along the zero lines. Inside each quadrant, illustrative shapes of any of the radial profiles from where the gradients may come from, when located in any of the four quadrants.}
\end{figure}

\begin{table*}
\footnotesize
\centering
    \begin{tabular}{ l c c c c c c c c }
    \hline
    \hline
 & \multicolumn{3}{c}{$\nabla_{in}$ [dex/[$R_{e}$]$^{\dagger}$} & \multicolumn{3}{c}{$\nabla_{out}$ [dex/[$R_{e}$]$^{\dagger}$} \\ 
& Mean & Median & $\sigma$ & Mean & Median & $\sigma$ & Quadrant(mean) & Quadrant(median)   \\
\hline
Log(Age$_{LW}$) & -0.12 & -0.9 & 0.25 & -0.16 & -0.12 & 0.20 & III & III\\
Log(Age$_{MW}$) & 0.04 & 0.05 & 0.15 & 0.02 & 0.02 & 0.12 & I & I \\
Log($Z_{LW}$)  & -0.07 & -0.07 & 0.18 & -0.07 & -0.07 & 0.15 & III & III \\ 
Log($Z_{MW}$)  & 0.13 & 0.10 & 0.31 & 0.09 & 0.10 & 0.24 & I  & I \\ 
$A_V$ & 0.18 & 0.13 & 0.19 & 0.19 & 0.14 & 0.20 & I & I \\ 
$D_{n4000}$ & -0.01 & 0.0 & 0.09 & -0.01 & -0.01 & 0.08 & III & III \\ 
Log($M/L$)  & -0.03 & 0.00 & 0.20 & -0.14 & -0.04 & 0.42 & III & IV \\ 
Log($\Sigma_{*}$) & -0.56 & -0.53 & 0.22 & -0.48 & -0.47 & 0.17 & III & III  \\ 
Log($\Sigma_{SFR_{SSP}}$) & -0.38 & -0.41 & 0.44 & -0.30 & -0.32 & 0.33 & III & III  \\ 
Log(sSFR$_{SSP}$)  & 0.16 & 0.10 & 0.43 & 0.16 & 0.13 & 0.30 & I & I \\ 
Log($\Sigma_{SFR_{H\alpha}}$) & -0.44 & -0.35 & 0.43 & -0.37 & -0.33 & 0.37 & III & III \\ 
Log(sSFR$_{H\alpha}$) & 0.01 & 0.00 & 0.37 & 0.04 & 0.02 & 0.30 & I & I \\ 
\hline  
\hline  
    \end{tabular}
\caption{ Statistical properties of the internal ($\nabla_{in}$) and external ($\nabla_{out}$) gradients, measured between 0-1 $R_{e}$ and 0.75-1.5 $R_{e}$ respectively, obtained with the median of all the tested methods. The first three columns show the mean, median and standard deviation of each distribution for the internal gradients, while the next three, show the same statistical properties but for the external gradients. Finally, using the mean and median values of inner and outer gradients, the last two columns specify their corresponding quadrants. Notice that this simple statistics show that dwarf galaxies in MaNDALA avoid quadrant II. $^{\dagger}$ Units change to [mag/$R_{e}$] and [$\AA$/$R_{e}$] for the gradients of $A_V$ and $D_{n4000}$ respectively.}\label{Table:MedianGrads_Stats}
\end{table*}

\subsection{Ages and $D_{n4000}$ index}

 Both the inner and outer luminosity-weighted age gradients are mostly slightly negative, with 73\% falling in the quadrant \textit{III}, with many dwarfs having outer gradients slightly more negative than the inner ones (medians of $-0.10$ and $-0.16$, respectively, see Table \ref{Table:MedianGrads_Stats}). However, 25\% of galaxies have positive inner gradients, with a significant fraction of them showing flat or negative outer gradients (quadrant \textit{IV}). For the mass-weighted ages, both inner and outer gradients are closer to 0, with a small dispersion around this value. Most dwarfs have slightly positive gradients, lying in the quadrant \textit{I} (median inner and outer gradients are 0.05 and 0.01, respectively, see Table \ref{Table:MedianGrads_Stats}). Galaxies with inner negative gradients, also tend to have negative outer gradients, lying in quadrant \textit{III}. In conclusion, age gradients in MaNDala galaxies are diverse, showing a distribution around the $0$ value, with the mass-weighted ones biased towards positive values, while the luminosity-weighted ones towards negative values; see also Figure \ref{Fig:all_profiles_medianprof}. 

Since $Age_{MW}$ reflects more long-lived stellar populations that dominate in the mass, while $Age_{LW}$ reflects more younger stars, the fact that the $Age_{MW}/Age_{LW}$ ratio increases with radius (as can be inferred from Fig. \ref{Fig:all_profiles_medianprof}) suggests that outer regions experienced late episodes of SF. As seen in this Figure, $Age_{MW}\gg{}Age_{LW}$ at all radii. In \citet{CanoDiaz2022}, we showed that the global median $Age_{MW}/Age_{LW}$ ratio for MaNDala galaxies is high, $\sim 14$, with significant dispersion, indicating recent episodes of SF and diverse SF histories. This suggest that most MaNDala galaxies formed the majority of their stars early on at all radii, but later, or the most, recent episodes of SF (which do not contribute significantly to the total stellar mass) also occurred mostly in the outskirts. The fact that the $Age_{MW}$ gradient is near to $0$ or even positive in many cases, may suggest stellar migration processes and feedback-driven gas ejection from the center to the outskirts. In Section \ref{Sec:implications} we discuss in more detail these possibilities. 

As for the $D_{n4000}$ gradients, almost all of them fall in quadrant \textit{III} of Figure \ref{Fig:CompareInVsOutGrad}, lying along the one-to-one line and showing minimal dispersion. The median radial profile is slightly decreasing (Figure \ref{Fig:all_profiles_medianprof}), similar to the $Age_{LW}$ profile. This is expected since $D_{n4000}$ is a proxy for luminosity-weighted age.  However, the $D_{n4000}$ radial profiles are more regular and show less variation than those of $Age_{LW}$.

\subsection{Metallicities}

The inner and outer $Z_{LW}$ gradients are generally negative, with 65\% of galaxies falling in quadrant \textit{III}, lying nearly along the one-to-one line, and with values not too far from the $0$ value (median of $-0.07$ for both, see Table \ref{Table:MedianGrads_Stats}). In contrast, the inner and outer $Z_{MW}$ gradients tend to be positive, with 60\% in quadrant \textit{I} with a median value of $0.10$ for both, but with a large dispersion, especially for the inner gradients (see the histograms in the corresponding panel of Fig. \ref{Fig:CompareInVsOutGrad}). Dwarf galaxies tend to have flatter outer $Z_{MW}$ gradients compared to inner ones, see also Figure \ref{Fig:all_profiles_medianprof}.  
Inner regions ($R<R_e$) have on average $Z_{LW}>Z_{MW}$, while at larger radii, both metallicities are similar. In \citet{CanoDiaz2022} we showed that the median of the overall $Z_{MW}/Z_{LW}$ ratio of MaNDala galaxies is $\sim 0.7$, with a large dispersion. The SP of these galaxies have larger differences in the $Z_{MW}/Z_{LW}$ ratio in the inner regions (on average with values of about $0.5$ in the innermost radii) suggesting that younger SPs in the inner regions formed from chemically enriched gas likely due to local SF feedback and with little accretion of pristine gas. 
In the outer regions, the similarity between $Z_{MW}$ and $Z_{LW}$ may be due to a number of reasons. For example, the enriched gas ejected by the SF feedback may not be reaccreted in the outer regions because the gravitational potential is too shallow to retain this gas gravitationally bound at these radii, or because less enriched stars formed in the centre and migrate to the outskirts. The significant scatter of mass-weighted metallicities and their gradients suggest that dwarf galaxies evolved through a variety of paths.

\subsection{Dust attenuation $A_V$}

 The inner and outer gradients of $A_V$ mostly fall in quadrant $I$ of Figure \ref{Fig:CompareInVsOutGrad} and lie along the one-to-one line, i.e., most of the MaNDala galaxies have nearly flat or even slightly positive $A_V$ gradients (median value of inner gradients is 0.13 and while for the outer one is 0.14), with large scatter variations (see Table \ref{Table:MedianGrads_Stats}, see also Figure \ref{Fig:all_profiles_medianprof}). The low $A_V$ values for these dwarf galaxies agree well with findings from \citet[][using GALEX, SDSS, and WISE photometry calibrated on the Herschel ATLAS]{Salim+2018}  in the $10^8-10^9$ stellar mass range. It is notable that for these galaxies, the $A_V$ attenuation tends to increase with radius, similar to $Z_{MW}$, but opposite to $Z_{LW}$.

\subsection{$M/L$ ratio}

 The inner and outer gradients of the  $M/L$ ratio tend to fall in quadrant \textit{III}, 43\% of the sample, with generally steeper negative gradients in the outer regions. However, in general, most of the  $M/L$ gradients are near to $0$ spread across the four quadrants around the 0-0 point, as seen in the histograms of Figure \ref{Fig:CompareInVsOutGrad}. Given the wide variety in the  $M/L$ radial profiles (Figure \ref{Fig:all_profiles_medianprof}) and the uncertainty in determining their gradients (see Section \ref{Sec:MethodsComparisson}) we can conclude that the $M/L$ radial profiles of our sample of dwarf galaxies tend to be nearly flat. However, overall $M/L$ ratio vary significantly, with the diversity in $Age_{MW}/Age_{LW}$ ratios discussed above. These results suggest that dwarf galaxies are strongly affected by late episodes of SF.


\subsection{$\Sigma_{SFR}$ and $\Sigma_{*}$}

 For the SFR, measured via SSPs and the $H\alpha$ line, almost all galaxies fall in quadrant \textit{III} of decreasing inner and outer gradients. The outer gradients are slightly flatter than the inner ones, placing them just above the one-to-one line (see also Fig.~\ref{Fig:all_profiles_medianprof}). In general, the SFR surface density, $\Sigma_{\rm SFR}$, decreases with radius, as does the stellar surface density, $\Sigma_{*}$.   From Table  \ref{Table:MedianGrads_Stats}, we infer that in the inner regions, $\Sigma_{{\rm SFR},{SSP}} \propto \Sigma\ast^{0.77}$, while at the outer radii, $\Sigma_{{\rm SFR},{SSP}} \propto \Sigma\ast^{0.68}$. These values are generally aligned with the results from \citet{Cano-Diaz+2019} where the spatially resolved star-forming regions follow the relation $\Sigma_{\rm SFR} \propto \Sigma_\ast^{0.95}$. Similar slopes are observed when using the gradient values of $\Sigma_{{\rm SFR},{H\alpha}}$.  Thus, for our dwarf galaxies, the $\Sigma_{\rm SFR}$ inner and outer profiles are, on average, shallower than the respective $\Sigma_{*}$ profiles showing that SF is not suppressed in the outer regions since the sSFR grows in the outskirts, and suggesting an assembly from the inside out. In both cases, our results show that both the $\Sigma_{*}$ and $\Sigma_{\rm SFR}$ profiles tend to be sub-exponential.


\subsection{sSFR}
 
The specific SFR profiles, $\Sigma_{SFR}(R)/\Sigma_{*}(R)$, give insight into a galaxy's radial SF history and its cessation. As shown in Figure \ref{Fig:CompareInVsOutGrad}, the sSFR gradients of our dwarf galaxies, whether using SSPs or the $H\alpha$ line, tend to be flat or even slightly positive (see also Fig.~\ref{Fig:all_profiles_medianprof}), especially in the outer regions. For example, using SSPs, the median inner and outer gradients are 0.10 and 0.15 (Table \ref{Table:MedianGrads_Stats}). This is expected, given the previously mentioned relationship between the $\Sigma_{\rm SFR}$ and $\Sigma_*$ profiles. This suggest that either the SF is slowing down in the inner regions, or it has become more efficient in the outer ones. The tendencies of our dwarfs to increase the $Age_{MW}/Age_{LW}$ ratio and dust attenuation with radius point to the latter case. We will discuss this more in Section \ref{Sec:implications}.

\section{Discussion}\label{Sec:discussion}

This paper, examines 124 local dwarf galaxies from the MaNDALA dataset, which includes spatially resolved spectroscopic data from MaNGA and photometric information from DESI \citep[see][for more details]{CanoDiaz2022}. We explore different methods for characterizing the radial profiles of various galaxy properties, and propose four methods, detailed in Section \ref{Sec:RadialProfilesMethods}, for computing these radial profiles. The first two methods use concentric elliptical rings that aggregate information into a single value per radius: $i)$ the median of sample points within a defined ring based on DESI photometry, and $ii)$ the mass-weighted integrated values within the ring. The other two methods use non-linear information from the profiles: $i)$ cubic splines, and $ii)$ Gaussian process regression fits. For each method, we define three approaches to obtain inner ($0 \leq R/R_e < 1$) and outer ($0.75 \leq R/R_e < 1.5$) SP properties gradients.

In this section, we discuss the effects of the PSF on estimating the gradients, examine systematic differences between methods, and explores the implications of our results in the context of galaxy evolution.

\subsection{The effects of the PSF}\label{Sec:PSF_impact}

The PSF primarily affects the central regions of galaxies, which in turn influences the central parts of the radial profiles, and therefore the calculations of inner gradients. \citet{Belfiore2017} extensively discusses this effect in the gradients of a sample of MaNGA galaxies specifically those with $9.0 < \log (M_{*}/M_{\odot}) < 11.5$ and axis ratios $b/a>0.4$. They conclude that the PSF leads to an artificial flattening in the central parts of the metallicity radial profiles. This flattening also affects the measured gradients, which are determined by fitting a linear model to azimuthally averaged radial bins, similar to our collapsed median profiles. 

\begin{figure*}
    \includegraphics[width=\textwidth]{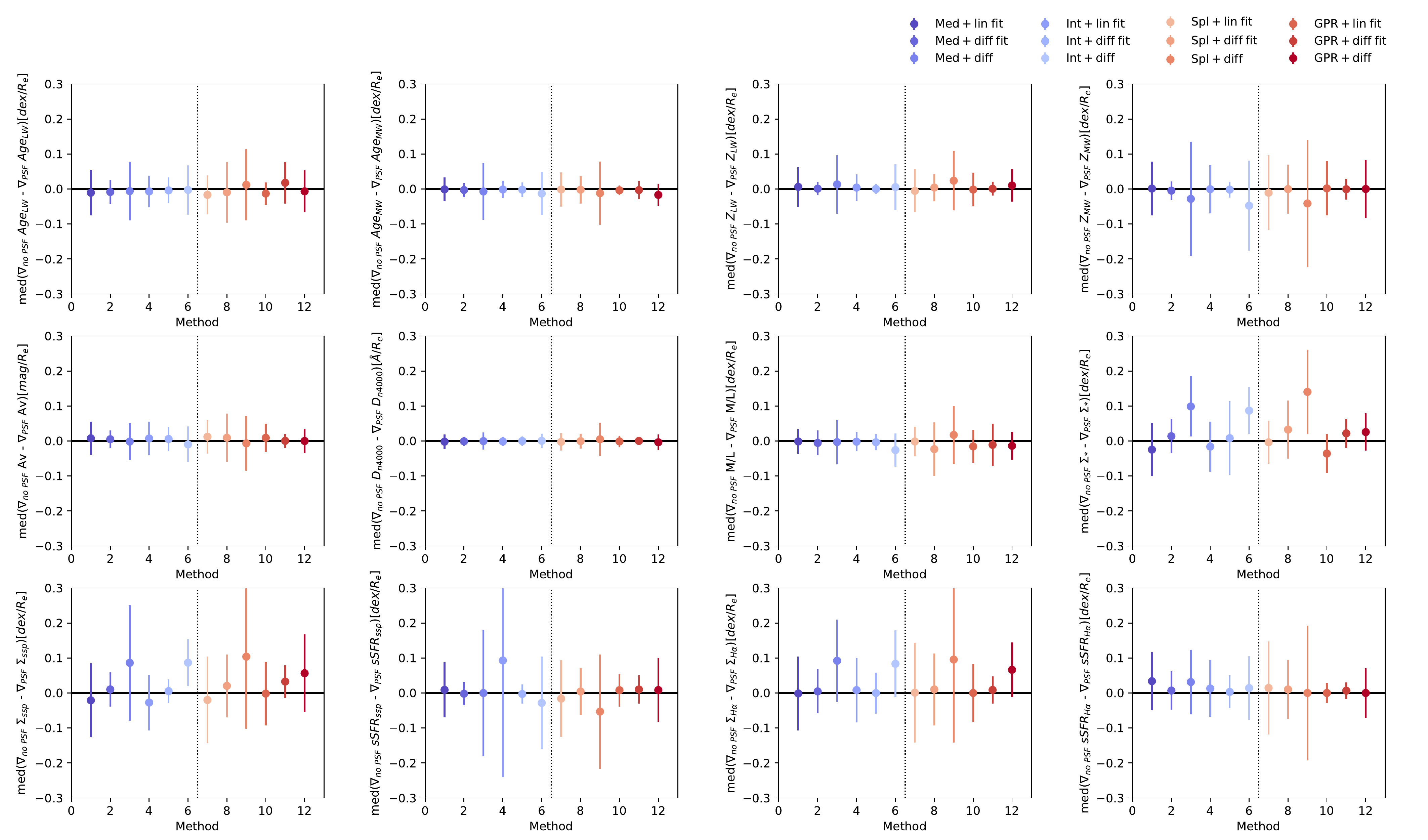}
  \caption{Median offsets between the gradients derived without and with points below the $R_{PSF}$. Color code is the same as in Figure \ref{Fig:GradComparisson_grads_vs_median}.}
  \label{Fig:median_offset_psf}
\end{figure*}

To test the effect of the PSF on our sample of dwarf galaxies, we compare the inner gradients derived using all bins/points including those within the PSF radius ($R_{PSF}$), with those derived by excluding points/bins inside this radius. Figure \ref{Fig:median_offset_psf} shows these comparisons, using the median value of the differences in gradients obtained with and without removing the bins/points within the PSF radius  ($\nabla G_{no PSF}$-$\nabla G_{PSF}$). The color code corresponds to the different methods used to derive the radial profiles and the gradients.

Figure \ref{Fig:median_offset_psf} shows that the median offsets between the gradients obtained, including and excluding the bins/points within the $R_{PSF}$ are $\sim 0$, except for the $Z_{MW}$, $\Sigma_{*}$, $\Sigma_{SFR}$ and the sSFRs properties. Exceptions arise when using the difference between two points method to calculate inner gradients, with the medians registering a value of $\sim 0.1$ dex/$R_{e}$). Notably, the median value minimizes when this method is paired with the non-linear GPR method for radial profiles, In contrast, it tends to maximize when paired with the non-linear splines method. This variation is particularly significant \textit{depending on the methods} used to derive radial profiles and gradients, especially when employing the difference between two points method. This result is intuitive, since it depends greatly on the measurements performed in the central region of the galaxies, where the PSF has a greater influence. {This effects is minimized and maximized} when pairing the difference of two points gradient derivation method with the two non-linear methods to derive the radial profiles. This may be due to the very different approaches that these methods use. On one hand, the Spline method uses a high-order interpolation approach, which may be more sensitive to non-linear variations in the profiles introduced by the effect of the PSF. On the other, hand the GPR method uses a non-linear regression approach, which optimizes the size of the kernels it uses, see Section \ref{Sec:ProfilesGPR}. Therefore, we may be witnessing is the result of this optimization, which translates into the minimization of the PSF effect in the central region of the profiles.

The fact that this effect is more relevant for specific galaxy properties is also expected, since these properties usually have more complex shapes at the internal radii in the radial profiles. Thus if the PSF affects the radial profiles, this will be more important in these profiles. Appendix \ref{Append:AllProfiles} shows that the overall profiles are generally quite flat in the cases of the $Z_{MW}$ and $sSFR_{H\alpha}$ properties. However when looking closely at the individual profiles, it reveals a variety of internal shapes, which may be why the difference between two points gradient derivation method is particularly sensitive to the PSF effects.

\subsection{Systematics}\label{Sec:systematics}

As discussed and emphasized in Section \ref{Sec:MethodsComparisson}, the derived gradients of certain SP properties are sensitive to the methodology used for characterizing radial profiles. In this section, we examine how these gradients vary across different methods. Here we quantify systematic trends by computing the following difference  
\begin{equation}
   \Delta \mathcal{G}_i = \nabla \mathcal{G}_i - \text{med}\left(\{\nabla \mathcal{G}\}_i\right).
   \label{eq:difference_gradient_i_and_median_grad}
\end{equation}
This equation quantifies the difference between the gradient property $\mathcal{G}$ obtained using method $i$ and the median gradient from all available methods for a given galaxy. Tables \ref{Table:systmatics_grads_GPR} and \ref{Table:systmatics_grads_int} present metrics respectively for the GPR and integrated methods, such as the median, and the difference between the 84th and 16th percentiles of all properties studied here, $\mathcal{G}$. In addition to computing these two statistical estimators, we calculate the Pearson correlation coefficient, $\rho_{X,Y}$,  between the gradient of property $X =\nabla \mathcal{G}_i$ obtained using method $i$ and the difference given by Eq. \ref{eq:difference_gradient_i_and_median_grad}, $Y = \Delta \mathcal{G}_i$.

The above Pearson correlation indicates the degree of correlation between $\Delta \mathcal{G}$ and the gradient $\nabla \mathcal{G}$ from a specific method. If all methods yield Pearson correlation coefficients $\rho \sim 0$, we expect $\Delta \mathcal{G} \rightarrow 0$ and all methods contributed equally to the median gradient value $\text{med}\left(\{\nabla \mathcal{G}\}_i\right)$. In other words, all the methods are nearly identical. Otherwise, it suggests that the value $\text{med}\left(\{\nabla \mathcal{G}\}_i\right)$ is dominated by variations from the methods for determining the gradients. This represents a rough diagnostics towards understanding systematic differences in the methods. Other quantities are the median of $\Delta \mathcal{G}$ which quantifies the systematic differences between all methods, and $(P_{84} - P_{16})/2$ which quantifies the 1$\sigma$ distribution of $\Delta \mathcal{G}$.

\begin{figure*}
    \includegraphics[width=\textwidth]{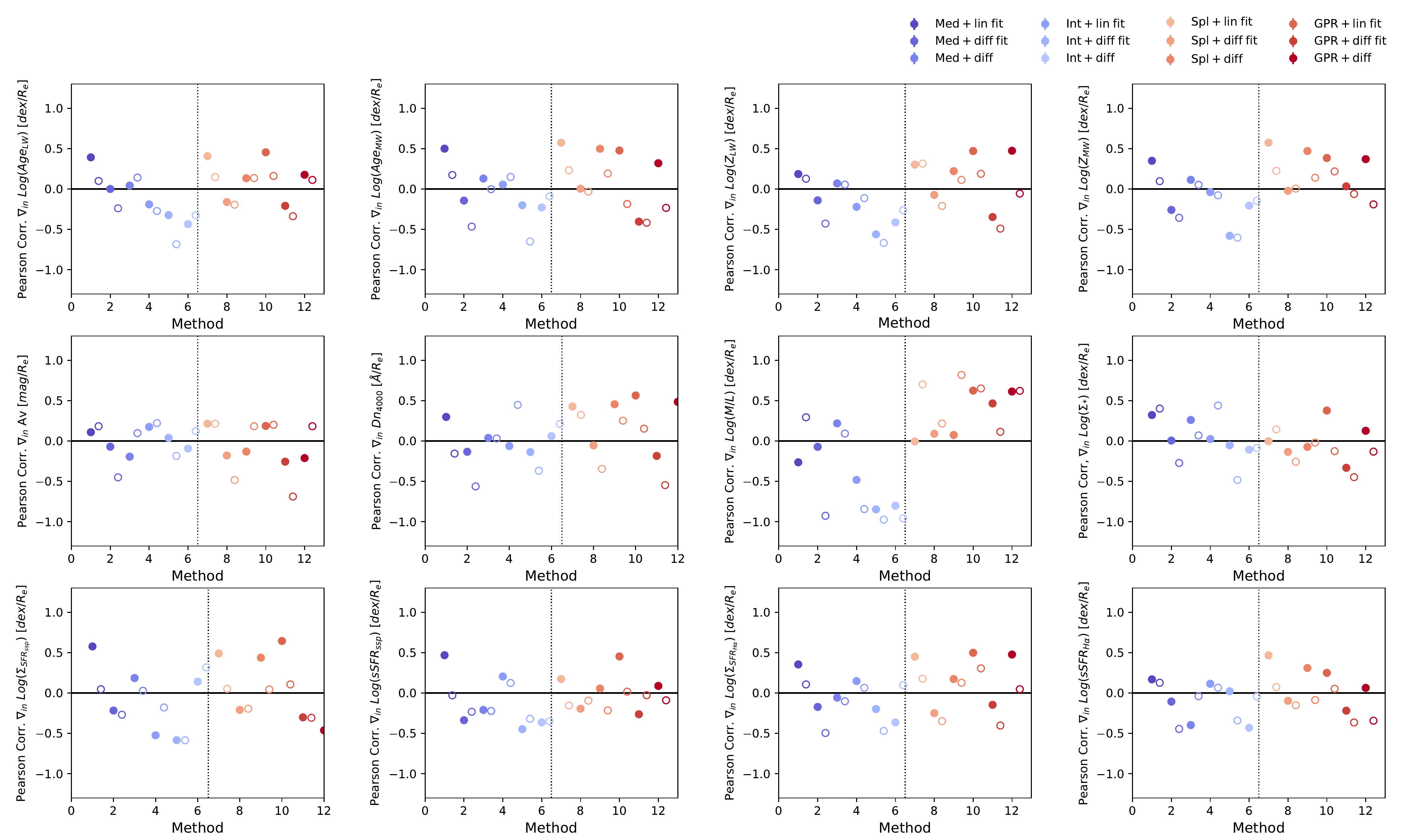}
  \caption{Pearson Correlation Factors. Color and symbol codes are the same as in Figure \ref{Fig:GradComparisson_grads_vs_median}}
  \label{Fig:pearson_methods}
\end{figure*}

\begin{figure*}
    \includegraphics[width=\textwidth]{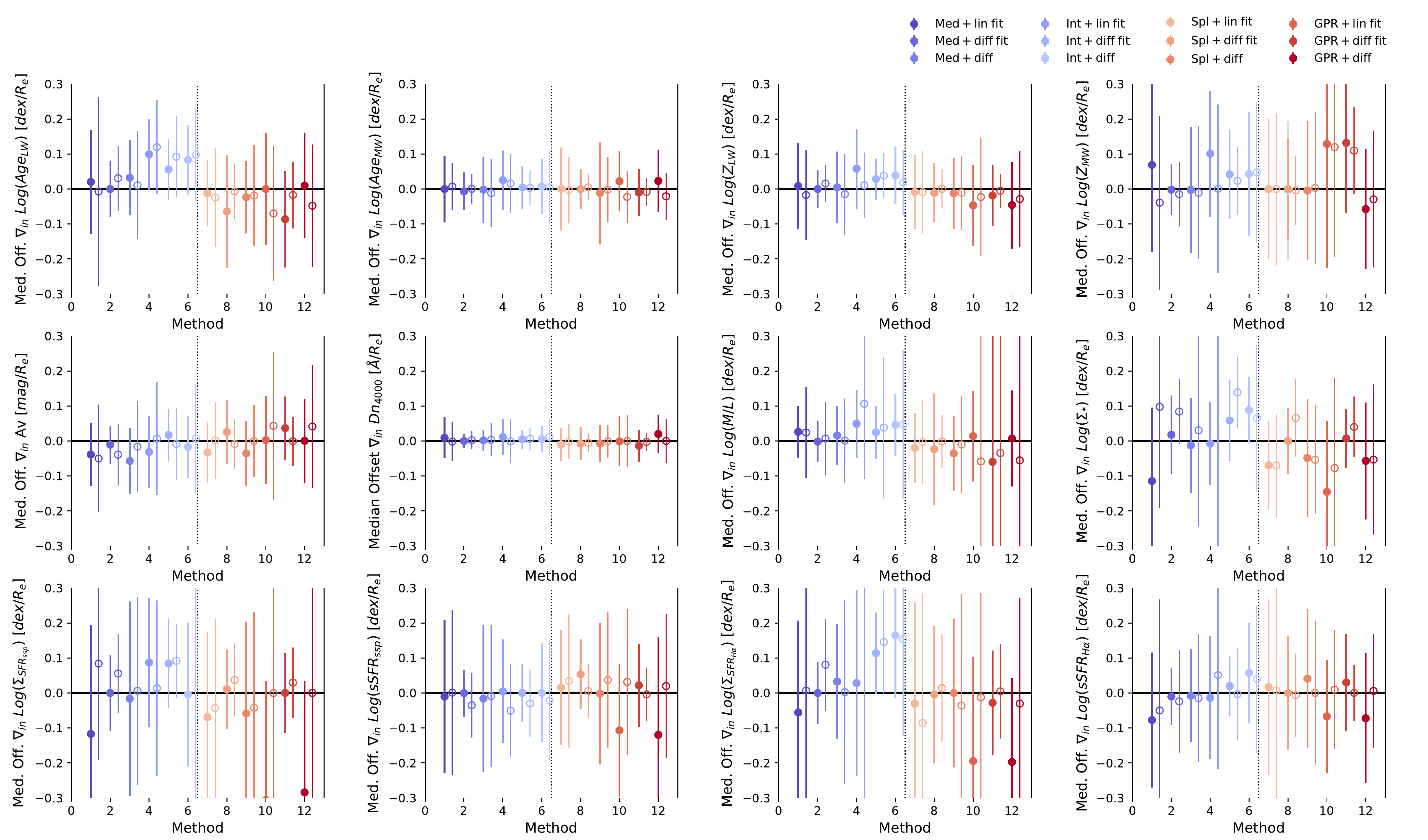}
  \caption{Median Offsets. Color and symbol codes are the same as in Figure \ref{Fig:GradComparisson_grads_vs_median}.}
  \label{Fig:median_offset_methods}
\end{figure*}

Figures \ref{Fig:pearson_methods} and \ref{Fig:median_offset_methods} show respectively the Pearson correlation coefficient and the systematic offset, $\Delta \mathcal{G}$, for each method described in Section \ref{Sec:grad_methods}. Filled circles represent inner gradients values, while empty circles represent outer gradients. Error bars in Figure \ref{Fig:median_offset_methods} represent the 1$\sigma$ distribution of $\Delta \mathcal{G}$ quantified as $(P_{84} - P_{16})/2$. In both figures, the dotted line separate the division between collapsed and non-linear radial profiles in both figures.

In general, inner gradients from collapsed radial profiles present lower Pearson coefficients compared to those based on non-linear profiles, suggesting that non-linear profiles introduce larger fluctuations in $\Delta \mathcal{G}$. This result is reasonable because collapsed profiles smooth out, to some degree, the complexity of dwarf galaxy profiles, see Figure \ref{Fig:example_all_profiles}, summarizing them into one number per bin. In contrast, non-linear profiles exploit the spatially resolved nature of the MaNDALA sample data, making them more prone to noisy data. However, see \ref{Sec:ProfilesGPR}, where we discuss how to mitigate the above. Noisy profiles, such as those for SFRs and sSFRs, based on both SSP and H$_\alpha$, produce a strong Pearson correlation in both methods: the collapsed and the non-linear ones. Again, this is the result of the noisy data in those profiles.

It is unclear which method of characterizing radial gradients has the smaller Pearson coefficient. Linear fits generally show a higher Pearson correlation across most gradient estimation methods for median collapsed profiles. In contrast, for integrated collapsed profiles, linear fits tend to have the lowest Pearson correlation.

Figure \ref{Fig:median_offset_methods} shows the median offset between different gradient estimation methods and the median of all methods. Filled circles represent inner gradients within $R_e$, while empty circles represent outer gradients between $0.75 < R/R_e < 1.5$. The error bars indicate the $1\sigma$ distribution, estimated as $(P_{84} - P_{16}) / 2$.

Our results show that light-weighted ages have a systematic median effect no more significant than $\sim 0.2$ dex/$R_e$ for both inner and outer gradients. Light weighted ages have a median inner and outer value of $\nabla_{Age_{LW},in}=-0.1$ and $\nabla_{Age_{LW},out}=-0.14$, respectively, well within the systematic value. For mass-weighted ages, all methods display a value of approximately $\sim 0$, indicating consistent gradient values. The $1\sigma$ distribution is around the maximum systematic effect for light-weighted ages, while for mass-weighted ages is $\sim 0.2$ dex/$R_e$. For light-weighted metallicity gradients, we observe some dispersion, though not larger than $\sim 0.1$ dex/$R_e$, with an average dispersion of the same order across methods. Mass-weighted metallicity gradients show maximum values of $\sim 0.15$ dex/$R_e$ and a large dispersion ($\sigma \sim 0.2$ dex/$R_e$). Again, these maximum deviations are of the same order or even more significant than the median values obtained for their gradients. The maximum values for $\nabla A_V$ are $\sim 0.1$ dex/$R_e$, with an average dispersion of $\sim 0.15$ dex/$R_e$. $\nabla D_{n4000}$ is the most stable quantity, with all methods yielding a value of $\sim0$ and very small dispersion. Mass-to-light ratios exhibit little systematic effect, though non-linear profile methods show a broader dispersion. Gradients for stellar mass, SFR, and sSFR show larger differences, with maximum values of $\sim 0.2$, $\sim 0.4$, and $\sim 0.15$ dex/$R_e$, respectively, and significant dispersion, often exceeding the systematic differences.

In general, these results demonstrate that a single method can provide a rough estimate of the gradient for a given galaxy property. However, using multiple methods helps to robustly estimate errors in measurements, especially if the galaxies have complex spatial distributions, such as those of dwarf galaxies.

Tables \ref{Table:systmatics_grads_GPR} and \ref{Table:systmatics_grads_int} provide P-values, Pearson coefficients, median shifts, and the 1$\sigma$ distribution values for both GPR and the integrated profile in concentric elliptical rings. These tables are helpful to compare different methods for characterizing profiles where gradients have been derived, allowing one to apply the shifts in the tables to facilitate the comparison between methods.

Finally, analyzing the gradients obtained from different methods, particularly those that deviate most from the median gradient, we find that they have little effect on the median itself. Instead, using different methods to measure gradients, captures the natural complexity of the radial distribution of a given property for each galaxy, which introduces an intrinsic uncertainty in gradient determination that should be taken into account. 
 
\begin{table*}
\centering
\text{Comparison of gradient method estimations from profiles based on the gaussian process regression fit}\\
    \begin{tabular}{c c c c c c c c c c c c c}
    \hline
    \hline
 & \multicolumn{11}{c}{Inner gradient} \\ 
 GPR & \multicolumn{4}{c}{Linear Fit} & \multicolumn{4}{c}{Difference Fit} & \multicolumn{4}{c}{Difference} \\ 
\footnotesize{Gal. Prop. $\Delta$} & \footnotesize{P-value} & $\rho$ & \footnotesize{$P_{50}$} & \footnotesize{$P_{84}-P_{16}$} & \footnotesize{P-value} & $\rho$ & \footnotesize{$P_{50}$} & \footnotesize{$P_{84}-P_{16}$} & \footnotesize{P-value} & $\rho$ & \footnotesize{$P_{50}$} & \footnotesize{$P_{84}-P_{16}$} \\
\hline
\footnotesize{Log($Age_{LW}$)} & 0.000 & 0.463 & 0.000 & 0.317 & 0.023 & -0.204 & -0.076 & 0.265 & 0.051 & 0.175 & 0.012  & 0.293 \\
\footnotesize{Log($Age_{MW}$)} & 0.000 & 0.477 & 0.023 & 0.171 & 0.000 & -0.419 & -0.013 & 0.133 & 0.000 & 0.321 & 0.023  & 0.176 \\
\footnotesize{Log($Z_{LW}$)} & 0.000 & 0.479 & -0.036 & 0.228 & 0.000 & -0.347 & -0.019 & 0.160 & 0.000 & 0.473 & -0.046  & 0.234 \\
\footnotesize{Log($Z_{MW}$)} & 0.000 & 0.380 & 0.129 & 0.710 & 0.754 & 0.028 & 0.130 & 0.407 & 0.000 & 0.370 & -0.057  & 0.342 \\
\footnotesize{$A_V$} & 0.042 & 0.183 & 0.003 & 0.264 & 0.005 & -0.253 & 0.036 & 0.188 & 0.018 & -0.211 & 0.000  & 0.240 \\
\footnotesize{$D_{n4000}$} & 0.000 & 0.548 & 0.000 & 0.147 & 0.076 & -0.160 & -0.010 & 0.071 & 0.000 & 0.469 & 0.020 & 0.108 \\
\footnotesize{Log($M/L$)} & 0.000 & 0.616 & 0.014 & 0.240 & 0.000 & 0.464 & -0.055 & 1.509 & 0.000 & 0.604 & 0.010  & 0.274 \\
\footnotesize{Log($\Sigma_\ast$)} & 0.000 & 0.389 & -0.154 & 0.437 & 0.000 & -0.378 & 0.001 & 0.157 & 0.113 & 0.143 & -0.057  & 0.336 \\
\footnotesize{Log($\Sigma_{SFR_{ssp}}$)} & 0.000 & 0.644 & -0.306 & 0.674 & 0.001 & -0.293 & 0.000 & 0.230 & 0.000 & 0.470 & -0.284 & 0.636 \\
\footnotesize{Log(${sSFR_{ssp}}$)} & 0.000 & 0.457 & -0.106 & 0.380 & 0.004 & -0.255 & 0.022 & 0.240 & 0.333 & 0.088 & -0.120 & 0.557 \\
\footnotesize{Log($\Sigma_{SFR_{H_\alpha}}$)} & 0.000 & 0.513 & -0.194 & 0.589 & 0.175 & -0.130 & -0.030 & 0.301 & 0.000 & 0.488 & -0.197  & 0.482 \\
\footnotesize{Log($sSFR_{H_\alpha}$)} & 0.005 & 0.253 & -0.062 & 0.316 & 0.016 & -0.220 & 0.015 & 0.277 & 0.492 & 0.064 & -0.072  & 0.372 \\
    \hline
     & \multicolumn{11}{c}{Outer gradient} \\ 

\footnotesize{Log($Age_{LW}$)} & 0.070 & 0.163 & -0.070 & 0.385 & 0.000 & -0.336 & -0.015 & 0.183 & 0.205 & 0.115 & -0.048  & 0.350 \\
\footnotesize{Log($Age_{MW}$)} & 0.039 & -0.185 & -0.022 & 0.148 & 0.000 & -0.471 & -0.009 & 0.085 & 0.010 & -0.231 & -0.021  & 0.133 \\
\footnotesize{Log($Z_{LW}$)} & 0.041 & 0.184 & -0.022 & 0.341 & 0.000 & -0.484 & -0.003 & 0.095 & 0.521 & -0.058 & -0.027  & 0.274 \\
\footnotesize{Log($Z_{MW}$)} & 0.015 & 0.218 & 0.119 & 0.628 & 0.526 & -0.057 & 0.114 & 0.243 & 0.034 & -0.191 & -0.029  & 0.391 \\
\footnotesize{$A_{V}$} & 0.026 & 0.200 & 0.043 & 0.421 & 0.000 & -0.689 & 0.000 & 0.142 & 0.041 & 0.183 & 0.041  & 0.350 \\
\footnotesize{$D_{n4000}$} & 0.100 & 0.148 & 0.000 & 0.149 & 0.000 & -0.545 & -0.003 & 0.055 & 0.696 & 0.035 & 0.000  & 0.129 \\
\footnotesize{Log($M/L$)} & 0.000 & 0.659 & -0.077 & 1.527 & 0.207 & 0.114 & -0.026 & 0.713 & 0.000 & 0.633 & -0.056  & 1.293 \\
\footnotesize{Log($\Sigma_{\ast}$)} & 0.069 & -0.164 & -0.074 & 0.503 & 0.000 & -0.465 & 0.044 & 0.173 & 0.144 & -0.132 & -0.051  & 0.432 \\
\footnotesize{Log($\Sigma_{SFR_{ssp}}$)} & 0.238 & 0.107 & 0.000 & 0.765 & 0.001 & -0.302 & 0.030 & 0.196 & 0.761 & 0.028 & 0.000  & 0.637 \\
\footnotesize{Log($sSFR_{ssp}$)} & 0.816 & 0.021 & 0.032 & 0.411 & 0.856 & -0.016 & -0.006 & 0.155 & 0.326 & -0.089 & 0.020  & 0.409 \\
\footnotesize{Log($\Sigma_{SFR_{H_\alpha}}$)} & 0.001 & 0.305 & -0.014 & 0.596 & 0.000 & -0.371 & 0.008 & 0.253 & 0.610 & 0.047 & -0.030  & 0.603 \\
\footnotesize{Log($sSFR_{H_\alpha}$)} & 0.556 & 0.053 & 0.010 & 0.344 & 0.000 & -0.371 & 0.000 & 0.157 & 0.000 & -0.343 & 0.006  & 0.323 \\

\hline  
\hline  
    \end{tabular}
\caption{ P-values, person coefficients and median shifts of the difference between inner gradients and the median gradient, see Figure \ref{Fig:GradComparisson_grads_vs_median}.}\label{Table:systmatics_grads_GPR}
\end{table*}

\begin{table*}
\centering
\text{Comparison of gradient method estimations from profiles based on the integration of concentric rings}\\
    \begin{tabular}{c c c c c c c c c c c c c}
    \hline
    \hline
     & \multicolumn{11}{c}{Inner gradient} \\ 

 Int & \multicolumn{4}{c}{Linear Fit} & \multicolumn{4}{c}{Difference Fit} & \multicolumn{4}{c}{Difference} \\ 

\footnotesize{Gal. Prop. $\Delta$} & \footnotesize{P-value} & $\rho$ & \footnotesize{$P_{50}$} & \footnotesize{$P_{84}-P_{16}$} & \footnotesize{P-value} & $\rho$ & \footnotesize{$P_{50}$} & \footnotesize{$P_{84}-P_{16}$} & \footnotesize{P-value} & $\rho$ & \footnotesize{$P_{50}$} & \footnotesize{$P_{84}-P_{16}$} \\
\hline

\footnotesize{Log($Age_{LW}$)} & 0.870 & -0.023 & 0.121 & 0.269 & 0.000 & -0.682 & 0.087 & 0.232 & 0.000 & -0.323 & 0.095  & 0.263 \\
\footnotesize{Log($Age_{MW}$)} & 0.286 & 0.149 & 0.016 & 0.166 & 0.000 & -0.649 & 0.003 & 0.097 & 0.321 & -0.090 & 0.004  & 0.211 \\
\footnotesize{Log($Z_{LW}$)} & 0.441 & -0.108 & 0.011 & 0.186 & 0.000 & -0.667 & 0.041 & 0.131 & 0.004 & -0.258 & 0.018  & 0.182 \\
\footnotesize{Log($Z_{MW}$)} & 0.589 & -0.076 & 0.001 & 0.481 & 0.000 & -0.603 & 0.023 & 0.194 & 0.107 & -0.146 & 0.047  & 0.401 \\
\footnotesize{$A_V$} & 0.119 & 0.221 & 0.007 & 0.324 & 0.040 & -0.185 & -0.008 & 0.204 & 0.180 & 0.122 & 0.007  & 0.314 \\
\footnotesize{$D_{n4000}$} & 0.001 & 0.443 & -0.002 & 0.127 & 0.000 & -0.368 & 0.006 & 0.060 & 0.018 & 0.214 & 0.012  & 0.105 \\
\footnotesize{Log($M/L$)} & 0.000 & -0.843 & 0.106 & 0.429 & 0.000 & -0.973 & 0.038 & 0.403 & 0.000 & -0.955 & 0.047  & 0.425 \\
\footnotesize{Log($\Sigma_\ast$)} & 0.001 & 0.440 & -0.801 & 3.157 & 0.000 & -0.475 & 0.139 & 0.205 & 0.401 & -0.076 & 0.065  & 0.422 \\
\footnotesize{Log($\Sigma_{SFR_{ssp}}$)} & 0.021 & 0.317 & -1.060 & 3.730 & 0.000 & -0.587 & 0.092 & 0.212 & 0.080 & -0.158 & 0.014  & 0.506 \\
\footnotesize{Log($sSFR_{ssp}$)} & 0.370 & 0.126 & -0.051 & 0.264 & 0.000 & -0.321 & 0.029 & 0.191 & 0.000 & -0.354 & -0.022  & 0.330 \\
\footnotesize{Log($\Sigma_SFR_{H_\alpha}$)} & 0.641 & 0.068 & 0.331 & 2.004 & 0.000 & -0.471 & 0.147 & 0.297 & 0.324 & 0.096 & 0.154  & 0.553 \\
\footnotesize{Log($sSFR_{H_\alpha}$)} & 0.659 & 0.064 & 0.051 & 0.538 & 0.000 & -0.341 & -0.003 & 0.262 & 0.654 & -0.044 & 0.044 & 0.421 \\

    \hline
     & \multicolumn{11}{c}{Outer gradient} \\ 
     
\footnotesize{Log($Age_{LW}$)} & 0.060 & -0.190 & 0.099 & 0.200 & 0.000 & -0.324 & 0.056 & 0.175 & 0.000 & -0.437 & 0.081 & 0.201 \\
\footnotesize{Log($Age_{MW}$)} & 0.585 & 0.056 & 0.025 & 0.170 & 0.025 & -0.201 & 0.004 & 0.120 & 0.011 & -0.228 & 0.007 & 0.157 \\
\footnotesize{Log($Z_{LW}$)} & 0.027 & -0.225 & 0.058 & 0.226 & 0.000 & -0.567 & 0.027 & 0.117 & 0.000 & -0.423 & 0.040  & 0.164 \\
\footnotesize{Log($Z_{MW}$)} & 0.702 & -0.039 & 0.101 & 0.359 & 0.000 & -0.580 & 0.042 & 0.256 & 0.022 & -0.206 & 0.043  & 0.355 \\
\footnotesize{$A_V$} & 0.095 & 0.174 & -0.031 & 0.208 & 0.675 & 0.038 & 0.017 & 0.148 & 0.318 & -0.094 & -0.017 & 0.176 \\
\footnotesize{$D_{n4000}$} & 0.444 & -0.079 & 0.011 & 0.099 & 0.058 & -0.171 & 0.005 & 0.055 & 0.544 & 0.055 & 0.006 & 0.076 \\
\footnotesize{Log($M/L$)} &  0.000 & -0.481 & 0.049 & 0.193 & 0.000 & -0.846 & 0.027 & 0.152 & 0.000 & -0.801 & 0.046 & 0.177 \\
\footnotesize{Log($\Sigma_\ast$)} & 0.715 & 0.038 & -0.006 & 0.236 & 0.635 & -0.043 & 0.068 & 0.237 & 0.334 & -0.087 & 0.090 & 0.186 \\
\footnotesize{Log($\Sigma_{SFR_{ssp}}$)} & 0.172 & 0.141 & -0.004 & 0.411 & 0.000 & -0.584 & 0.084 & 0.255 & 0.002 & -0.275 & 0.087 & 0.368 \\
\footnotesize{Log($sSFR_{ssp}$)} & 0.047 & 0.203 & 0.004 & 0.297 & 0.000 & -0.448 & 0.000 & 0.171 & 0.000 & -0.364 & 0.000 & 0.278 \\
\footnotesize{Log($\Sigma_{SFR_{H_\alpha}}$)} & 0.165 & 0.146 & 0.034 & 0.530 & 0.032 & -0.203 & 0.118 & 0.234 & 0.000 & -0.366 & 0.166 & 0.366 \\
\footnotesize{Log($sSFR_{H_\alpha}$)} & 0.278 & 0.114 & -0.014 & 0.351 & 0.823 & 0.021 & 0.023 & 0.174 & 0.000 & -0.431 & 0.057 & 0.288 \\

\hline  
\hline  
    \end{tabular}
\caption{ P-values, person coefficients and median shifts of the difference between inner gradients and the median gradient, see Figure \ref{Fig:GradComparisson_grads_vs_median}.}\label{Table:systmatics_grads_int}
\end{table*}

\subsection{Stellar population gradients: from massive to dwarf galaxies}\label{Sec:discussion_grads_vs_mass}

Previous studies have measured SP properties gradients in normal to massive galaxies, $M_\ast > 10^{9} M_\odot$, to understand how galaxies formed and evolved \citep[see e.g.,][]{GonzalezDelgado2015} focusing mainly within the galaxies' half-light radius, $R_e$. A key question that we may ask for dwarf galaxies is whether their stellar gradients are a natural extension of those observed in more massive galaxies or if they present a qualitative change compared to them. 

\begin{figure*}
    \includegraphics[width=\textwidth]{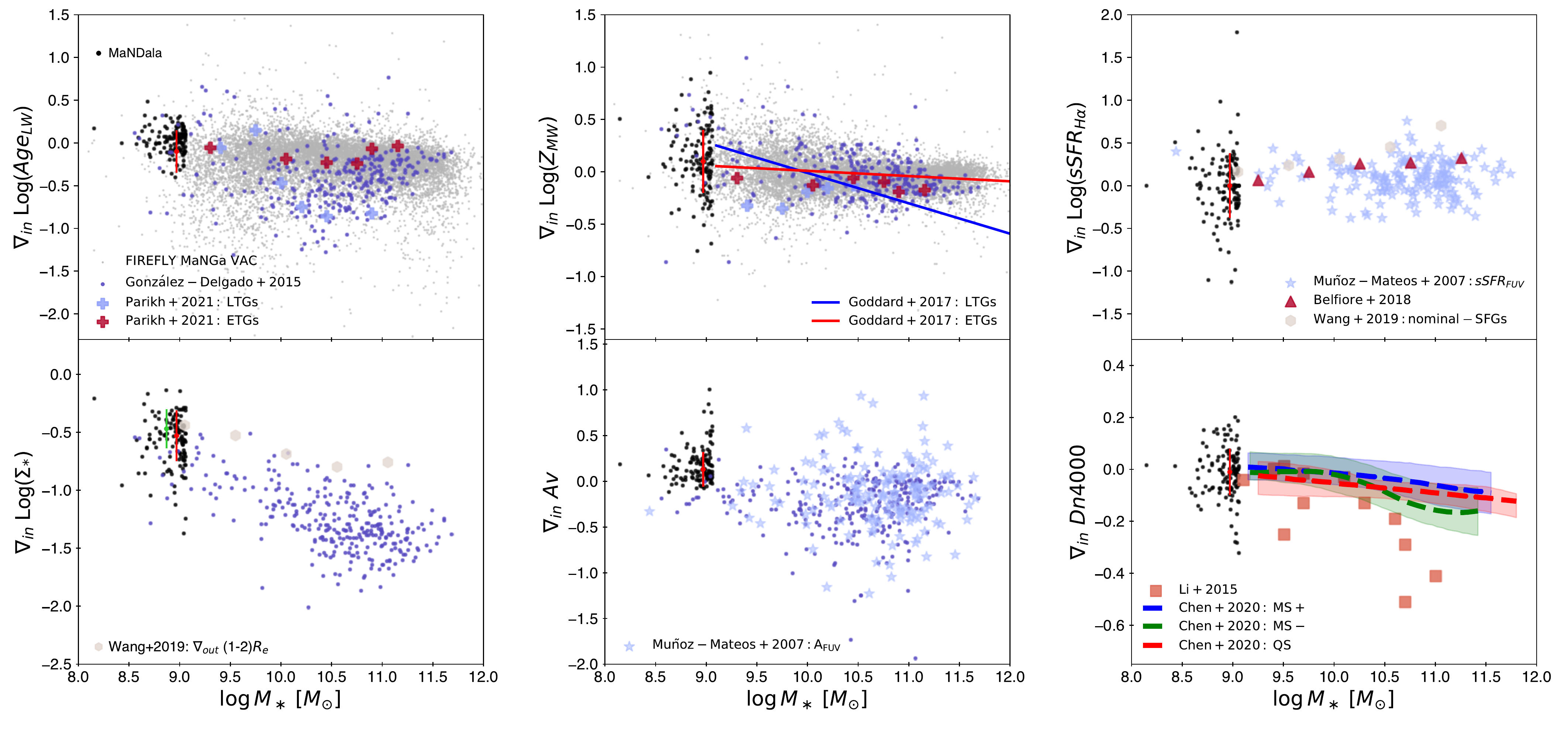}
  \caption{$\nabla_{G}$ Vs. $M_{*}$, this figure intends to compare the inner gradients of six galaxy properties $G$ for the MaNDala sample, with the gradients available in the literature for samples of galaxies that are predominantly massive. Over the MaNDala sample, a magenta point shows the median values of Log($M_{*}$) and of the gradient property being plotted. The error bar corresponds to $1\sigma$ for each property. In the case of the Log($\Sigma_{*}$) panel an equivalent point in cyan is shown for the outer gradients to make a better comparison with the outer gradients reported by \citet{Wang2019}.}
  \label{Fig:GradientsVsMass}
\end{figure*}

Figure \ref{Fig:GradientsVsMass} shows the inner gradients, $R<R_e$, of six properties of the MaNDala sample as a function of the $M_{*}$, compared to gradients of other galaxy samples available in the literature that are predominantly massive. In the case of the $Age_{LW}$ gradients, upper left panel, we can compare our results with those based on MaNGA \citep[][the latter corresponding to the entire FIREFLY VAC]{Parikh2021,Neumann2022}, and the CALIFA sample \citep{GonzalezDelgado2015}. \citet{Parikh2021} stacked the spectra of various $M_{*}$ bins separated into late-type (LTG) and early-type galaxies (ETG), obtaining a stacked radial profile for each bin. They measure the gradients as the linear fit to these profiles up to $R_{e}$. For the FIREFLY VAC, \citet{Neumann2022} derived median radial profiles with ten equally separated bins within 1.5 $R_{e}$ and by performing a linear fit to those profiles (a method that is comparable to our collapsed median profiles paired with the linear fit method to derive the gradients). \citet{GonzalezDelgado2015} obtained their gradients by deriving azimuthally averaged 2D maps of the studied property, and performing elliptical apertures of 0.1 $R_{e}$ of size to extract the radial profiles. They measure the gradients as the difference between 0 and $R_{e}$. The plot shows that at the low-mass end the $Age_{LW}$ gradients tend to increase, on average, with the MaNDala sample being the natural extension of this trend. According to \citet{Parikh2021}, the gradients between LTGs and ETGs are very different at high masses, leading to a large scatter, but at lower masses gradients become more similar. The small scatter in the age gradients of our sample of dwarf galaxies relative to the higher-mass galaxies are consistent with this result, suggesting that bright dwarf galaxies have naturally nearly flat (and even positive) age gradients.

Since $D_{n4000}$ is a tracer of $Age_{LW}$, Figure \ref{Fig:GradientsVsMass} also compares our $D_{n4000}$ gradients with those reported by \citet{Chen2020}\footnote{We made use of the \url{ https://automeris.io/} tool to extract these data sets.} and \citet{Li2015} for 3654 and 12 MaNGA galaxies, respectively. Both works derived median radial profiles from the $D_{n4000}$ maps, and performed linear fits to obtain the gradients; in the case of \citet{Chen2020} the linear fits are performed within 1.5 $R_{e}$, while \citet{Li2015} performed them within $R_{e}$. Similar to $Age_{LW}$ gradients, the bulk of our $D_{n4000}$ gradients seem to follow the tendencies shown by the comparison data, in which the values of the gradients tend towards zero at lower $M_{*}$. 

For the $Z_{MW}$ gradients, upper middle panel, we can compare them with the gradients of the same massive galaxy samples used above for $Age_{LW}$, as well as with the results derived by \citet{Goddard2017}, who analyzed a sample of $\sim700$ galaxies observed by MaNGA. They derived the radial profiles using the galactocentric distances of each measured point within the galaxies and then performing a linear fit over them to obtain the gradients. The MaNDala gradients appear to be a natural extension of the tendency shown by the other samples, with some evidence of a larger scatter for the dwarf sample. As explained in the last subsection, we can see diverse inner radial profiles in our sample for this particular property.

The bottom left panel of Figure \ref{Fig:GradientsVsMass} shows the gradients in $\Sigma_{*}$. The $\Sigma_{*}$ gradients of MaNDala galaxies seem to be a natural extension of the tendency of those of larger galaxies, reported by CALIFA \citep{GonzalezDelgado2015} and a sample of $\sim 1000$ MaNGA galaxies \citep{Wang2019}. The latter derived 5 median radial profiles for 5 $M_{*}$ bins from a sample of only nominal star forming Galaxies. They finally performed linear fits between 1 and 2 $R_{e}$ of these profiles to obtain their gradients per each $M_{*}$ bin. This is an expected behavior since the dwarf galaxies tend to be less concentrated than the massive ones, which explains the shallower values of their inner gradients. In other words, dwarf galaxies have smaller Sersic indices than the most massive galaxies. As discussed in Sections \ref{Sec:general_fits} and \ref{Sec:diff_2points} the gradient, defined as the difference between the central part and $R_e$ of the galaxy in a Sersic-like profile, see Section \ref{eq:general_fit_grads}, is given by $\nabla_{G} = G_1 / \ln 10$. If $\Sigma_\ast$ follows a Sersic profile, this changes to $\nabla_{\Sigma_\ast} = -b_n / \ln 10$, where the term $b_n$ is defined in terms of the Sersic index $n$ \citep[see e.g.,][]{Capaccioli_1989}. For a Sersic index of $n=1$ and using the value $b_1 = 1.678$, the predicted gradient is $\nabla_{\Sigma_\ast} = -0.73$. The mean inner gradient we report in this paper, $\nabla_{\Sigma_\ast} = -0.56$ with a dispersion of $0.22$, is consistent with the fact that most dwarf galaxies have sub-exponential inner $\Sigma_{*}$ profiles.

Gradients of $A_{V}$ for the MaNDala galaxies, they are significantly different to the measurements by \citet[][CALIFA survey, see above]{GonzalezDelgado2015} and \citet{MunozMateos2007}, who analyzed higher-mass samples, with few galaxies in the same mass range as the MaNDala sample. \citet{MunozMateos2007} analyzed 161 galaxies observed with GALEX, deriving radial profiles from photometric data, and obtaining gradients through linear fits to the disk components, excluding bulges. As noted in previous sections, dust attenuation in our dwarf galaxies tends to increase with radius, resulting in mostly positive gradients, in contrast to the mostly negative $A_{V}$ gradients reported at higher masses, as shown in Figure \ref{Fig:GradientsVsMass}. That dust attenuation increases with radius for a large fraction of our dwarf galaxies is an unexpected result that will require further research to be fully understood.

As for the $sSFR_{H\alpha}$, on average, gradients in MaNDala are consistent with those reported by \citet{Wang2019} (within $R_{e}$) and \citet{Belfiore2018} for higher-mass MaNGA galaxies. Both authors used H$_\alpha$ luminosity to convert it into SFR. \citet{Belfiore2018} stacked radial profiles using elliptical annuli with a semi-major axis of 0.15 $R_{e}$ for five stellar mass bins (we then computed the difference between their central bin and the one centered at $R\approx R_e$) which tend to be slightly positive or zero. We also show for comparison the gradients presented by \citet{MunozMateos2007}, who use $(FUV-K)$ color profiles, the \citet{Kennicutt98} calibrator to convert FUV luminosity into SFR and a constant $M/L_K$ ratio to estimate the sSFR profile, assuming exponential discs, finding their gradients to be both, positive and negative at all stellar masses. In Figure \ref{Fig:GradientsVsMass} we show their gradients after converting them to the same units as our gradients. However, a non-negligible fraction of our galaxies show negative gradients, indicating that their sSFR tend to decline with radii. It is important to notice that the properties inferred with the $H\alpha$ line tend to be the noisiest ones in our analysis. Conversely, the sSFR inner gradients obtained from the SSP analysis are less scattered and tend to be positive (see Fig. \ref{Fig:CompareInVsOutGrad}).

In summary, for the MaNDala bright dwarf galaxy sample, the gradients of $Age_{LW}$, $D_{n4000}$, $Z_{MW}$ and $\Sigma_{*}$ seem to be a natural extension of the tendencies shown by the gradients of more massive galaxies towards the low-mass side. As for the inner gradients of sSFR, based on $H_\alpha$ luminosity as SFR tracer, the scatter is too large with many negative values, in rough agreement with the comparison data. The gradients that do not consistently extend to lower masses from those at higher masses, are the dust attenuation gradients $\nabla_{A_V}$. While it is important to note that the comparison studies use different methods and radial ranges to derive their gradients, making direct comparison difficult, we have verified that the variations among our 12 different determinations of $\nabla_{A_V}$ are consistent with each other, indicating that the above conclusion is robust and independent of the method used. In any case, a fraction of the $\nabla_{A_V}$ reported by \citet{MunozMateos2007} also reach positive values, comparable to the ones obtained for our sample. 

The exploratory nature of Figure \ref{Fig:GradientsVsMass} helps to visualize possible trends across masses for several galaxy property gradients. However, this Figure is also useful to highlight the problem of comparing gradient results across the literature, since currently there is not a standardized way to measure the gradients. In fact it would be of major importance to perform more studies that seek to homogenize the derivation of the gradients for large samples of galaxies with a large $M_{*}$ range, and to compare them with the different methodologies used in the literature. This could help to better understand the true nature of the shapes and scatter of the trends visible in Figure \ref{Fig:GradientsVsMass}, since at this moment it is not possible to objectively know how much they are affected by the introduction of several methodologies to derive the gradients.

\subsubsection{Some implications for the assembly of bright dwarf galaxies}
\label{Sec:implications}

The scope of this paper is to explore in detail different methods to derive radial profiles of several SP properties of bright dwarf galaxies from the IFS MaNGA survey, and their characterization through simple quantities such as inner and outer gradients. From our results, but mainly using Figure \ref{Fig:GradientsVsMass}, we can discuss some generic implications for the radial assembly of bright dwarf galaxies concerning higher mass galaxies.

 After decades of observational and theoretical studies of normal (giant) galaxies, the consensus is that their measured gradients are consistent with in-situ stellar mass growth from the inside out for the disk component, while when ex-situ growth (mergers) or secular dynamical processes intervene (interactions, bars, etc.), disks thicken or transform into spheroids with gradients that tend to flatten mainly due to radial mixing \citep[see e.g.][]{Pessa2023}. Does this scenario hold for bright dwarf galaxies?
 
 As discussed above and shown in Fig. \ref{Fig:GradientsVsMass}, the different gradients become flatter as $M_\ast$ becomes lower. For our dwarf galaxies ($M_\ast \sim 10^8-10^9$ M$_\odot$), some gradients, especially those for the $Age_{MW}$ and $Z_{MW}$, even tend to be positive, on average, see Section \ref{Sec:InVstGradiens}, and Figs. \ref{Fig:CompareInVsOutGrad} and \ref{Fig:all_profiles_medianprof}. This could be interpreted as evidence for `outside-in' formation, with a phase of early active SF and chemical enrichment throughout the entire galaxy until some processes intervene to lower the SF in the outer regions of dwarf galaxies implying older and more metallic stars. At the same time, stars continually form in the inner regions from a mixture of enriched and less-metalic gas (contributing to these regions with younger, less metallic stars). However, when analyzing $Age_{LW}$ and $Z_{LW}$, which are sensitive to the contribution of recently formed stellar populations, show flat to negative gradients, consistent with moderate inside-out formation. On the other hand, flat to even positive sSFR gradients, an increasing $Age_{MW}/Age_{LW}$ ratio with radius, indicate that currently the SF has not been decreased in the outer regions of most of our dwarfs. It is more efficient than in the inner regions consistent with the inside-out formation scenario. Thus, it is more likely that MaNDala dwarfs experienced moderate inside out growth, while older, more enriched stellar populations underwent massive outward radial migration, leading to the flattening or inversion of the $Age_{MW}$ and $Z_{MW}$ profiles, making more extended the $\Sigma_{*}$ profiles. 

The above scenario is consistent with results from zoom-in hydrodynamical cosmological simulations, where bursty SF and SN-driven feedback in shallow gravitational potentials can cause two (possibly related) effects: $i)$ strong fluctuations of the infalling/outflowing gas in the inner regions, which induce quick variations in the central gravitational potential and bulk motions of the gas concerning the halo center; both are efficient mechanisms of kinetic energy transfer to the collisionless particles (dark matter and stars), producing their expansion, which leads to halo core formation and migration of the oldest stars towards the outskirts \citep[see e.g.,][]{Mashchenko+2006,Governato+2010,Pontzen+2012,Gonzalez-Samaniego+2016,El-Badry+2016,Graus2019,Riggs+2024}; $ii)$ outflowing gas that later falls back into the galaxy and forms stars (the so-called ``breathing mode''), leading to an outflowing/infalling cycle that produces young enriched stars with inherent radial velocities, which causes some to migrate outwards \citep[e.g.,][but see \citealp{Riggs+2024}]{El-Badry+2016}. 

An unmistakable feature of the stellar populations of our bright dwarf galaxy sample is their diversity, with a variety of SF histories, many of them with recent bursts, which are more efficient towards the outskirts: $Age_{MW}/Age_{LW}$ ratio, sSFR and $A_V$\footnote{Since the amount of molecular gas correlates with dust attenuation \citep[and $Z$, e.g.,][]{YesufHo2019}, a positive $A_V$ gradient may imply a positive $H_2$ gas gradient.} 
increase with radius. Our results point generically to a significant early SF from low-metallicity gas, outward migration of a fraction of these old low-$Z$ stars (expansion), and some sustained SF (or late bursts) from chemically enriched gas in the central regions and from less enriched gas in the outskirts. This suggests that gas in the central regions, where the gravitational potential is stronger, is more recycled than in the outer ones, where $Z_{LW}\sim Z_{MW}$ (while $Z_{LW}> Z_{MW}$ in the center) or that radial mixing of stars has been long-range. These results strongly constrain on the feedback mechanisms of these galaxies, and the gas infall/outfall processes, which may depend on the environment and merging history. To study this question in detail, in future work we will explore our results as a function of morphology, environment, and various global properties of our dwarf galaxy sample.

\section{Summary and Conclusions}\label{Sec:conclusions}

We analyzed a sample of 124 galaxies from the MaNDala catalog of bright dwarf galaxies, with a median redshift and mass of the sample of $z=0.019$ and $M_\ast=9.33\times10^{8}M_\odot$, using spatially resolved spectroscopic data from MaNGA, which provides coverage up to at least $\sim R_e$. Specifically we used maps of various SP properties, and emission lines from the \textsc{Pipe3D} Value Added Catalogue. From this data, we derived the radial projected distributions of several key SP properties: luminosity and mass weighted age $Age$ and metallicity $Z$, dust attenuation $A_V$, the $4000$ \AA\ break index $D_{n4000}$, the mass-to-light ratio $M/L$, stellar mass surface density $\Sigma_{*}$, star formation rate density $\Sigma_{SFR}$ and $sSFR =\Sigma_{SFR}/\Sigma_\ast $, with the last two derived from both SSPs and the $H\alpha$ line.

Due to the wide range of star formation histories, spatial irregularities and clumpiness of dwarf galaxies, the spatial distribution of their stellar populations is complex, making the determination of radial profiles non-trivial. To address this, we explored four different methods to derive the profiles of the above properties,(Section \ref{Sec:RadialProfilesMethods}). The first two methods define a series of concentric elliptical rings, each separated by $\sim R_{PSF}$. The first  characterizes each ring using the median of the values of all the pixels within it, while the second integrates the values within the ring. The other two methods fully exploit the spatially resolved nature of the data by performing a non-linear cubic spline interpolation to the spatial distribution of the data, while the other uses a non-linear Gaussian regression algorithm to fit the profiles. 

Additionally, we tested four approaches to derive inner, \innergrad, and outer, \outergrad, gradients (in the 0-$R_e$ and 0.75-1.5$R_e$ radial ranges, respectively) for each of the four methods to derive radial profiles, Section \ref{Sec:grad_methods}. These four approaches include:
\begin{itemize}
    \item[1] Fitting a linear regression to the radial profiles within the defined radial ranges,

    \item[2] Measuring the difference between two radial points directly from the profiles,

    \item[3] Measuring the differences between two radial points from a generalized exponential fit to the profiles,

    \item[4] Computing the local derivative of the generalized exponential fit. 
\end{itemize}
By applying four gradient estimation approaches to each radial profile method, resulting in a total of 12 gradient estimations, we obtained the following results:

\begin{itemize}
    \item We find that the PSF effect is generally negligible, except when using the two-point difference to derive  gradients. On one hand this effect is minimized when this approach is paired with the non-linear GPR method for radial profile determination. On the other hand, is maximized when paired with the non-linear Spline method.

    \item Based on the comparisons among the 12 gradient estimations, we conclude that the choice of methodology for characterizing radial profiles significantly affect the results. 
    
    \item While it is not possible to select a single preferred method or approach for determining SP gradients (due to the very irregular spatial distribution of properties in dwarf galaxies), using the median of all determined gradients for each SP property provides a quantity quantity to nominally describe the radial distributions and with its dispersion reflecting the natural uncertainty due to the intrinsic spatial irregularity of the studied properties. 
\end{itemize}

Based on our fiducial gradients and the median of all radial profiles, we conclude the following for the entire MaNDala sample:

\begin{itemize}
    \item The median of the inner and outer gradients of luminosity-weighted ages and stellar metallicities are negative, though with significant dispersion, while their mass-weighted counterparts tend to be flat or even positive, with smaller dispersion. On average, the $Age_{\rm MW}/Age_{\rm LW}$ ratios are high exhibiting large scatter, and increase with radius, whereas the $Z_{\rm MW}/Z_{\rm LW}$ ratios are $<1$ in the center and $\sim 1$ in the outskirts. Additionally, we find negative median gradients for $Dn_{4000}$, though they are shallower than those of $Age_{\rm LW}$.
    
    \item The inner and outer gradients of $\Sigma_{*}$ and $\Sigma_{SFR}$ (both SSP- and $H\alpha$-based) are negative and, on average, shallower than an exponential decline. The SFR distribution is systematically shallower than the stellar mass distribution. We find positive median gradients for the sSFR, both based on the SSPs and $H\alpha$, as well as for the dust attenuation $A_V$. The latter exhibiting the steepest positive gradient among all properties. Regarding the $M/L$ ratio gradients, they are highly uncertain, with values closer to 0; the median inner gradient is slightly positive while the outer is slightly negative.
    
    \item For all the studied SP properties, except for $\Sigma_{*}$ and $\Sigma_{SFR}$, the inner and outer gradients are consistent among them, showing no significant variations between these two broad radial regions. In contrast, for $\Sigma_{*}$ and $\Sigma_{SFR}$, the outer gradients tend to be systematically shallower than the inner ones, i.e., the $\Sigma_{*}$ and $\Sigma_{SFR}$ profiles tend to become flatter in the outer regions. 
    

    \item For those gradients that were compared with those of more massive galaxies ($M_\ast>10^{9}M_\odot$) from the literature, the trends in $Age_{LW}$, $D_{n4000}$, $Z_{MW}$ and $\Sigma_{*}$ appear to be a natural extensions towards the lower masses. In contrast, the inner gradients of  $H_\alpha$-based sSFR exhibit significant scatter, with many negative values, roughly consistent with trends at higher masses. Finally, the dust attenuation gradients in our dwarf sample, are significantly steeper than expected if extrapolated from more massive galaxies.
\end{itemize}

The results obtained here from robust determinations of projected radial profiles and inner/outer gradients for various SP properties of bright dwarf galaxies show that they have a variety of SF histories, with sustained activity and/or recent bursts, especially towards the outskirts. These findings suggest a moderate inside-out formation, with a significant fraction of stars likely forming early from low-metallicity gas, followed by outward migration. Late SF in the inner regions, where the gravitational potential is stronger, appears to originate from chemically-enriched recycled gas In contrast, the outer regions, where $Z_{LW}\sim Z_{MW}$, some of the recycled gas may have been lost, while inflows of low-metallicity gas may have contributed to continued SF in the galaxy. Our results provide constraints on the feedback mechanisms in dwarf galaxies, as well as on the gas inflow/outflow processes, which may be influenced by the environment and merger history.

\begin{acknowledgments}

The authors acknowledge financial support from DGAPA-PAPIIT grant IN106823 and CONAHCYT ``Ciencia de Frontera'' grant G-543.
MCD, ACRO and LPA acknowledge financial support from CONAHCYT "Ciencia de Frontera" grant 320199. ARP acknowledges financial support from DGAPA-PAPIIT grant IN106924. HMHT acknowledges support from CONAHCYT CF-2023-G-1052 grant.
JAVM acknowledges support from the CONAHCYT Postdoctoral program \textit{Estancias Postdoctorales por M\'exico}. HJIM acknowledge financial support from CONAHCYT "Ciencia de Frontera" grant CBF-2023-2024-1418. Some of the calculations for this work were carried out on the HPC clusters Atocatl and Tochtli at LAMOD-UNAM. LAMOD is a collaborative project of DGTIC and IA, ICN, and IQ Institutes at UNAM. This work was also done with the support of human and processing resources by Grid UNAM, which is a collaborative effort promoted by DGTIC, and by the Institutes of Astronomy, Nuclear Sciences as well as Atmospheric Sciences and Climate Change in UNAM. 

Funding for the Sloan Digital Sky 
Survey IV has been provided by the 
Alfred P. Sloan Foundation, the U.S. 
Department of Energy Office of 
Science, and the Participating 
Institutions. 

SDSS-IV acknowledges support and 
resources from the Center for High 
Performance Computing  at the 
University of Utah. The SDSS 
website is www.sdss4.org.

SDSS-IV is managed by the 
Astrophysical Research Consortium 
for the Participating Institutions 
of the SDSS Collaboration including 
the Brazilian Participation Group, 
the Carnegie Institution for Science, 
Carnegie Mellon University, Center for 
Astrophysics | Harvard \& 
Smithsonian, the Chilean Participation 
Group, the French Participation Group, 
Instituto de Astrof\'isica de 
Canarias, The Johns Hopkins 
University, Kavli Institute for the 
Physics and Mathematics of the 
Universe (IPMU) / University of 
Tokyo, the Korean Participation Group, 
Lawrence Berkeley National Laboratory, 
Leibniz Institut f\"ur Astrophysik 
Potsdam (AIP),  Max-Planck-Institut 
f\"ur Astronomie (MPIA Heidelberg), 
Max-Planck-Institut f\"ur 
Astrophysik (MPA Garching), 
Max-Planck-Institut f\"ur 
Extraterrestrische Physik (MPE), 
National Astronomical Observatories of 
China, New Mexico State University, 
New York University, University of 
Notre Dame, Observat\'ario 
Nacional / MCTI, The Ohio State 
University, Pennsylvania State 
University, Shanghai 
Astronomical Observatory, United 
Kingdom Participation Group, 
Universidad Nacional Aut\'onoma 
de M\'exico, University of Arizona, 
University of Colorado Boulder, 
University of Oxford, University of 
Portsmouth, University of Utah, 
University of Virginia, University 
of Washington, University of 
Wisconsin, Vanderbilt University, 
and Yale University.

This project makes use of the MaNGA-Pipe3D dataproducts. The dataproducts used in this project got benefit of computational and human resources provided by the LAMOD- UNAM project through the clusters Atocatl and Tochtli. LAMOD is a collaborative effort between the IA, ICN, and IQ institutes at UNAM and DGAPA UNAM grants PAPIIT IG101620 and IG10122  We thank the IA-UNAM MaNGA team for creating this catalog, and the CONACyT-180125 project for supporting them. 

\end{acknowledgments}

%






\appendix

\section{Surface Density of Star Formation Rates}\label{Append:SFRs}

As stated in Section \ref{Sec:data}, the only properties for which the \textsc{Pipe3D} VAC does not provide maps are the $\Sigma_{SFRs}$, nor the ones estimated with the SSPs nor with the $H_{\alpha}$ line. For these cases we have independently estimated them.

To estimate the $\Sigma_{SFRs}$ using the SSPs, we integrate the stellar surface density of the populations corresponding to the last 32 Myr ($\Sigma_{SFR,ssp,32}$). To do this we follow the methodology described by \citet{Sanchez2022} in their Section 5.1.1. In particular, in order to calculate the $\Sigma{*}$ in the desired time interval we make use of the maps of V-band flux, $A_V$ and light fraction/weight of the corresponding SSPs provided by the \textsc{Pipe3D} VAC\footnote{Beside these maps, the  $M/L$ values for each SSP are required. These were obtained via private communication with H.J. Ibarra-Medel. A constant value of the wavelength range of 3500 $\AA$ has to be multiplied to these values, following the implementation provided in the public version of the \textsc{pyPipe3D} Code.}. Finally in order to obtain a map of the SFR density we apply the following equation:

\begin{equation}
 \Sigma_{SFR,ssp,t} = \frac{\Delta\Sigma_{*,t}}{\Delta{t}},
\end{equation}

as stated in Equation 8 of \citet{Sanchez2022}.

In the case of the $H_{\alpha}$ tracer, we make use of the flux intensity, and Equivalent Width (EW) maps of the aforementioned emission line provided by the \textsc{Pipe3D} VAC. Before calculating the SFR density with this tracer ($\Sigma_{SFR,H\alpha}$), we exclude all the pixels in the flux map, regardless of their galactocentric distance, that are not consistent with an ionizing source consistent with the SF activity, using our criteria based on the BPT diagnostic diagram \citep{Baldwin81}, according the Kewley demarcation line \citep{Kewley01} and the EW value of the line \citep[see details of our ionization classification criteria in: ][]{Cano-Diaz+2019, Cano-Diaz+2016}. Briefly, the usage of the demarcation limit in the BPT diagram allows to separate ionization sources from central activity and star formation, while the EW value allows to avoid confussion beween star formation and ionization from old stars. Once left with the pixels that are consistent with the SF activity, we simply apply the linear transformation provided by \citet{Kennicutt98}, in order to obtain the SFR values from the $H\alpha$ fluxes, as follows:

\begin{equation}
SFR(M_{\odot}\text{ year}^{-1})=7.9\times 10^{-42} L(H\alpha) (\text{ergs s}^{-1}),
\end{equation} which corresponds to Equation 2 of the aforementioned  reference. For this calculation the extinction is being considered. 

Finally, to obtain the sSFR for both tracers we use the total $\Sigma_{*}$ (i.e. the final stellar density integrated over the whole time interval allowed by \textsc{pyPipe3D}), which is the one provided in the $\Sigma_{*}$ by the \textsc{Pipe3D} VAC.

\section{Full set of profiles derived with four methods}\label{Append:AllProfiles}

\begin{figure}
  \centering
    \includegraphics[width=0.95\textwidth,height=0.44\textheight]{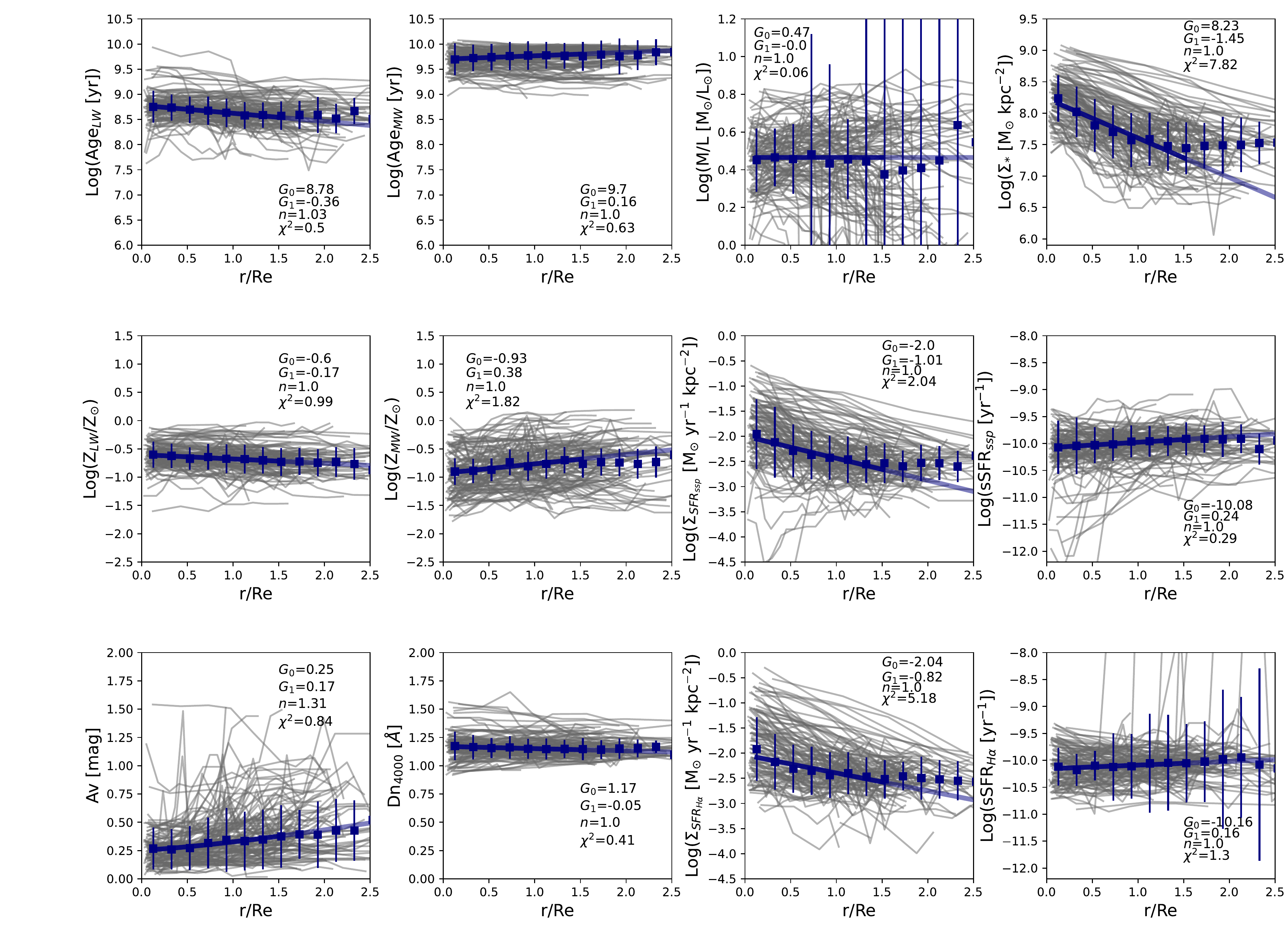}
  \caption{Full set of profiles for the MaNDala working sample, derived with the collapsed median method described in Section \ref{Sec:ProfilesMedians}. Grey lines represent the profiles, while the blue squares show a binned median profile. Their associated error bars correspond to $1\sigma$ within each radial bin. We also show a fit to the binned median profile up to 1.5 $R_{e}$ with a blue line, while the extended transparent blue line shows its extrapolation. The details of the fit are inside the plot areas, where the coefficients correspond to those of Equation \ref{eq:general_fit_grads} and to the $\chi^{2}$ of the fit.}\label{Fig:all_profiles_medianprof}
\end{figure}

\begin{figure}
  \centering
    \includegraphics[width=0.95\textwidth,height=0.44\textheight]{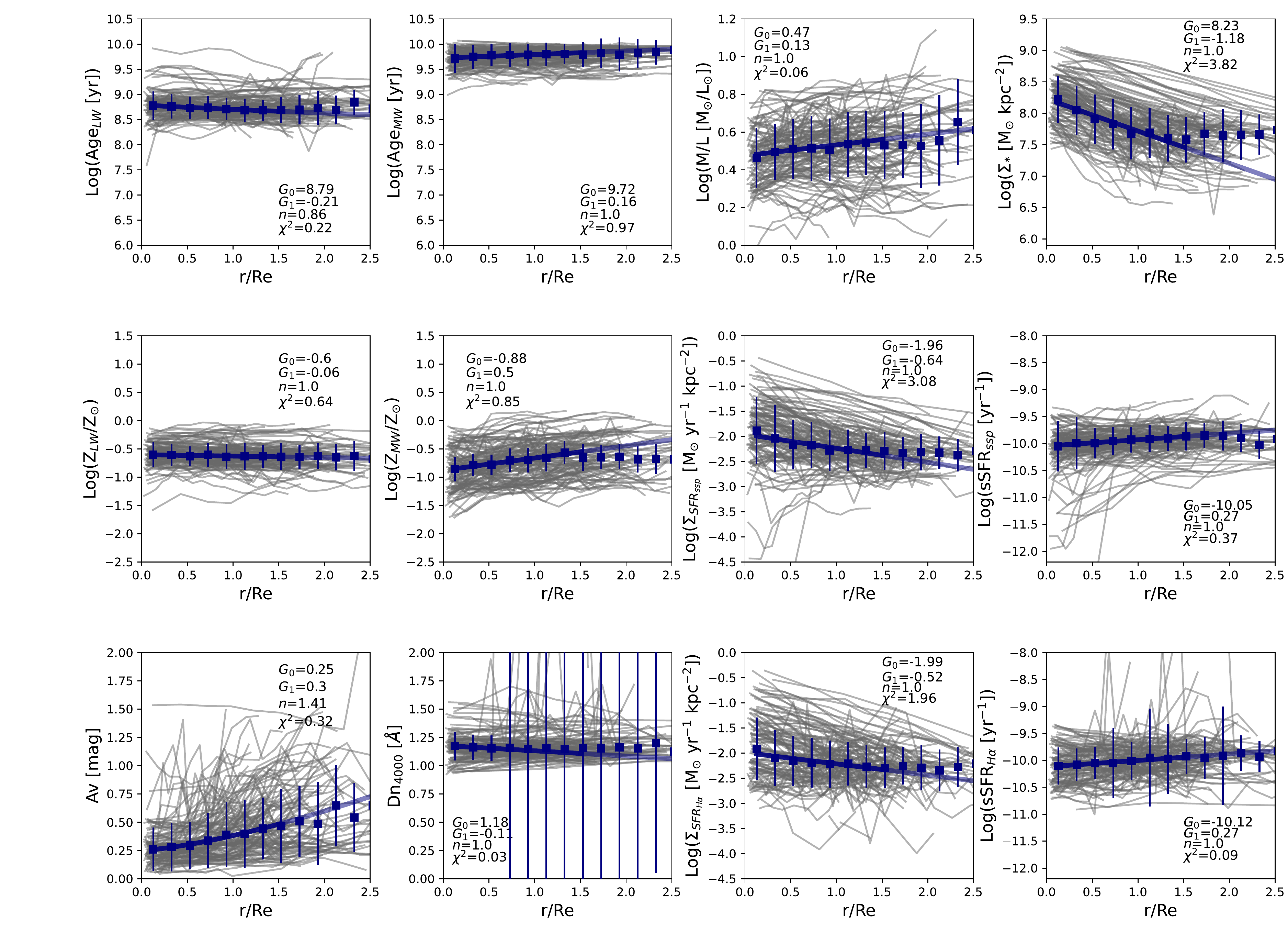}
  \caption{Same as Figure \ref{Fig:all_profiles_medianprof}, but for the collapsed integrated method described in Section \ref{Sec:ProfilesIntegrated}.}\label{Fig:all_profiles_integratedprof}
\end{figure}

\begin{figure}
  \centering
    \includegraphics[width=0.95\textwidth,height=0.44\textheight]{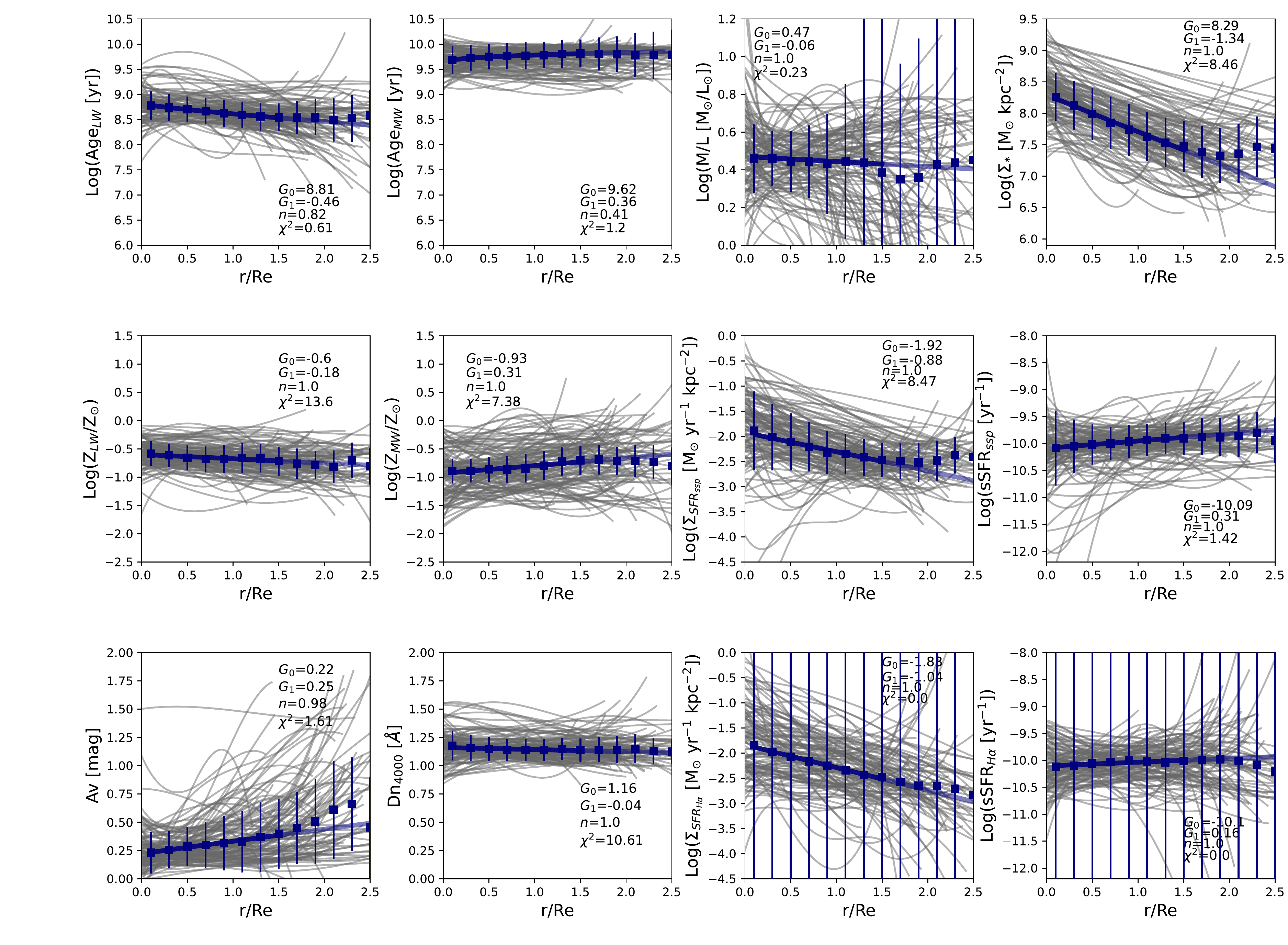}
  \caption{Same as Figure \ref{Fig:all_profiles_medianprof}, but for the non-linear spline method described in Section \ref{Sec:ProfilesSplines}.}\label{Fig:all_profiles_splineprof}
\end{figure}

\begin{figure}
  \centering
    \includegraphics[width=0.95\textwidth,height=0.44\textheight]{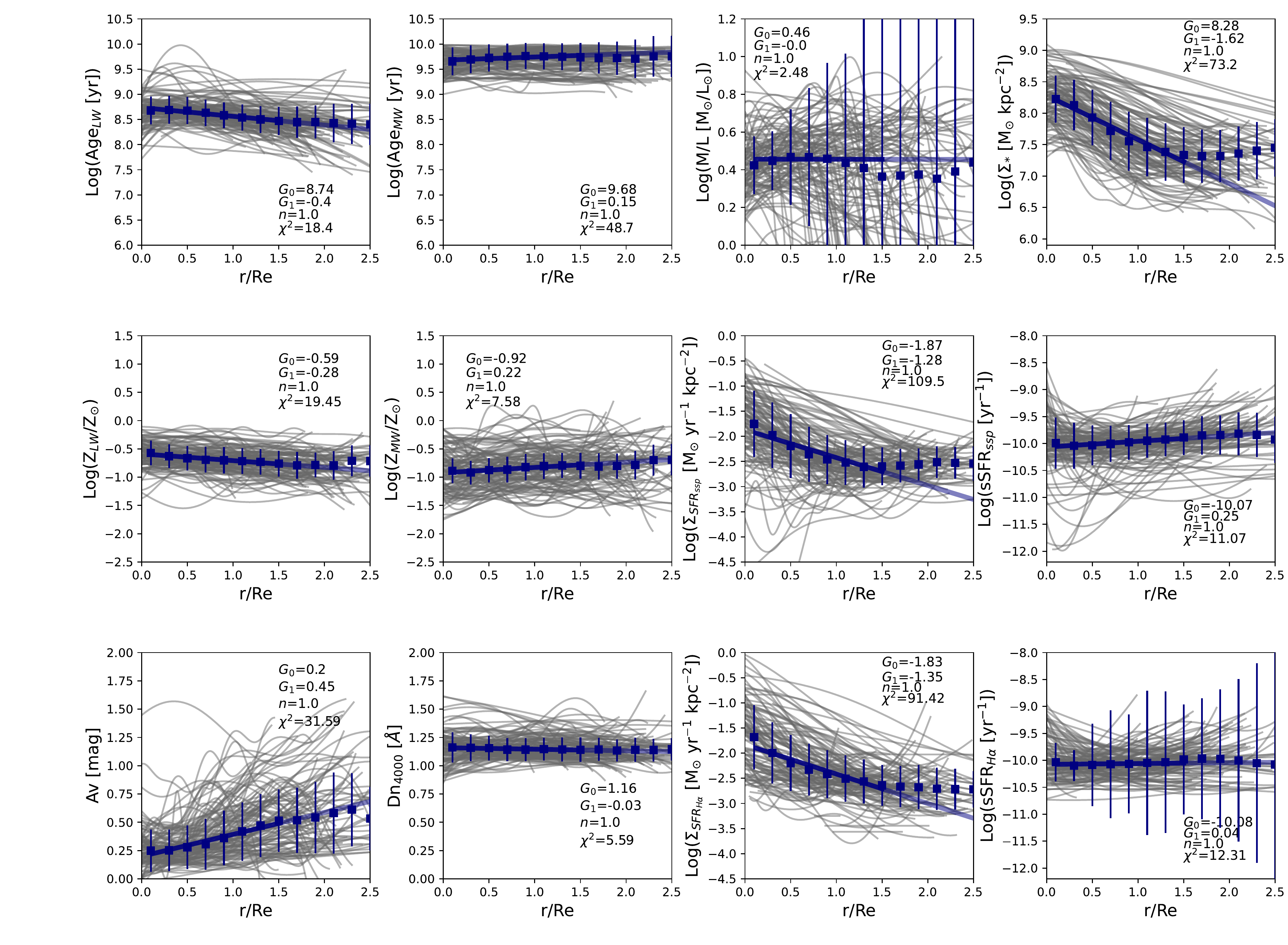}
  \caption{Same as Figure \ref{Fig:all_profiles_medianprof}, but for the non-linear GPR method described in Section \ref{Sec:ProfilesGPR}.}\label{Fig:all_profiles_gprprof}
\end{figure}

In this Appendix we show the profiles for the full MaNDala working sample, derived with the four described methods in Section \ref{Sec:RadialProfilesMethods}. In Figures \ref{Fig:all_profiles_medianprof}, \ref{Fig:all_profiles_integratedprof}, \ref{Fig:all_profiles_splineprof}, \ref{Fig:all_profiles_gprprof}, the profiles derived with the collapsed median, collapsed integrated, non-lineal spline and non-linear-GPR methods are shown respectively. 

\section{manga-7815-6101 profiles}\label{Append:ExampleProfiles}

\begin{figure*}
\centering
    \subfloat{%
	   \includegraphics[width=0.4\textwidth,height=0.4\textheight]{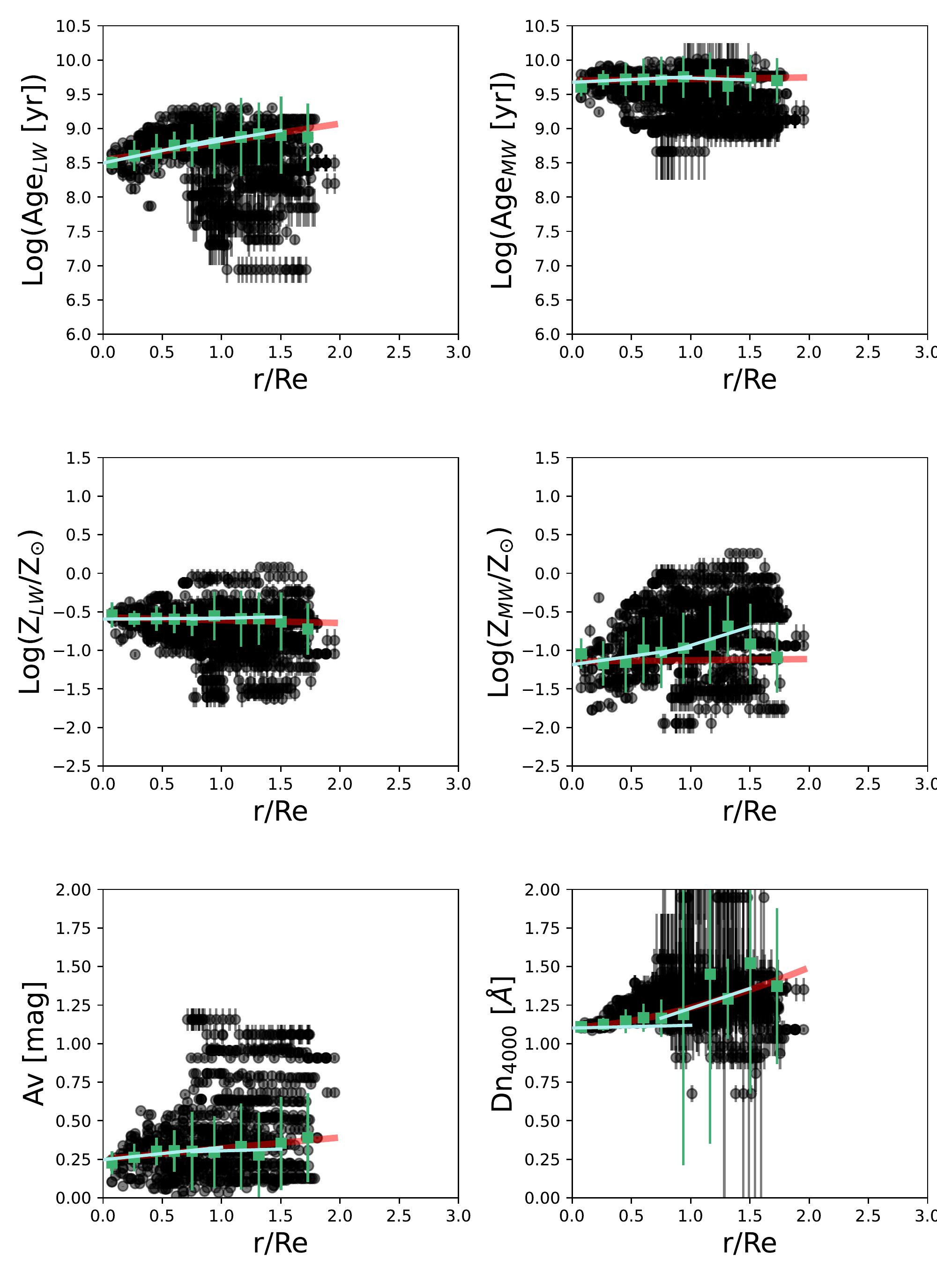}
    }\qquad
    \subfloat{%
	   \includegraphics[width=0.4\textwidth,height=0.4\textheight]{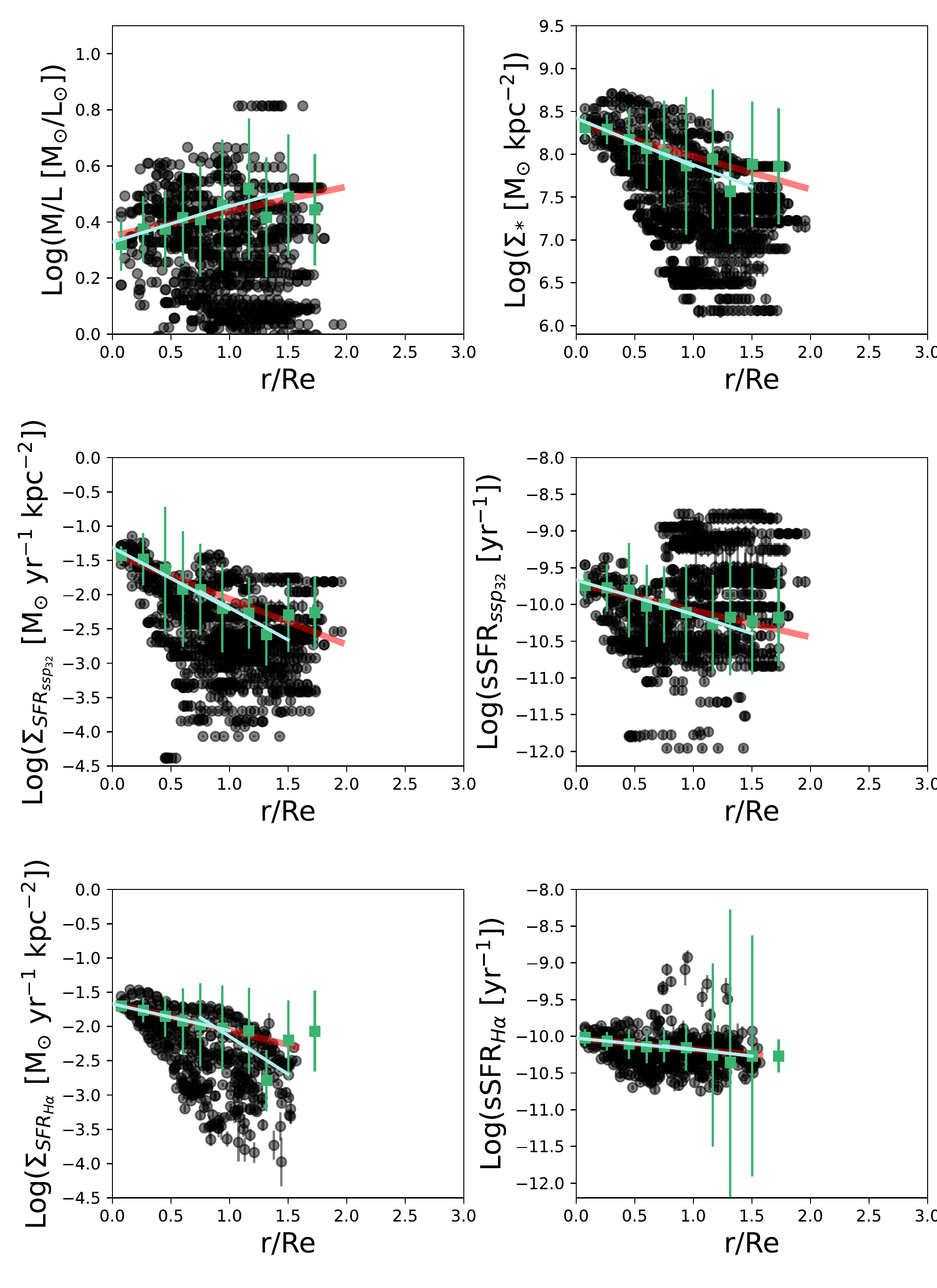}
    }\qquad
\caption{Spatially resolved profiles for all the studied properties of the manga-7815-6101 galaxy, shown in black circles. On top with green squares the collapsed integrated profiles are displayed, while the solid lines show the fit performed to them in the two radial ranges (0 -1 $R_{e}$ and 0.75 - 1.5 $R_{e}$). The red solid lines represent the generalized fit performed over the collapsed integrated profiles.}
\label{Fig:apena_7815-6101_IntProfiles} 
\end{figure*}

In this Appendix we show the complete set of the profiles for the galaxy manga-7815-6101. In all figures the black circles show the spatially resolved profiles. In Figures \ref{Fig:apena_7815-6101_IntProfiles} and \ref{Fig:apena_7815-6101_IntProfiles} we show on top of the spatially resolved profiles, the median and integrated ones (green squares) respectively, and the fits done to them (blue solid lines).

\begin{figure*}
\centering
    \subfloat{%
	   \includegraphics[width=0.4\textwidth,height=0.4\textheight]{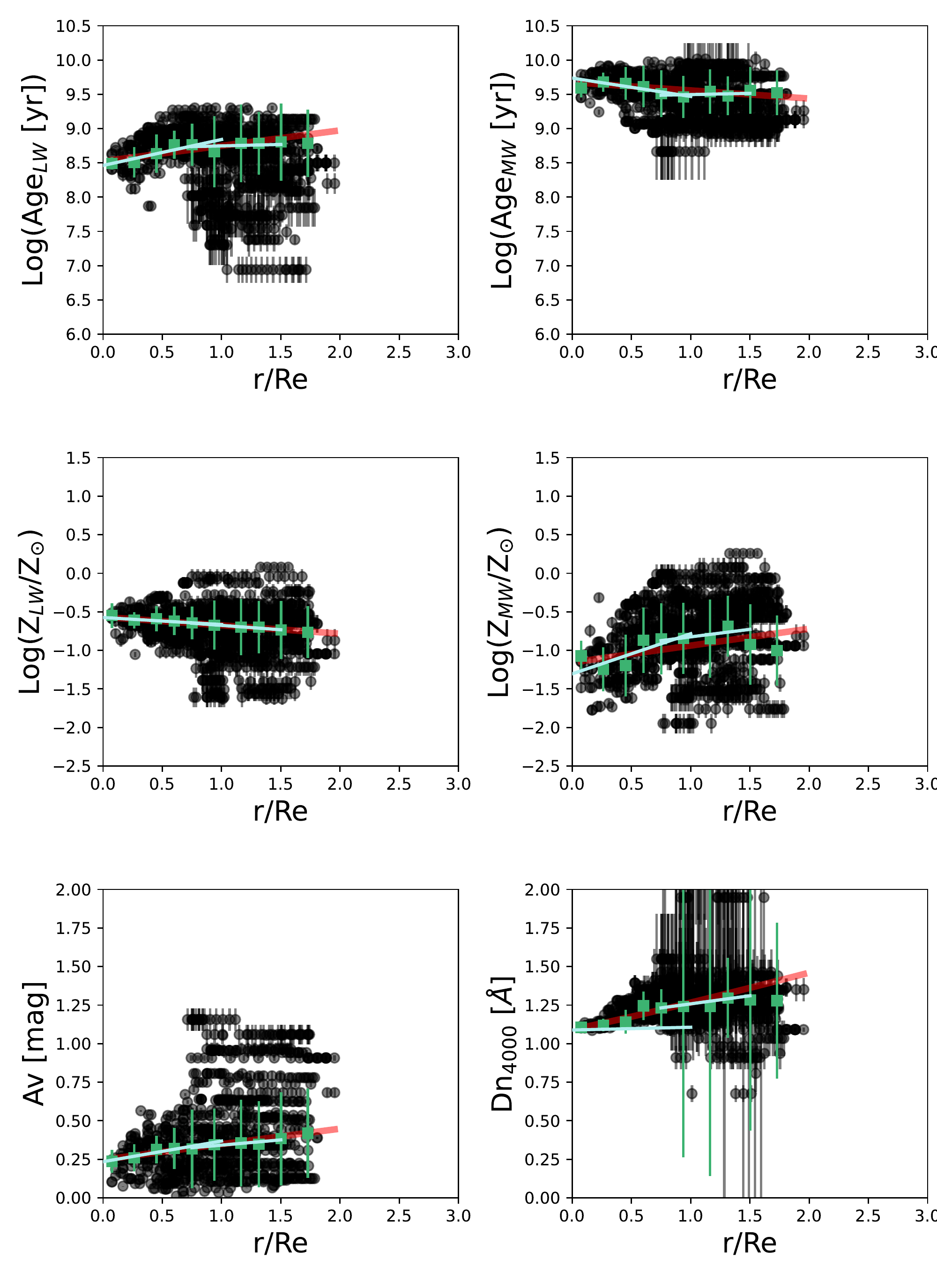}
    }\qquad
    \subfloat{%
	   \includegraphics[width=0.4\textwidth,height=0.4\textheight]{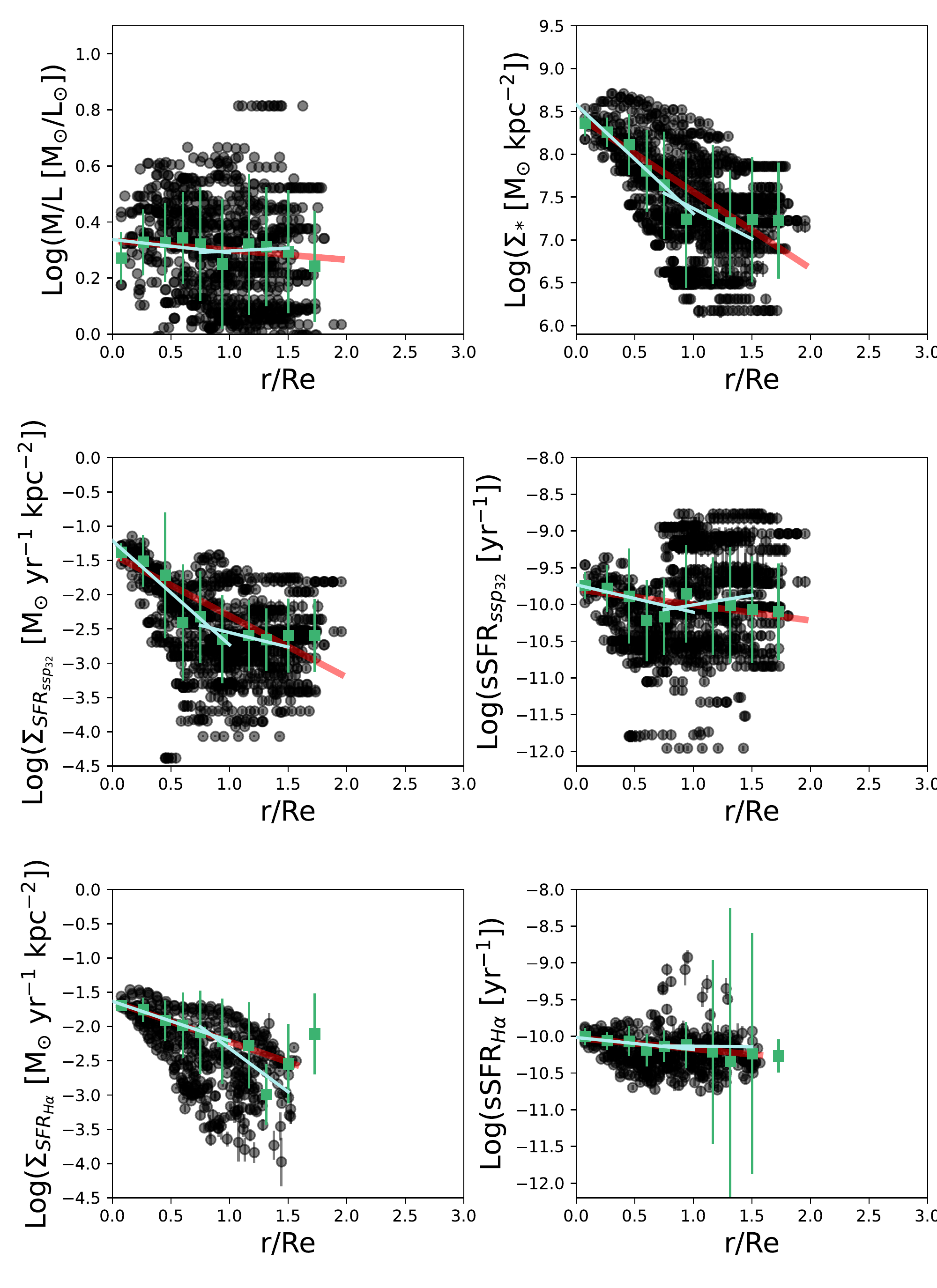}
    }\qquad
\caption{Spatially resolved profiles for all the studied properties of the manga-7815-6101 galaxy, shown in black circles. On top with green squares the collapsed median profiles are displayed, while the solid lines show the fit performed to them in the two radial ranges (0 -1 $R_{e}$ and 0.75 - 1.5 $R_{e}$. The red solid lines represent the generalized fit performed over the collapsed median profiles.)}
\label{Fig:apena_7815-6101_MedProfiles}
\end{figure*}

\begin{figure*}
\centering
    \subfloat{%
	   \includegraphics[width=0.4\textwidth,height=0.4\textheight]{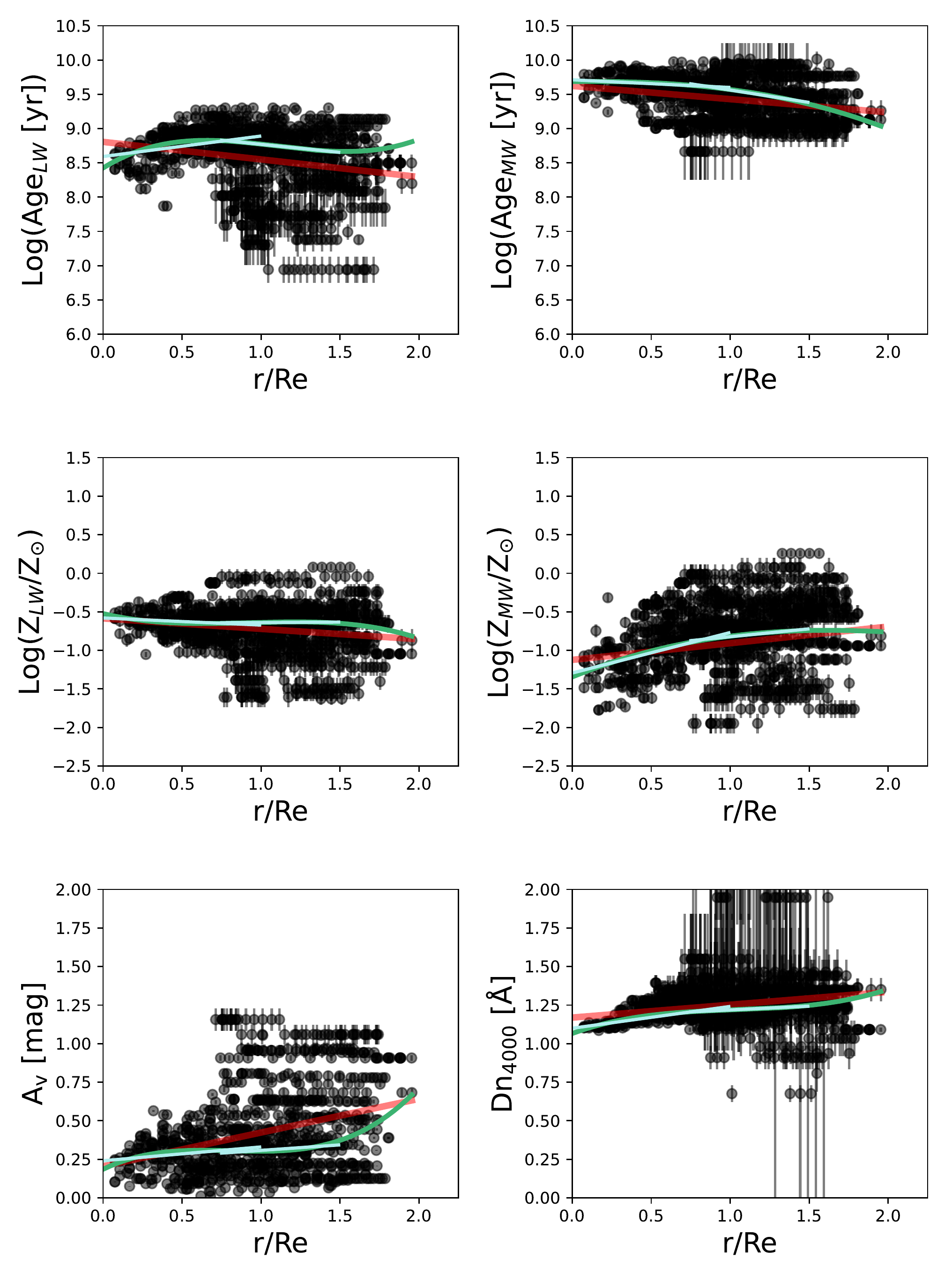}
    }\qquad
    \subfloat{%
	   \includegraphics[width=0.4\textwidth,height=0.4\textheight]{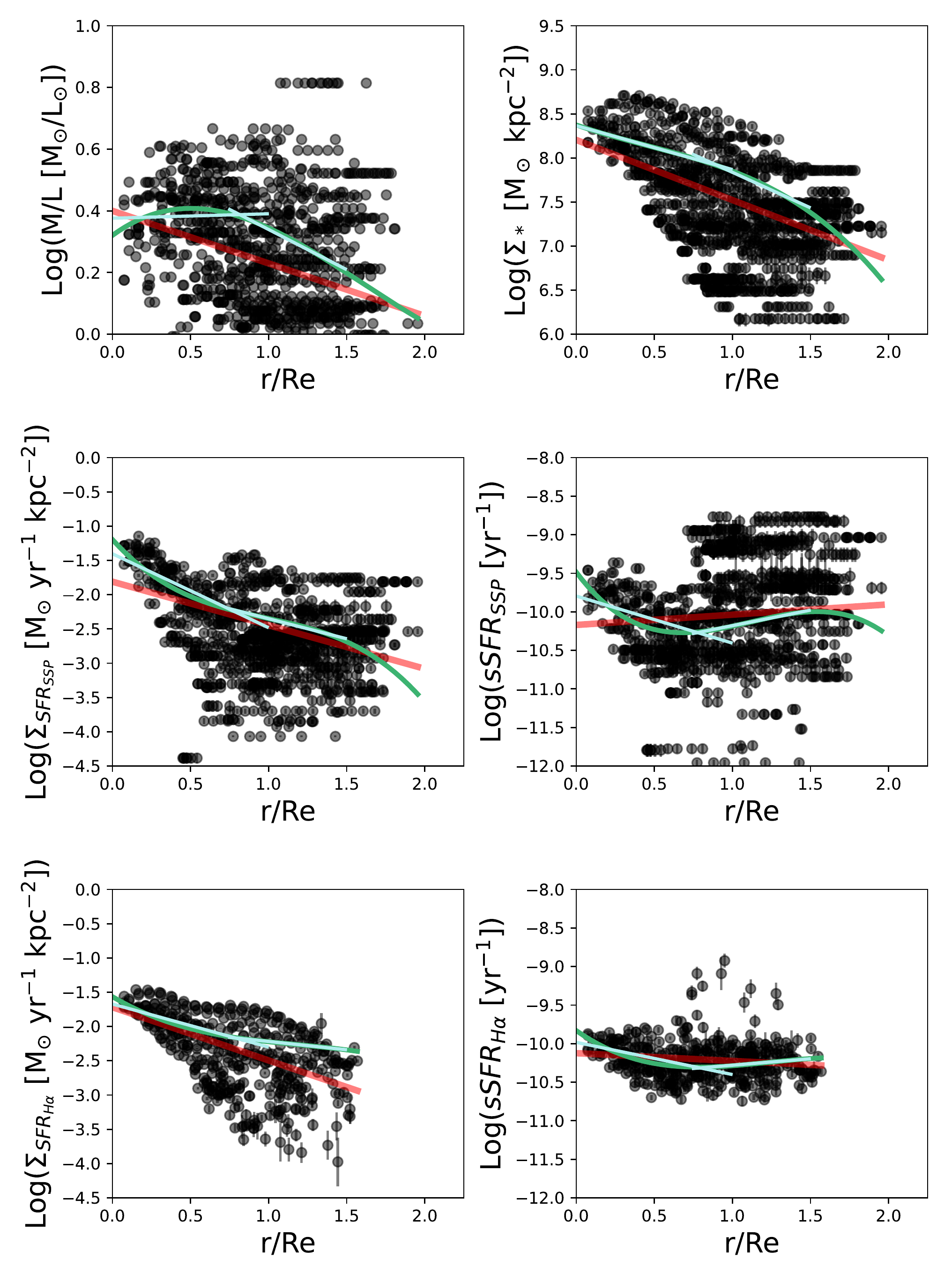}
    }\qquad
\caption{As in the two previous figures, the black points in all panels show the spatially resolved profiles of all the studied galaxy properties for manga-7815-6101. The green line is the non linear fit delivered by the Spline method. The blue solid lines represent the linear fits performed to the mentioned Spline fits at the two adopted radial ranges. The red solid lines represent the generalized fit performed over the non-linear Spline profiles.}
\label{Fig:apena_7815-6101_SplineProfiles}
\end{figure*}

\begin{figure*}
\centering
    \subfloat{%
	   \includegraphics[width=0.4\textwidth,height=0.4\textheight]{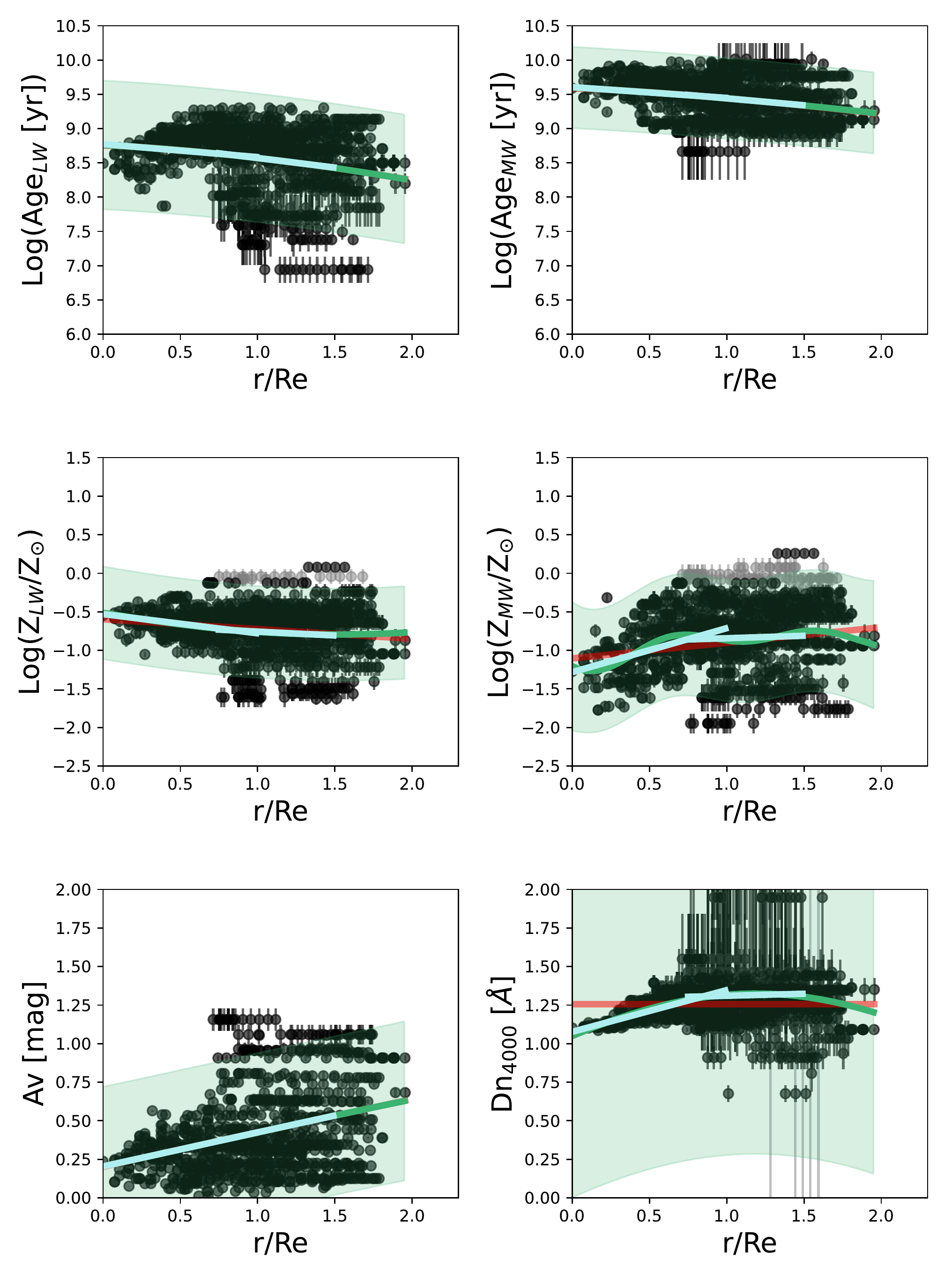}
    }\qquad
    \subfloat{%
	   \includegraphics[width=0.4\textwidth,height=0.4\textheight]{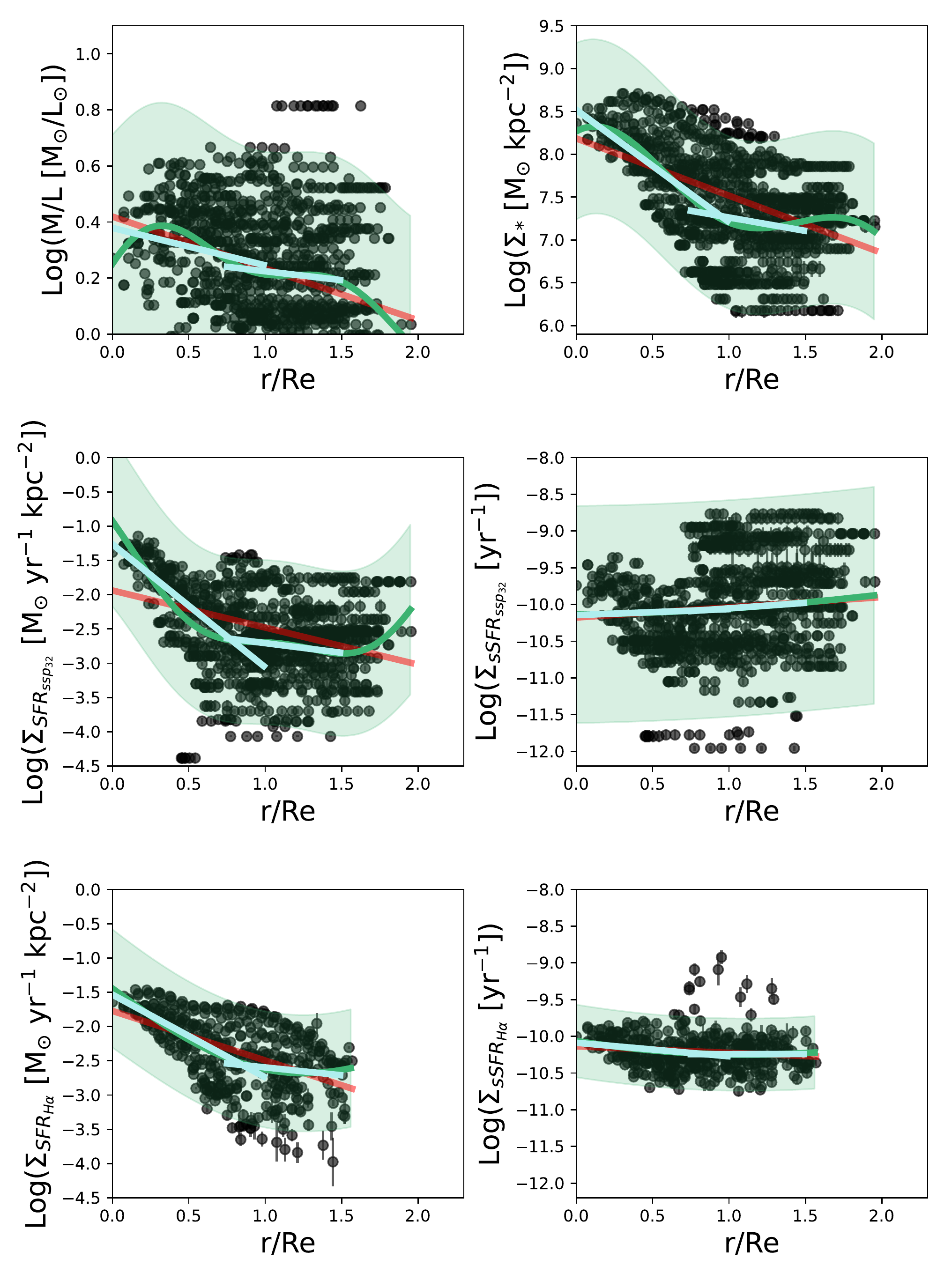}
    }\qquad
\caption{As in the three previous figures, the black points in all panels show the spatially resolved profiles of all the studied galaxy properties for manga-7815-6101. The green line is the probability distribution delivered by the GPR method, while the green shadow is its correspondent 95\% confidence interval. The blue solid lines represent the linear fits performed to the mentioned probability distribution at the two adopted radial ranges. The red solid lines represent the generalized fit performed over the non-linear GPR profiles.}
\label{Fig:apena_7815-6101_GPRProfiles}
\end{figure*}

In Figure \ref{Fig:apena_7815-6101_SplineProfiles}, the green lines shows the non-linear fit performed by the double Spline method, on top in blue, we show the linear fit done over the Splines functional form in the two radial ranges. Meanwhile in Figure \ref{Fig:apena_7815-6101_GPRProfiles}, the green lines represent the smoothed probability distribution delivered by the GPR method, while the green shadow is the 95\% confidence interval. The blue lines show the linear fits performed to the probability distributions for the two radial ranges.

\section{Online table description}\label{Append:OnlineTable}

A machine readable table is provided as online material of this work, in which the entire set of gradients are provided for the MaNDala sample. In the table one can find the gradients obtained using the four methods to derive the radial profiles: $i)$ the collapsed median profiles, $ii)$ the collapsed integrated profiles, $iii)$ the non-linear spline profiles and  $iv)$ the non-linear GPR profiles, in combination with the four methods to derive the gradients: $i)$ the slope of linear fits, $ii)$ the general fits derivative, $iii)$ the difference between two points from the general fit and $iv)$ the difference between two points from the radial profiles. All of these methods are described in Sections \ref{Sec:RadialProfilesMethods} and \ref{Sec:grad_methods}. 

This is a wide table with 721 columns, for which in Table \ref{Table:OnlineTable} we are only describing the columns that are necessary to understand the syntax of the column names. The interested reader should notice that we are only releasing the inner gradients (0-$R_{e}$) without the contribution of the PSF, since the outer ones (0.75-1.5 $R_{e}$) are unaffected by it. In the case of the gradients derived by the difference of two points are without errors for the reasons explained in Section \ref{Sec:diff_2points}. This is also the case for the gradients derived from the general fit (see Section \ref{Sec:general_fits} for the details). Finally, all the cells that have the value -9999.0 are to be discarded due to the failure of the fits (see details in Sections \ref{Sec:grad_methods}).

The Astronomical Journal will provide the permanent version of the online table once this manuscript is published. Meanwhile a temporal version of the table is available in a Drive\footnote{\url{https://drive.google.com/file/d/1btmnDA-17NwIEyZQrAveyXTCrkioGDT6/view?usp=sharing}}.

\begin{longtable*} {c c c}
\hline
\hline
 Column Index & Column Name & Column Description   \\
\hline
1 & Plateifu & Unique plate and IFU indicator for MaNGA galaxies \\
2 & LogAgeLW\_LinFitMed0to1 & Collapsed median prof. + slope of linear fit $\nabla_{in}$Log(Age$_{LW}$) \\
3 & LogAgeLW\_LinFitMed0to1\_err & Collapsed median prof. + slope of linear fit $\nabla_{in}$Log(Age$_{LW}$) Error \\
4 & LogAgeLW\_LinFitMed075to15 & Collapsed median prof. + slope of linear fit $\nabla_{out}$Log(Age$_{LW}$) \\
5 & LogAgeLW\_LinFitMed075to15\_err & Collapsed median prof. + slope of linear fit $\nabla_{out}$Log(Age$_{LW}$) Error \\
6 & LogAgeMW\_LinFitMed0to1 & Collapsed median prof. + slope of linear fit $\nabla_{in}$Log(Age$_{MW}$) \\
... & ... & ... \\
10 & LogMetLW\_LinFitMed0to1 & Collapsed median prof. + slope of linear fit $\nabla_{in}$Log($Z_{LW}$) \\ 
... & ... & ... \\
14 & LogMetMW\_LinFitMed0to1 & Collapsed median prof. + slope of linear fit $\nabla_{in}$Log($Z_{MW}$) \\ 
... & ... & ... \\
18 & $A_v$\_LinFitMed0to1 & Collapsed median prof. + slope of linear fit $\nabla_{in}$$A_V$ \\ 
... & ... & ... \\
22 & Dn4000\_LinFitMed0to1 & Collapsed median prof. + slope of linear fit $\nabla_{in}D_{n4000}$ \\ 
... & ... & ... \\
26 & LogML\_LinFitMed0to1 &  Collapsed median prof. + slope of linear fit $\nabla_{in}$Log( $M/L$) \\ 
... & ... & ... \\
30 & LogSigmaSte\_LinFitMed0to1 & Collapsed median prof. + slope of linear fit $\nabla_{in}$Log($\Sigma_{*}$) \\ 
... & ... & ... \\
34 & LogSigmaSFRssp\_LinFitMed0to1 & Collapsed median prof. + slope of linear fit $\nabla_{in}$Log($\Sigma_{SFR_{SSP}}$) \\ 
... & ... & ... \\
38 & LogsSFRssp\_LinFitMed0to1 & Collapsed median prof. + slope of linear fit $\nabla_{in}$sSFR$_{SSP}$ \\ 
... & ... & ... \\
42 & LogSigmaSFRha\_LinFitMed0to1 & Collapsed median prof. + slope of linear fit $\nabla_{in}\Sigma_{SFR_{H\alpha}}$ \\ 
... & ... & ... \\
46 & LogsSFRha\_LinFitMed0to1 & Collapsed median prof. + slope of linear fit $\nabla_{in}$sSFR$_{H\alpha}$ \\ 
... & ... & ... \\
50 & LogAgeLW\_LinFitInt0to1 & Collapsed integrated prof. + slope of linear fit $\nabla_{in}$Log(Age$_{LW}$) \\
... & ... & ... \\
98 & LogAgeLW\_LinFitSpl0to1 & Non-linear spline prof. + slope of linear fit $\nabla_{in}$Log(Age$_{LW}$) \\
... & ... & ... \\
146 & LogAgeLW\_LinFitGPR0to1 & Non-linear GPR prof. + slope of linear fit $\nabla_{in}$Log(Age$_{LW}$) \\
... & ... & ... \\
194 & LogAgeLW\_LinFitMed0to1\_noPSF & {\shortstack{Collapsed median prof. + slope of linear fit \\ $\nabla_{in}$Log(Age$_{LW}$) without PSF}} \\
195 & LogAgeLW\_LinFitMed0to1\_err\_noPSF & {\shortstack{Collapsed median prof. + slope of linear fit \\ $\nabla_{in}$Log(Age$_{LW}$) Error without PSF}} \\
... & ... & ... \\
218 & LogAgeLW\_LinFitInt0to1\_noPSF & {\shortstack{Collapsed integrated prof. + slope of linear fit \\ $\nabla_{in}$Log(Age$_{LW}$) without PSF}} \\
... & ... & ... \\
242 & LogAgeLW\_LinFitSpl0to1\_noPSF & {\shortstack{Non-linear spline prof. + slope of linear fit \\ $\nabla_{in}$Log(Age$_{LW}$) without PSF}} \\
... & ... & ... \\
266 & LogAgeLW\_LinFitGPR0to1\_noPSF & {\shortstack{Non-linear GPR prof. + slope of linear fit \\ $\nabla_{in}$Log(Age$_{LW}$) without PSF}} \\
... & ... & ... \\
290 & LogAgeLW\_DiffMed0to1 & {\shortstack{Collapsed median prof. + diff. between two points \\ $\nabla_{in}$Log(Age$_{LW}$)}} \\
... & ... & ... \\
314 & LogAgeLW\_DiffInt0to1 & {\shortstack{Collapsed integrated prof. + diff. between two points \\ $\nabla_{in}$Log(Age$_{LW}$)}} \\
... & ... & ... \\
338 & LogAgeLW\_DiffSpl0to1 & {\shortstack{Non-linear spline prof. + diff. between two points \\ $\nabla_{in}$Log(Age$_{LW}$)}} \\
... & ... & ... \\
362 & LogAgeLW\_DiffGPR0to1 & {\shortstack{Non-linear GPR prof. + diff. between two points \\ $\nabla_{in}$Log(Age$_{LW}$)}} \\
... & ... & ... \\
386 & LogAgeLW\_DiffMed0to1\_noPSF & {\shortstack{Collapsed median prof. + diff. between two points \\ $\nabla_{in}$Log(Age$_{LW}$) without PSF}} \\
... & ... & ... \\
398 & LogAgeLW\_DiffInt0to1\_noPSF &  {\shortstack{Collapsed integrated prof. + diff. between two points \\ $\nabla_{in}$Log(Age$_{LW}$) without PSF}} \\
... & ... & ... \\
410 & LogAgeLW\_DiffSpl0to1\_noPSF &  {\shortstack{Non-linear spline prof. + diff. between two points \\ $\nabla_{in}$Log(Age$_{LW}$) without PSF}} \\
... & ... & ... \\
422 & LogAgeLW\_DiffGPR0to1\_noPSF &  {\shortstack{Non-linear GPR prof. + diff. between two points \\ $\nabla_{in}$Log(Age$_{LW}$) without PSF}} \\
... & ... & ... \\
434 & LogAgeLW\_GenFit\_DiffMed0to1 &  {\shortstack{Collapsed median prof. + diff. between two points\\from general fit $\nabla_{in}$Log(Age$_{LW}$)}} \\
... & ... & ... \\
458 & LogAgeLW\_GenFit\_DiffInt0to1 &  {\shortstack{Collapsed integrated prof. + diff. between two points\\from general fit$\nabla_{in}$Log(Age$_{LW}$)}} \\
... & ... & ... \\
482 & LogAgeLW\_GenFit\_DiffSpl0to1 &  {\shortstack{Non-linear spline prof. + diff. between two points\\from general fit$\nabla_{in}$Log(Age$_{LW}$)}} \\
... & ... & ... \\
506 & LogAgeLW\_GenFit\_DiffGPR0to1 &  {\shortstack{Non-linear GPR prof. + diff. between two points\\from general fit$\nabla_{in}$Log(Age$_{LW}$)}} \\
... & ... & ... \\
530 & LogAgeLW\_GenFit\_DerMed0to1 &  {\shortstack{Collapsed median prof. + derivative from general fit\\$\nabla_{in}$Log(Age$_{LW}$)}} \\
... & ... & ... \\
554 & LogAgeLW\_GenFit\_DerInt0to1 &  {\shortstack{Collapsed integrated prof. + derivative from general fit\\$\nabla_{in}$Log(Age$_{LW}$)}} \\
... & ... & ... \\
578 & LogAgeLW\_GenFit\_DerSpl0to1 &  {\shortstack{Non-linear spline prof. + derivative from general fit\\$\nabla_{in}$Log(Age$_{LW}$)}} \\
... & ... & ... \\
602 & LogAgeLW\_GenFit\_DerGPR0to1 &  {\shortstack{Non-linear GPR prof. + derivative from general fit\\$\nabla_{in}$Log(Age$_{LW}$)}} \\
... & ... & ... \\
626 & LogAgeLW\_GenFit\_DiffMed0to1\_noPSF &  {\shortstack{Collapsed median prof. + diff. between two points\\from general fit $\nabla_{in}$Log(Age$_{LW}$) without PSF}} \\
... & ... & ... \\
638 & LogAgeLW\_GenFit\_DiffInt0to1\_noPSF &  {\shortstack{Collapsed integrated prof. + diff. between two points \\ from general fit $\nabla_{in}$Log(Age$_{LW}$) without PSF}} \\
... & ... & ... \\
650 & LogAgeLW\_GenFit\_DiffSpl0to1\_noPSF &  {\shortstack{Non-linear spline prof. + diff. between two points\\ from general fit $\nabla_{in}$Log(Age$_{LW}$) without PSF}} \\
... & ... & ... \\
662 & LogAgeLW\_GenFit\_DiffGPR0to1\_noPSF &  {\shortstack{Non-linear GPR prof. + diff. between two points \\ from general fit $\nabla_{in}$Log(Age$_{LW}$) without PSF}} \\
... & ... & ... \\
674 & LogAgeLW\_GenFit\_DerMed0to1\_noPSF &  {\shortstack{Collapsed median prof. + derivative from general fit\\$\nabla_{in}$Log(Age$_{LW}$) without PSF}} \\
... & ... & ... \\
686 & LogAgeLW\_GenFit\_DerInt0to1\_noPSF &  {\shortstack{Collapsed integrated prof. + derivative from general fit\\$\nabla_{in}$Log(Age$_{LW}$) without PSF}} \\
... & ... & ... \\
698 & LogAgeLW\_GenFit\_DerSpl0to1\_noPSF &  {\shortstack{Non-linear spline prof. + derivative from general fit\\$\nabla_{in}$Log(Age$_{LW}$) without PSF}} \\
... & ... & ... \\
710 & LogAgeLW\_GenFit\_DerGPR0to1\_noPSF &  {\shortstack{Non-linear GPR prof. + derivative from general fit\\$\nabla_{in}$Log(Age$_{LW}$) without PSF}} \\

\hline  
\hline  
\caption{Outline of the columns of the online table with the entire set of gradients for the MaNDala sample.}
\label{Table:OnlineTable}
\end{longtable*}


\bibliography{references}{}
\bibliographystyle{aasjournal}



\end{document}